%% file: Anisotropy.tex
\documentclass[aps,12pt,longbibliography]{revtex4-1}

\usepackage[lofdepth,lotdepth]{subfig}
\usepackage{graphicx}
\usepackage{algorithm}
\usepackage{algpseudocode}
\usepackage{amssymb}
\usepackage{gensymb}

\usepackage[latin1]{inputenc}
\usepackage{tikz}
\usetikzlibrary{shapes,arrows,positioning,calc}

\input commands.tex

\tikzstyle{startstop} = [rectangle, rounded corners, minimum width=3cm, minimum height=1cm,text centered, draw=black, fill=red!30]
\tikzstyle{io} = [trapezium, trapezium left angle=70, trapezium right angle=110, minimum width=3cm, minimum height=1cm, text centered, draw=black, fill=blue!30]
\tikzstyle{process} = [rectangle, minimum width=3cm, minimum height=1cm, align=left, draw=black, fill=orange!30]
\tikzstyle{empty} = [rectangle, align=left]
\tikzstyle{decision} = [diamond, minimum width=3cm, minimum height=1cm, align=center, draw=black, fill=green!30]
\tikzstyle{arrow} = [thick,->,>=stealth]

\begin{document}

\title{3D modelling of macroscopic force-free effects in superconducting thin films and rectangular prisms.}
\markboth{Modeling of screening currents in coated conductor magnets ...}{}

\author{M. Kapolka, E. Pardo}%

\address{Institute of Electrical Engineering, Slovak Academy of Sciences, Dubravska 9, 84104 Bratislava, Slovakia}


\date{\today}

\begin{abstract}

{When the magnetic field has a parallel component to the current density $\vJ$ there appear force-free effects due to flux cutting and crossing. This results in an anisotropic $\vE(\vJ)$ relation, being $\vE$ the electric field. Understanding force-free effects is interesting not only for the design of superconducting power and magnet applications but also for material characterization.}

{This work develops and applies a fast and accurate computer modeling method based on a variational approach that} can handle force-free anisotropic $\vE(\vJ)$ relations and perform {fully three dimensional (3D)} calculations. 
{We present a} systematic study of force-free effects in rectangular thin films and prisms with {several finite} thickness{es under applied magnetic fields with arbitrary angle $\theta$ with the surface}. The results are compared with the same situation with isotropic $\vE(\vJ)$ relation.

The thin film {situation} shows gradual {critical} current density penetration {and} {a general} increase of the magnitude of the magnetization with the angle $\theta$ {but a minimum at the remnant state of the magnetization loop.} The prism model presents current paths {with 3D bending} for all angles $\theta$. The average current density over the thickness agrees very well with the thin film model except for the {highest} {angles}. The prism hysteresis loops reveal a peak after the remnant state, which {is due to the parallel component of the self-magnetic-field and is implicitly neglected for thin films.}

The presented numerical method shows the capability to take force-free situations into account for general 3D situations with a high number of degrees of freedom. The results reveal new features of force-free effects in thin films and prisms.    

\end{abstract}

\maketitle


\section{Introduction}
\label{s.intro}

Type II superconductors are essential for large bore or high-field magnets \cite{larbalestier14NaM,Kim17SST,Park18IES,Liu16IES} and are promising for power applications, such as motors for air-plane 
\cite{Masson07IES,masson13IES} or ship propulsion \cite{Yanamoto17IES,Gamble11IES}, generators\cite{Jeong17IES,abrahamsen10SST,suprapower}, grid power-transmission cables \cite{Volkov16PCS,Yagi15IES}, 
transformers \cite{Hellmann17IES,Glasson17IES,Schwenterly99IES,Mehta11PhC}{, and} or fault-current limiters \cite{Pascal17IES,Xin13IES,Morandi13PhC,souc12SST,Kozak16SST}. The Critical Current Density, ${J_c}$, 
of type II superconductors depends on the magnitude and angle of the local magnetic field. There are {three} types of anisotropy which we call {``intrinsic",} ``de-pinning",  and ``force free" anisotropy. 

{The ``intrinsic" anisotopy is the following. Certain superconductors present an axis with suppressed superconductivity, where the critical current density is lower. In cuprates, for instance, the critical current density in the $c$ crystallographic axis is much smaller than in the $ab$ plane. There is also important anisotropy in REBCO vicinal films due to flux channeling \cite{durrell03JAP,Lao17SST}.}

The ``de-pinning" anisotropy of ${J_c}$ is due to anisotropic maximum pinning forces caused by either anisotropic pinning centres or {anisotropic vortex cores \cite{blatter94RMP}}. When the current density ${\vJ}$ is perpendicular to ${\vB}$ and the electric field ${\vE}$ is parallel to ${\vJ}$, the anisotropy of ${J_c}$ is always due to de-pinning anisotropy. 
This kind of anisotropy is important for {High-Temperature Superconductors} (HTS), {such as} (Bi,Pb)$_{2}$Sr$_{2}$Ca$_{2}$Cu$_{3}$O$_{10}$ {and $RE$}Ba$_{2}$Cu$_{3}$O$_{7-x}$, and iron-based superconductors. The magnetic field dependence and anisotropy has an impact on the performance of magnets and power applications. 

{Another} type of anisotropy is the ``force-free" anisotropy, which appears when the current density presents a substantial parallel component with the local magnetic field. 
The parallel ${\vJ}$ component does not contribute to the macroscopic driving force (or Lorentz force) on the vortices, ${\vF=\vJ\times\vB}$, being the microscopic vortex dynamics for ${\vB\parallel\vJ}$ 
a complex process that includes flux cutting and crossing \cite{Vlasko15FL,clem82PRB,clem11SST}. Many power applications with rotating applied fields are influenced with force-free effects. 
In principle, the force-free anisotropy also appears for intrinsically isotropic materials.

There are many macroscopical physical models on force-free anisotropy that regard both flux cutting and de-pinning, such as {the} Double Critical State Model\cite{clem82PRB}, 
the General Double Critical State Model\cite{clem84PRB}, Brant and Mikitik Extended Double Critical State Model \cite{Brandt07PRB} and the Elliptic Critical State Model 
\cite{romerosalazar03APL}. A valuable comparison of these models can be found in \cite{clem11SST}. There are many experimental works on de-pinning anisotropy, such as state of the art REBCO 
commercial tapes \cite{Chepikov17SST,Lijima15IES,Lee14IES,Rosii16SST,Xu17IES,Lin17AIM,Miura17SST}, {Bi2223 tapes \cite{Ayai08PCS, Goyal97JMR} and iron based \cite{Pallecchi15SST, Yi17APS, Ma17SST,Hecher18PRB} conductors, as well as} a database of anisotropic $J_c$ measurements \cite{HTSdatabase}. Correction of self-magnetic field in critical current, $I_c$, measurements is also important 
\cite{Pardo11SST,Zermeno16IES}.    

In this article, we focus on force-free effects, which cause anisotropy when ${\vJ}$ has a parallel component to ${\vB}$ (or ${\vE}$ is not parallel to ${\vJ}$). We also base our study in modelling only. 
The object of study are thin films and rectangular prisms of several thickness with various angles of the applied fields, with a especial focus on the current path and hysteresis loops. 
We compared results with the isotropic situation, in order to understand the observed behavior. The modelling is performed by Minimum Electro-Magnetic Entropy Production in 3D \cite{Pardo17JCP}, 
which is suitable for 3D calculations, and avoids spending variables in the air. Moreover, the method enables force-free anisotropic power laws \cite{badia15SST}, which is the core of this study.


\section{Mathematical model}


\subsection{MEMEP 3D method}

This study is based on the Minimum ElectroMagnetic Entropy Production in 3D (MEMEP 3D) \cite{Pardo17JCP}, which is a variational method. The method solves the effective magnetization ${\vT}$, 
defined as

\begin{equation}
\ \nabla\times\vT=\vJ,
\label{T_J}
\end{equation}
where ${\vJ}$ is the current density. In addition to the magnetization case, MEMEP 3D can also take transport currents into account, after adding an extra term in (\ref{T_J}) 
(see \cite{Pardo17JCP}). We take the interpretation that ${\vT}$ is an effective magnetization due to the screening currents. The ${\vT}$ vector is non-zero only inside the sample, and hence 
the method avoids discretization of the air around the sample. The advantages of MEMEP 3D are reduction of computing time, {enabling an} increase of total number of degrees of freedom 
in the sample volume, and efficient parallelization. The general equation of electric field ${\vE}$ is derived from Maxwell equations
\begin{equation}
\ \vE(\vJ)=-\dot\vA-\nabla\phi,
\label{EJ_gen}
\end{equation} 

\begin{equation}
\ \nabla\cdot\vJ=0,
\label{divJ0}
\end{equation} 
where ${\phi}$ is the scalar potential. 

In the Coulomb's gauge, we can split the vector potential ${\vA}$ to ${\vA_a}$ and ${\vA_J}$, where ${\vA_a}$ is the vector potential contributed by the applied field and ${\vA_J}$ is the vector potential 
contributed by the current density inside the sample. Then, ${\vA_J}$ is

\begin{equation}
\ \vA_J({\bf r})=\frac{\mu_{0}}{4\pi}\int_{V}d^{3}{{\bf r}'}\frac{\vJ({\bf r}')}{|{\bf r}-{\bf r}'|} = \frac{\mu_{0}}{4\pi}\int_{V}d^{3}{{\bf r}'}\frac{\nabla'\times\vT({\bf r}')}{|{\bf r}-{\bf r}'|},
\label{Coulomb}
\end{equation}    
where ${{\bf r}}$ and ${{\bf r}'}$ are position vectors.
	
According {to} the definition of ${\vT}$, we can rewrite equations {(\ref{EJ_gen}) and (\ref{divJ0})} into
\begin{equation}
\ \vE(\nabla\times\vT)=-\dot\vA-\nabla\phi,
\label{EJ_T}
\end{equation} 

\begin{equation}
\ \nabla\cdot\left(\nabla\times\vT\right)=0.
\label{}
\end{equation} 

The second equation is always satisfied, and hence we must solve only the first equation. As it was shown in \cite{Pardo17JCP}, minimizing the following functional, is the same as solving 
equation (\ref{EJ_T}).
 
The functional is     
\begin{equation}
\ L_{T}=\int_{V}d^{3}{\bf r}\left[\frac{1}{2}\frac{\Delta{\vA}_{J}}{\Delta t}\cdot(\rotDT)+\frac{\Delta{\vA}_{a}}{\Delta{t}}\cdot(\rotDT)+U(\rotT)\right],
\label{functional}
\end{equation}
where ${U}$ is the dissipation factor, defined as
\begin{equation}
\ U\left(\vJ\right)=\int_{0}^{J}\vE\left(\vJ'\right)\cdot d\vJ'.
\label{}
\end{equation}
The functional can include any ${\vE(\vJ)}$ relation with its corresponding dissipation factor. 
The functional is solved in the time domain in time steps like ${t=t_0+\Delta t}$, where ${t}$ is the present time, ${t_0}$ is the previous time step and ${\Delta t}$ is the time between two time steps. The magnetic moment ${\vm}$ is calculated by equation 
\begin{equation}
\ {\bf{m}}=\frac{1}{2}\int d^{3}{\bf r} \vJ\times{\bf r}, 
\label{}
\end{equation}  
where ${{\bf r}}$ is a position vector of interpolated ${\vJ}$ at the centre of the cell. The magnetization is ${\vM=\vm/V}$ and ${V}$ is volume of he sample. Then, we define ${\vT,\vA_J,\vA_a}$ as the value 
of the corresponding variables at the present time step; ${\Delta\vT,\Delta\vA_J,\Delta\vA_a}$ are the change of the variables between two time steps; and ${\vT_0,\vA_{J0},\vA_{a0}}$ 
are the variables from the previous time step. {In this work,} the applied magnetic field {${{\bf B}_a}$} is uniform {and} ${\Delta t}$ {is constant}, although the method enables non-uniform {${{\bf B}_a}$} and variable ${\Delta t}$.


\subsection{${\vE(\vJ)}$ relation}

In a previous study \cite{Pardo17SST}, we used the isotropic power law as ${\vE(\vJ)}$ relation in the functional (\ref{functional})  

\begin{equation}
\vE(\vJ)=E_c \left ( \frac{|\vJ|}{J_c} \right )^n\frac{\vJ}{|\vJ|},
\label{EJ}
\end{equation} 
where ${\vB\perp\vJ,\vE\parallel\vJ,}$ and ${E_c}$ is the critical electric field ${10^{-4}}$ V/m, ${J_c}$ is the critical current density, and ${n}$ is the power law exponent or ${n}$ factor. 
The ${n}$ factor depends on the quality of the superconducting materials, temperature and local magnetic field ${\vB}$. The Bean Critical State Model (CSM)\cite{bean62PRL,london63PhL} 
{corresponds to} ${n\to\infty}$, but real superconductors present smaller ${n}$ factors, ranging from around 10 to the order of 100. The case of ${n}$=100 is practically equivalent to the CSM. 
The dissipation factor for isotropic ${\vE(\vJ)}$ relation of (\ref{EJ}) is 

\begin{equation}
\ U\left(\vJ\right)=\frac{E_{c}J_{c}}{n+1}\left(\frac{|\vJ|}{J_c}\right)^{n+1}.
\label{UJ_iso}
\end{equation}  
In this article, we focus on the anisotropic case, in order to model the force-free effects with anisotropic power law \cite{badia15SST}.    

\begin{equation}
\vE(\vJ)=2m_0U_0\left[ \left( \frac{J_\parallel}{J_{c\parallel}}\right)^2 + \left( \frac{J_\perp}{J_{c\perp}}\right)^2\right]^{m_{0}-1}\cdot\left(\frac{J_\parallel}{J_{c\parallel}^2}{\ve}_\parallel
+\frac{J_\perp}{J_{c\perp}^2}{\ve}_\perp\right),
\label{EJani}
\end{equation}
where ${m_{0}=(n+1)/2}$, ${U_{0}=E_{c}J_{c\perp}/(n+1)}$, ${J_\parallel=\vJ\cdot\vB/|\vB|}${,} ${J_\perp=|\vJ\times\vB|/|\vB|}$, {and} ${J_{c\perp}}$ and ${J_{c\parallel}}$ are critical current densities parallel and perpendicular to ${\vB}${, respectively}. Vector ${\vB}$ is the local magnetic field and ${{\ve}_\perp}$, ${{\ve}_\parallel}$ are unit vectors of the current density, where ${{\ve}_\parallel=\vB/|\vB|}$, ${{\ve}_\perp=\vJ_\perp/|\vJ_\perp|}$ and ${\vJ_\perp=\vJ-J_\parallel\ve_\parallel}$. Notice that ${\vJ=J_\parallel\ve_\parallel+J_\perp\ve_\perp}$ and ${J_\perp}$ is always positive. The applied magnetic field ${\vB_a}$ is not always perpendicular to the sample surface {[}figure \ref{Anisotropy.fig} (a){]}. The corresponding anisotropic dissipation factor is 
\begin{equation}
\ U\left(\vJ,\vB\right)=U_0\left[\left(\frac{J_\parallel}{J_{c\parallel}}\right)^{2}+\left(\frac{J_\perp}{J_{c\perp}}\right)^{2}\right]^{m_0}.
\label{UJ_ani}
\end{equation}

The anisotropic power law becomes the {elliptic }CSM for large enough ${m_0}$ or ${n}$ with two critical current densities ${J_{c\perp},J_{c\parallel}}$, 
which apply according to direction of the local magnetic field ${\vB}$. The problem of the anisotropic ${U(\vJ,\vB)}$ relation is the uncertainty of the unit vector of local magnetic field ${\vB}$ with 
very low or zero ${|\vB|}$. In the samples there exist places where the local magnetic field vanishes. We suggest the following 
two options in order to remove this uncertainty. 

\begin{figure}[tbp]
\centering
 \subfloat[][]
{\includegraphics[trim=0 0 -20 0,clip,height=4.5 cm]{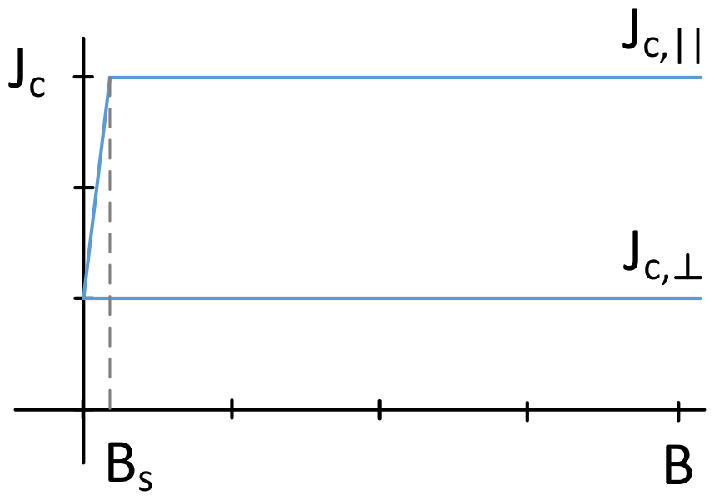}} 
 \subfloat[][]
{\includegraphics[trim=0 0 0 0,clip,height=4.5 cm]{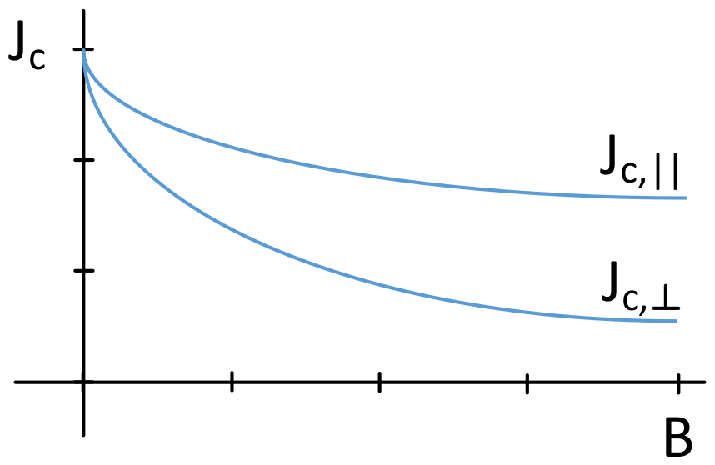}} 
\caption{{Using a magnetic field dependence for $J_{c\parallel}$ and $J_{c\perp}$ avoids indeterminations at $\vB=0$.} (a) {Elliptic} double critical-state model {(CSM)}. (b) Anisotropic Kim model.}
\label{sketch1.fig}
\end{figure}

The first option is to use a sharp ${J_{c\perp}(\vB)}$ and ${J_{c\parallel}(\vB)}$ dependence, where at ${|\vB|\to\infty}$ they follow ${J_{c\parallel}\neq J_{c\perp}}$ and at ${|\vB|= 0}$, 
${J_{c\parallel}=J_{c\perp}\equiv J_{c0}}$ with a linear transition between ${|\vB|=0}$ and a certain magnetic field ${|\vB|=B_s}$, being ${B_s}$ a small magnetic field [figure \ref{sketch1.fig}(a)]. 
The limit of ${B_s\to 0}$ corresponds to the {elliptic }critical state model \cite{Perezgonzalez85JAP}. For simplicity, we consider only this linear dependence  of ${B}$ 
for ${J_{c\parallel}}$, keeping ${J_c\perp}$ as constant. The reason is to reproduce the Bean CSM for perpendicular applied fields.    

The magnetic field is calculated from the current density after the functional is minimized. The functional is solved iteratively \cite{Pardo17JCP}: at the first iteration, ${\vT}$ is 
calculated with ${\vB_J=\vB_{J0}}$ and ${\vB_a\neq 0}$, being ${\vB_{J0}}$ the magnetic field generated by ${\vJ}$ at the previous time step, 
the second iteration starts with ${\vB_J\neq 0}$ calculated from ${\vJ}$ at the previous iteration, where ${\vJ=\nabla\times\vT}$; iterations are repeated until we find a solution 
with given tolerance in each component of ${\vJ}$. The sharp ${J_{c\parallel}(B)}$ dependence causes numerical problems in this iterative method, since a small error in ${\vB}$ causes a large error in 
${\vJ}$ in the next iteration step. 

In order to avoid this numerical problem, the functional is minimized in a certain time ${t}$ with the total magnetic field ${\vB}$ from the previous time step ${\vB(t-\Delta t)}$. The {vector potential ${\bf A}$ is still} calculated according the {present} time ${t}$. This is the reason why the remanent state is shifted by ${\Delta t}$ in the results. The negative effect of that assumption 
can be decreased by increasing the number of time steps in one period of applied magnetic field. 

Another option to avoid the problems at ${|\vB|\to 0}$ is to assume Kim's model for {both ${J_{c\parallel}(B)}$} and ${J_{c\perp}(B)}$ dependences, where ${J_{c\parallel}(B=0)=J_{c\perp}(B=0)\equiv J_{c0}}$ and ${J_{c\parallel}\neq J_{c\perp}}$ for ${\vB\to\infty}$ [figure \ref{sketch1.fig}(b)]. {This }Kim model is
 
\begin{equation}
\ J_{c\parallel}(B)=\frac{J_{c0}}{\left(1+\frac{|\vB|}{B_{0\parallel}}\right)^m},
\label{KimJcpa}
\end{equation} 

\begin{equation}
\ J_{c\perp}(B)=\frac{J_{c0}}{\left(1+\frac{|\vB|}{B_0\perp}\right)^m},
\label{KimJcpe}
\end{equation} 
where in this article we choose m=0.5, ${B_{0\perp}}$=20 mT, ${J_{c0}=3\cdot10^{10}}$ A/mm${^{2}}$ and ${B_{0\parallel}=9B_{c\perp}}$, so that 
${J_{c\parallel}\left(B\to\infty\right)=3J_{c\perp}\left( B\to\infty\right)}$. For this case, the ${J_{c\parallel}(B)}$ and ${J_{c\perp}(B)}$ dependences are not sharp, 
and hence we use the original iterative method for magnetic field dependent ${J_c}$. Then, ${\vJ}$ at time ${t}$ uses ${\vB}$ of the same time ${t}$. Moreover, this smooth ${J_{c\perp}(B), J_{c\parallel}(B)}$ 
dependence is more realistic than the {elliptical} CSM. 

\subsection{Sector minimization}

\begin{figure}[tbp]
\centering
 \subfloat[][]
{\includegraphics[trim=0 0 0 0,clip,width=5.5 cm]{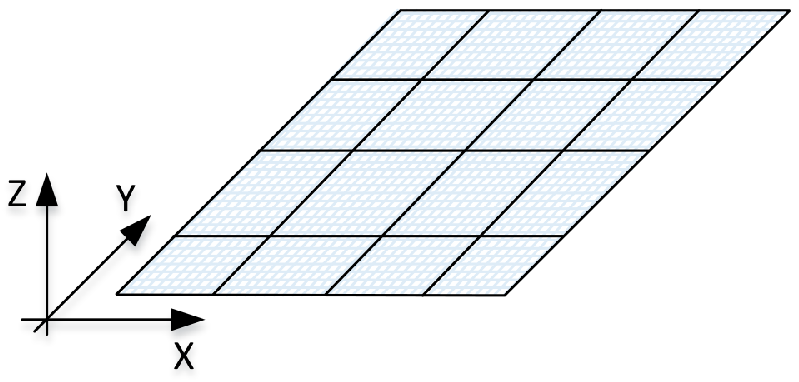}}
 \subfloat[][]
{\includegraphics[trim=0 0 0 0,clip,width=5.5 cm]{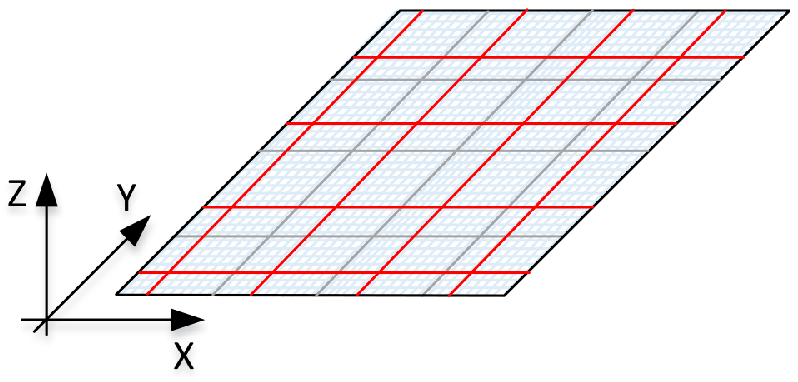}}
 \subfloat[][]
{\includegraphics[trim=0 0 0 0,clip,width=5.5 cm]{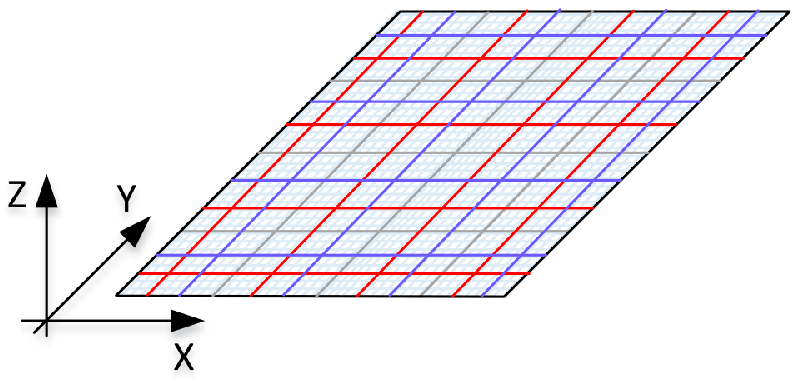}}
\caption{{Using three sets of sectors speeds up the calculations. (a) Boundaries of the first set only. (b) First and second set. (c) All three sets.}}
\label{set.fig}
\end{figure}

Reduction of computing time is of essential importance for 3D calculations. We already studied the case of parallel minimization by sectors, where sectors are overlapping by one cell \cite{Pardo17JCP}. 
In this article, we {increase} the overlapping of sectors in the following way. Now, the sectors are not overlapping to each other, and hence they share only the edge on the border, 
which are not solved (figure \ref{set.fig}(a)). Then, we added {other }2 sets of sectors, but the boundary in each set of sectors is shifted along the diagonal by 1/3 of the sector-diagonal size 
(figure \ref{set.fig}(b,c)). The edge in the boundary in the first set is solved at least once in some of the other two sets. The additional sets {increase} the memory usage, 
which is still low, but {they decrease the} computing time. Sets of the sectors are minimized in series one after the other, but sectors within each set are solved in parallel to achieve 
high efficiency of parallel computing. Although computing all three sets of sectors in parallel could further enhance parallelization, we have found that solving each set sequentially 
reduces computing time. The process of solving all 3 sets subsequently is repeated until the maximum difference in any component of ${\vT}$ between two iterations of the same set is below a certain 
tolerance. We use elongated cells, in order to improve the accuracy for a given number of cells, as detailed in appendix A.

\section{Results and discussion}

As a result of the variational model, we calculated two geometries like an infinitesimally thin film (or simply, ``thin film") and a thin prism with finite thickness ({or} ``thin prism") 
[figure \ref{Anisotropy.fig} (a),(b)]. The force-free effects are modelled with the anisotropic power-law in combination of either constant ${J_{c\perp}}$, ${J_{c\parallel}}$ or 
Kim model ${J_{c\perp}(B)}$, ${J_{c\parallel}(B)}$. We calculated as well the pure isotropic case of a thin film and thin prism for comparison. 
The calculations are performed with two values of {the $n$}-factor, 30 and 100, in order to have results close to the realistic values and analytical critical-state formulas, respectively. 


\subsection{Anisotropic force-free effects in films}

{In this section, we study square thin films of dimensions 12$\times$12 mm$^2$ and thickness 1 $\mu$m. We also take the common assumption of the thin film limit, which consists on averaging the electromagnetic properties over the sample thickness. For our method, this is achieved by taking only one cell along the sample thickness. We used a total number of degrees of freedom of 4200.
}

\subsubsection{Power device situation}

\begin{figure}[tbp]
\centering
 \subfloat[][]
{\includegraphics[trim=-30 0 0 0,clip,width=7 cm]{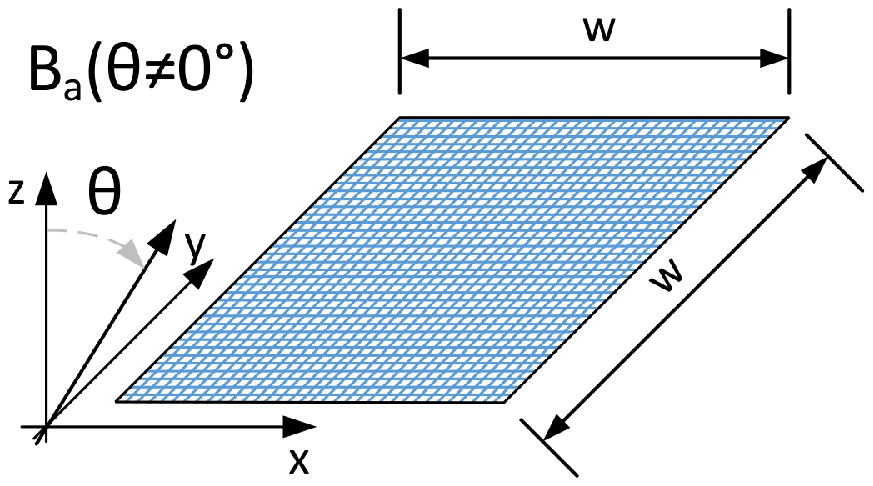}}
 \subfloat[][]
{\includegraphics[trim=-30 0 0 0,clip,width=7 cm]{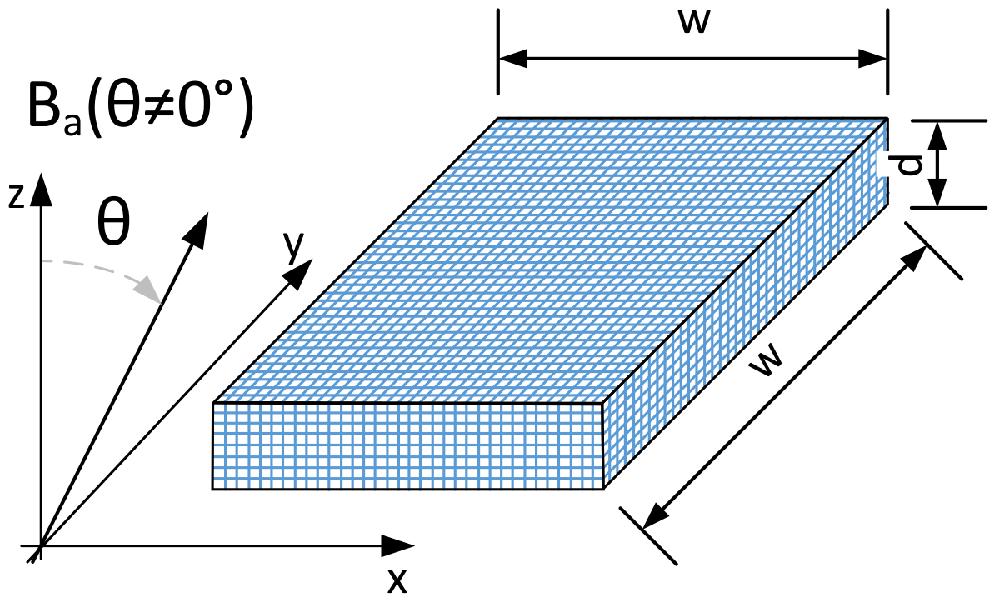}}
\caption{Sketches of the geometry with the variable angle ${\theta}$ of the applied magnetic field (a) thin strip (b) prism.}
\label{Anisotropy.fig}
\end{figure}

In this section, the magnetic applied field ${{\bf{B}}_a}$ has a sinusoidal waveform of 50 Hz and the same perpendicular, ${B_{a,z}}$, component for all angles ${\theta}$ with amplitude ${B_{a,z,m}=}$ 
50 mT [figure \ref{Anisotropy.fig}(a)]. The angle ${\theta=0\degree}$ is completely perpendicular to the surface of the thin film. We calculated the cases with 
${\theta=0\degree,45\degree,60\degree,80\degree}$. For this study, the perpendicular critical current density ${J_{c\perp}}$ is equal to ${3\cdot10^{10}}$ A/m${^{2}}$ and ${J_{c\parallel}}$ is 3 times higher. The dependence of ${J_c}$ on the magnetic field is on figure \ref{sketch1.fig}(a), where we choose ${B_s=}$ 1 mT. The $n$ factor of the anisotropic power law is equal 30, which is a realistic value for REBCO tapes in self-field. 

The first case, with ${\theta=0\degree}$, is shown on figure \ref{Jc(B,0).fig}. The penetration of the current density to the film strip is explained by colour maps of ${|\vJ|}$ normalized to ${J_{c\perp}}$, while the lines are current flux lines. The current density gradually penetrates to the sample after {increasing} the applied field [figure \ref{Jc(B,0).fig}(a)], 
until it reaches almost saturated state at the peak of applied field [figure \ref{Jc(B,0).fig}(b)]. During {the decrease} of the applied field, current starts penetrating again from the edges of the sample with opposite sign till the centre. The quasi remanent state, at ${B_a\approx}$0 mT, presents symmetric penetration of ${\vJ}$ along both ${x}$ and ${y}$ axis [figure \ref{Jc(B,0).fig}(c)]. 
We show the first time step after remanence, ${\vB=0}$, for comparison with the cases with ${\theta\neq 0\degree}$, where we use ${\vB}$ of the previous time step in order to obtain 
${J_{c\parallel}}$ [figure \ref{Jc(B,0).fig}{(a)}].           

The second case is for ${\theta=45\degree}$ and applied field amplitude ${B_{am}=}$70.7 mT (figure \ref{Jc(B,45).fig}). The force-free effects appear during {the increase} of the applied field 
[figure \ref{Jc(B,45).fig}(a)]. The current lines parallel to the ${x}$ axis are more aligned with the direction of the applied field. 
Therefore, ${J_{c\parallel}}$ becomes relevant, and hence current density at that direction is higher compared to the current density along the ${y}$ axis. 
The current penetration depth is smaller from top and bottom at the peak [figure \ref{Jc(B,45).fig}(b)] compared to that from the sides. The penetration depth of ${J_y}$ from right and left is the same 
as for ${\theta=0\degree}$, because ${J_y}$ is still perpendicular to ${B_a}$. The quasi remanent state [figure \ref{Jc(B,45).fig}(c)] with the applied field close to zero experiences the self-field 
as dominant component of the local magnetic field. Then, the self-field in the thin film approximation has only ${B_z}$ component, which is completely perpendicular to the surface and the current density. 
Therefore, only ${J_{c\perp}}$ is relevant and the maximum ${\vJ}$ in the sample is decreased back to that value. 

The last two cases, ${\theta=60\degree}$ and ${80\degree}$, present similar behavior. The penetration of ${J_x}$ to the sample is even smaller during {the increase} of the applied field 
[figures \ref{Jc(B,60).fig}(a), \ref{Jc(B,80).fig}(a)], because of the higher angles ${\theta=60\degree,80\degree}$. The maximum ${J_x}$ component at the peak of the applied field 
[figures \ref{Jc(B,60).fig}(b), \ref{Jc(B,80).fig}(b)] is reaching 2.5 and 3 times of ${J_{c\perp}}$, which is the value of ${J_{c\parallel}}$. Again, at remanent state [figures \ref{Jc(B,60).fig}(c), 
\ref{Jc(B,80).fig}(c)] the maximum ${J_x}$ component is decreased back to values around ${J_{c\perp}}$ because of the self-field without any parallel component of the local magnetic field.

\begin{figure}[htp]
\centering
 \subfloat[][]
{\includegraphics[trim=30 0 50 0,clip,height=5.5 cm]{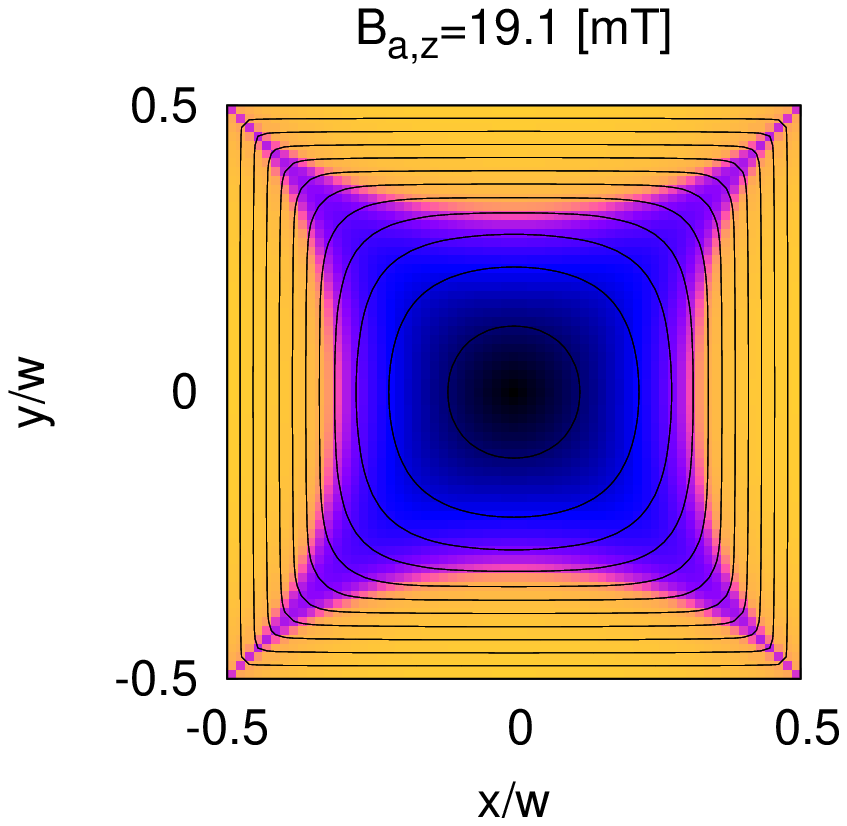}} 
 \subfloat[][]
{\includegraphics[trim=85 0 80 0,clip,height=5.5 cm]{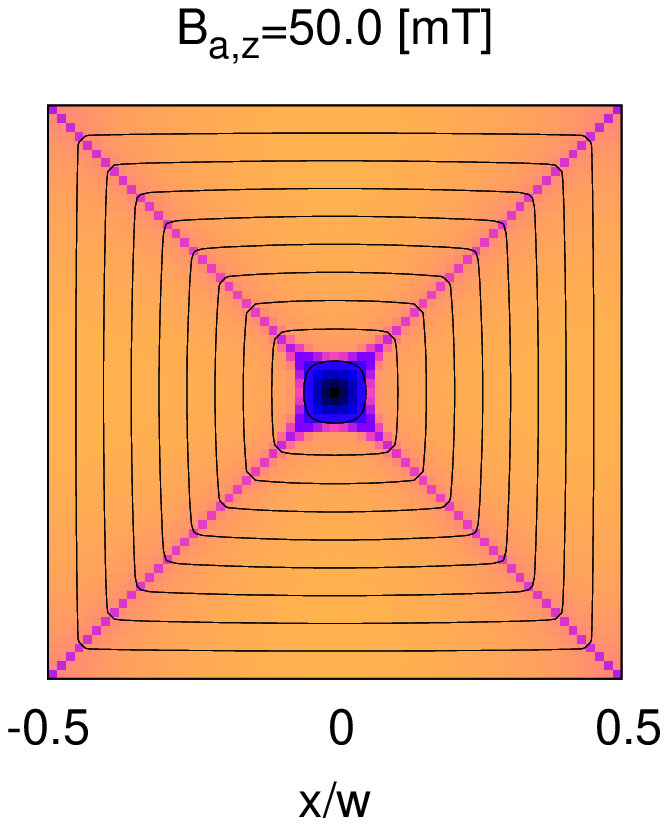}} 
 \subfloat[][]
{\includegraphics[trim=50 0 25 0,clip,height=5.5 cm]{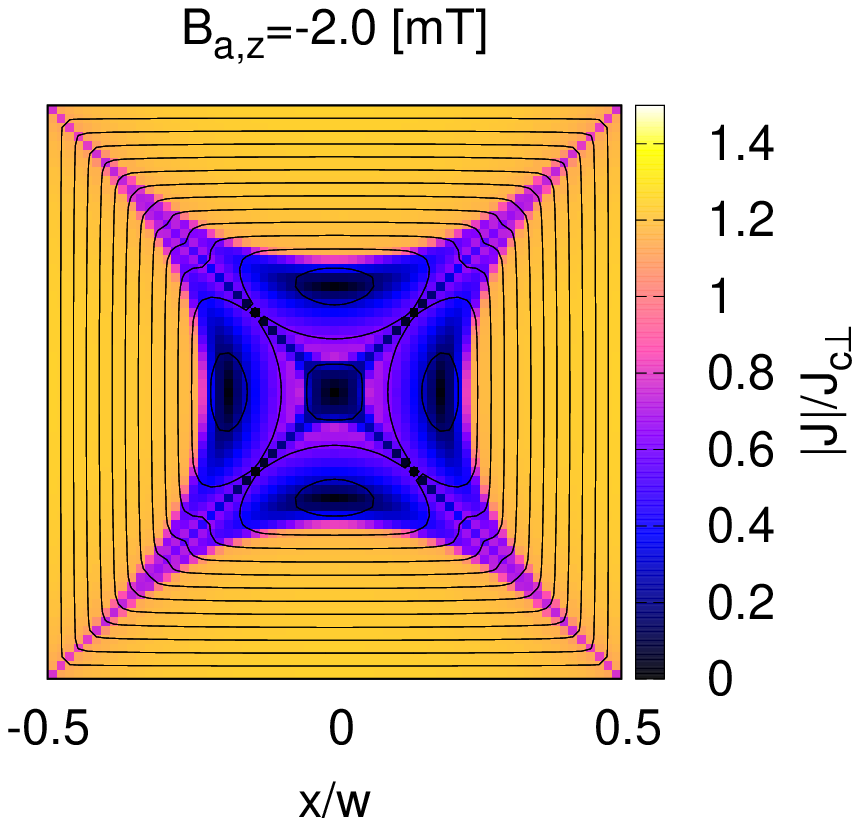}} 
\caption{{Penetration process of the critical-current density in the thin film with force-free anisotropic $\vE(\vJ)$ relation. The applied magnetic field is sinusoidal with $\theta=0$ (perpendicular to the surface), ${B_{am}=}$50 mT, and ${f=}$50 Hz. (a) Initial curve. (b) Peak of the applied field. (c) Quasi-remanent state. The figure also shows the current flux lines.} }
\label{Jc(B,0).fig}
\end{figure}

\begin{figure}[htp]
\centering
 \subfloat[][]
{\includegraphics[trim=30 0 50 0,clip,height=5.5 cm]{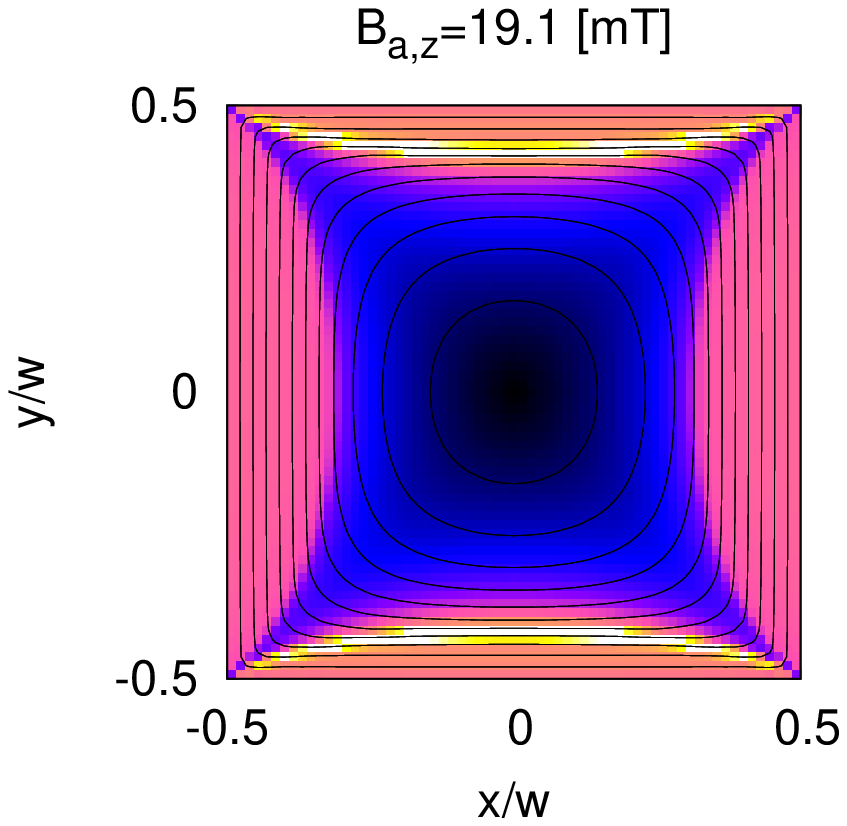}} 
 \subfloat[][]
{\includegraphics[trim=85 0 80 0,clip,height=5.5 cm]{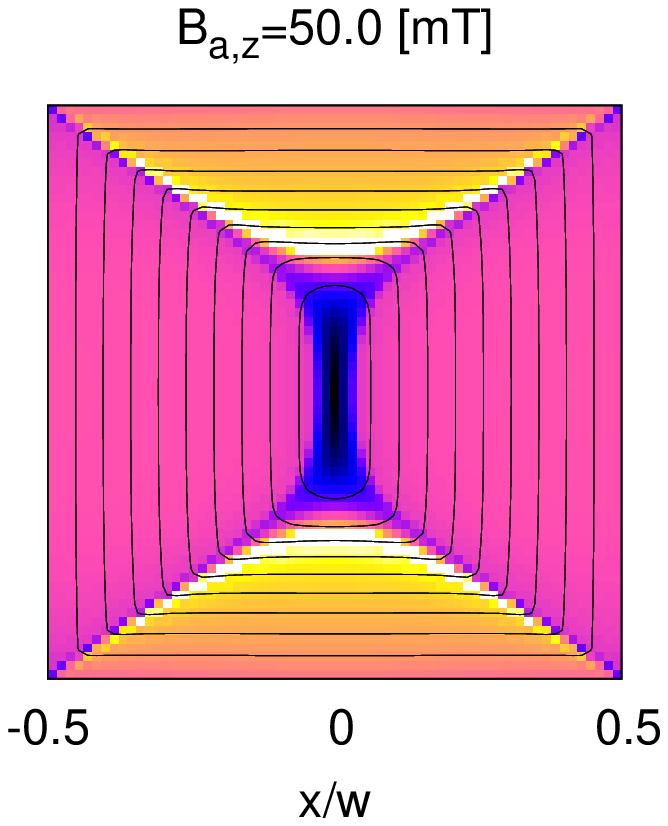}} 
 \subfloat[][]
{\includegraphics[trim=50 0 25 0,clip,height=5.5 cm]{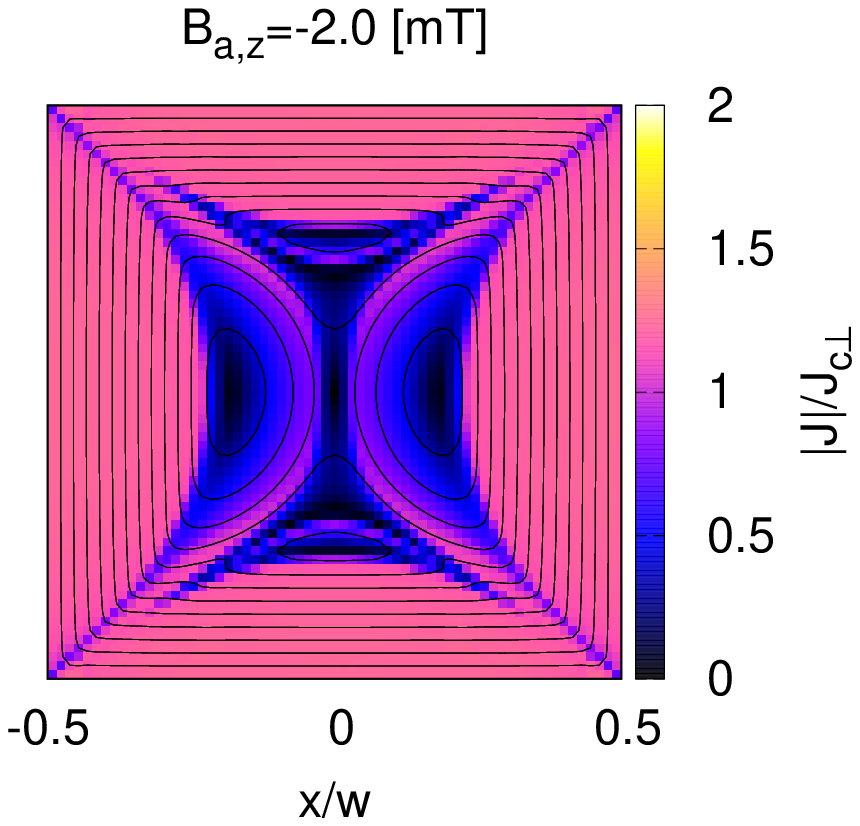}} 
\caption{{The same as figure \ref{Jc(B,0).fig} but for $\theta=45\dg$.} }
\label{Jc(B,45).fig}
\end{figure}

\begin{figure}[htp]
\centering
 \subfloat[][]
{\includegraphics[trim=30 0 50 0,clip,height=5.5 cm]{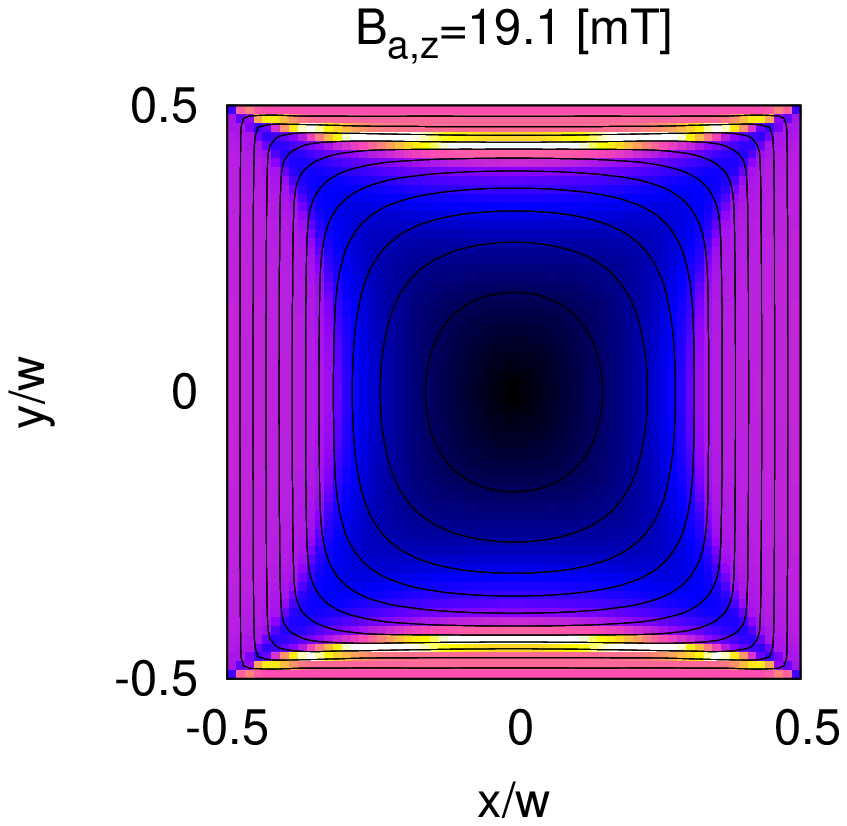}} 
 \subfloat[][]
{\includegraphics[trim=85 0 80 0,clip,height=5.5 cm]{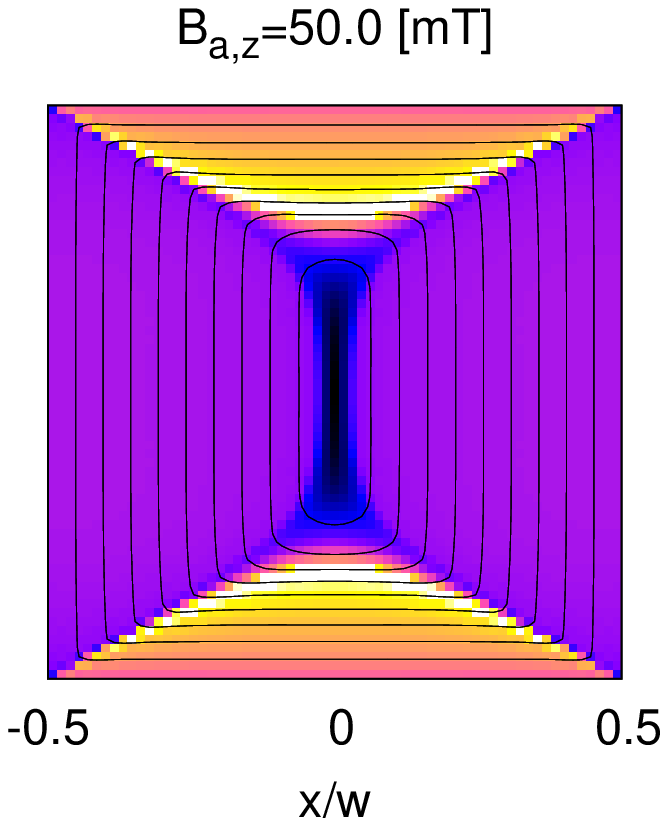}} 
 \subfloat[][]
{\includegraphics[trim=50 0 25 0,clip,height=5.5 cm]{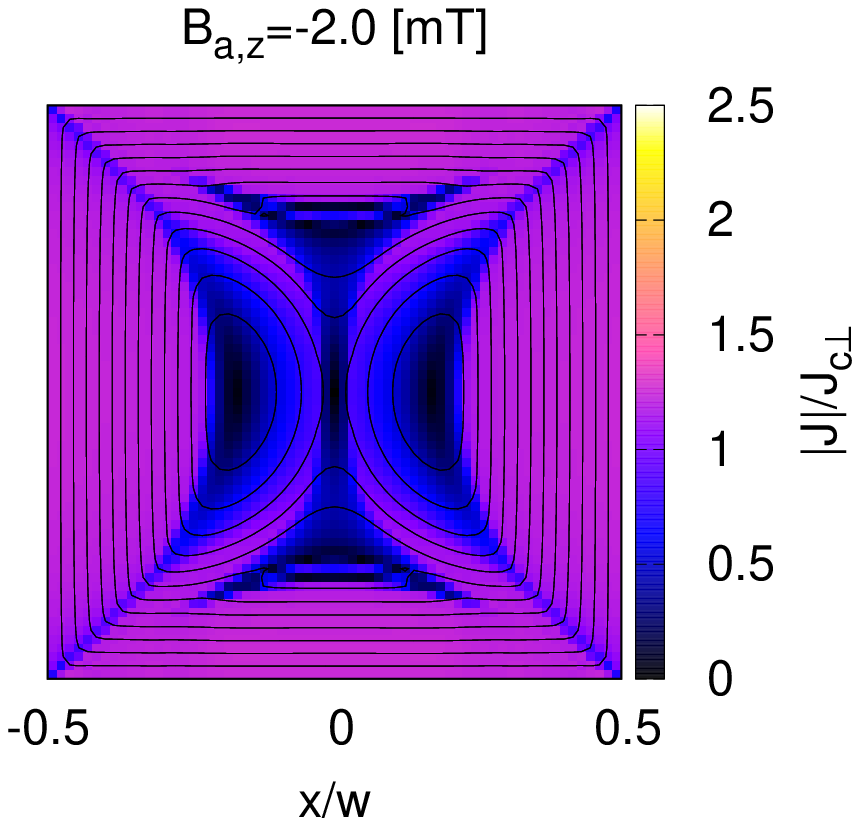}} 
\caption{{The same as figure \ref{Jc(B,0).fig} but for $\theta=60\dg$.} }
\label{Jc(B,60).fig}
\end{figure}

\begin{figure}[htp]
\centering
 \subfloat[][]
{\includegraphics[trim=30 0 50 0,clip,height=5.5 cm]{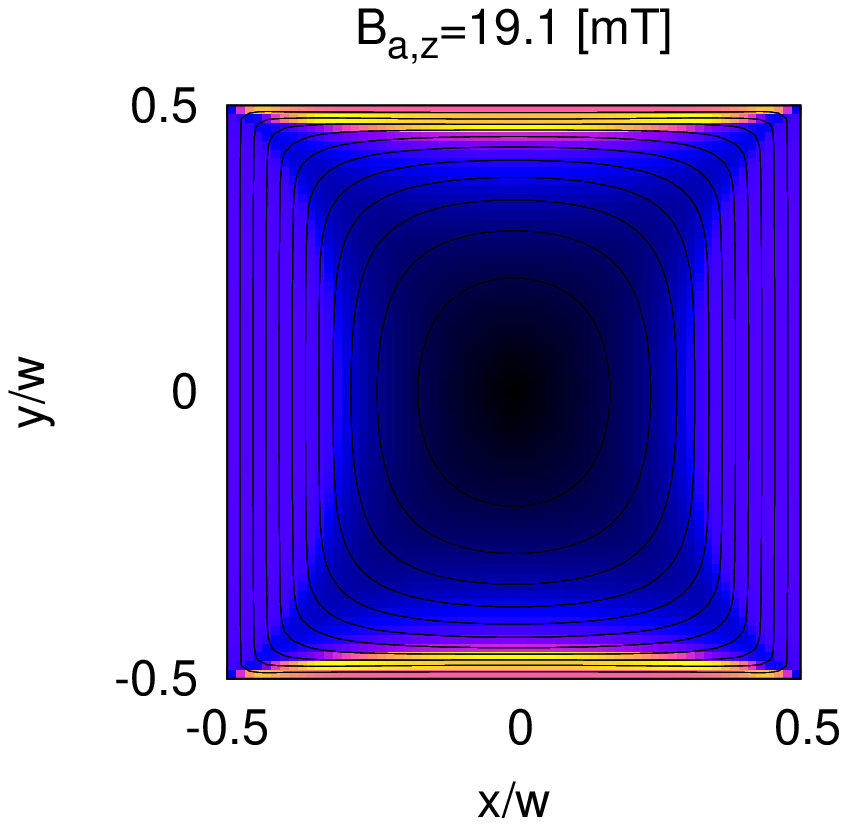}} 
 \subfloat[][]
{\includegraphics[trim=85 0 80 0,clip,height=5.5 cm]{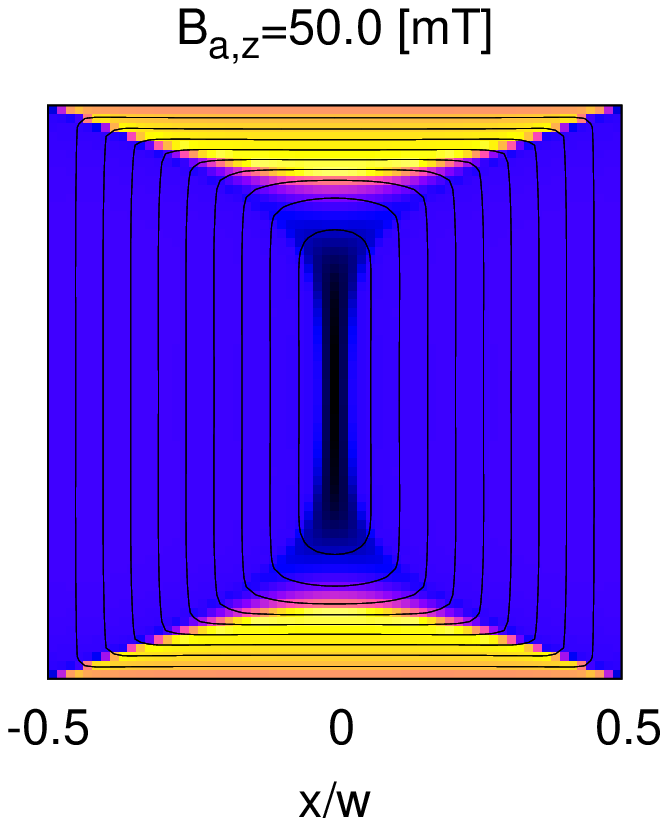}} 
 \subfloat[][]
{\includegraphics[trim=50 0 25 0,clip,height=5.5 cm]{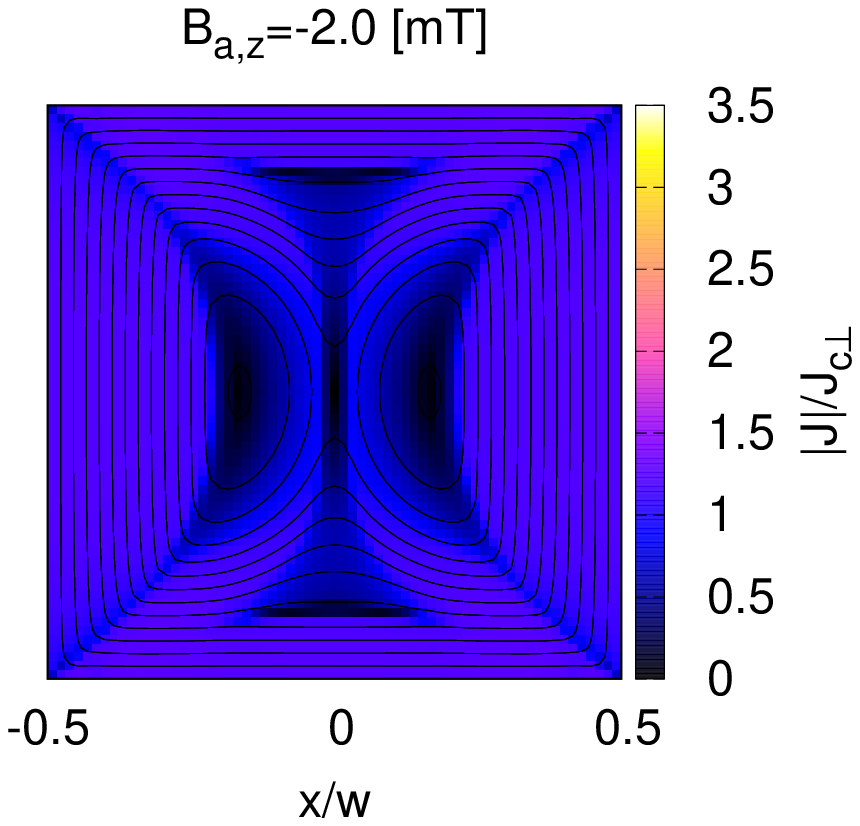}} 
\caption{{The same as figure \ref{Jc(B,0).fig} but for $\theta=80\dg$. For sufficiently large applied fields (a,b) there appear zones with $|\vJ|\approx J_{c\perp}$ and $|\vJ|\approx J_{c\parallel}$, while at the quasi-remanent state (c) $|\vJ|$ is limitted to $J_{c\perp}$}. }
\label{Jc(B,80).fig}
\end{figure}

\begin{figure}[tbp]
\centering
 \subfloat[][]
{\includegraphics[trim=70 0 70 0,clip,height=8 cm]{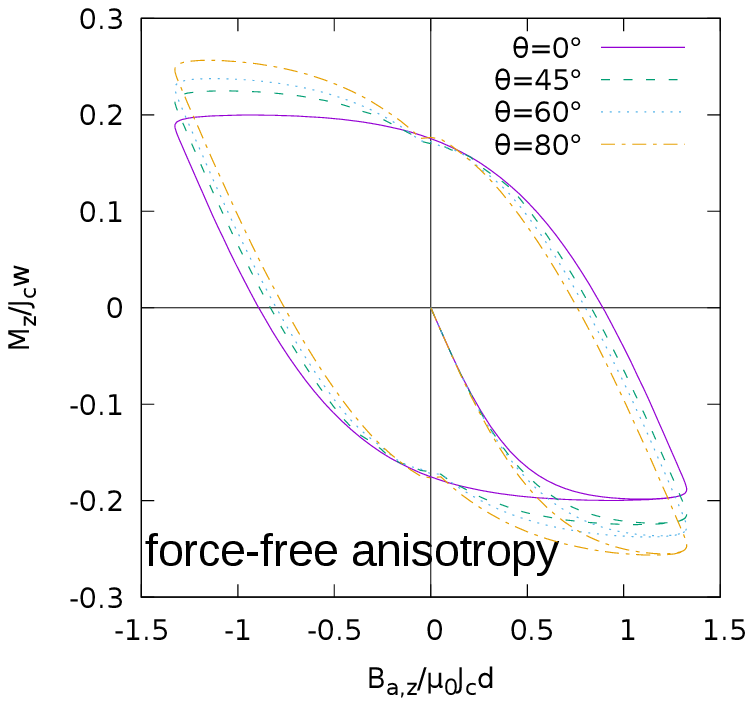}}
 \subfloat[][]
{\includegraphics[trim=70 0 70 0,clip,height=8 cm]{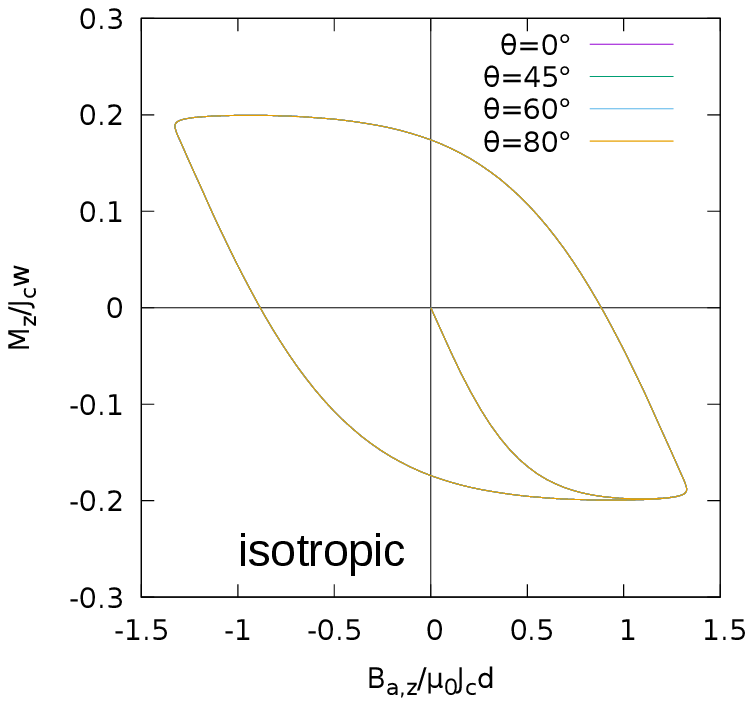}}
\caption{{Force free anisotropy increases the magnetization at the peak of the applied field. Case for thin film} with $n$ value 30, sinusoidal applied magnetic field ${B_{a,z}=}$50.0 mT and ${f=}$50 Hz{.} (a) {Force-free} anisotropic power law. (b) Isotropic power law.}
\label{hys3.fig}
\end{figure}

The hysteresis loops for all angles ${\theta}$ of the applied field are on figure \ref{hys3.fig}(a). The larger the applied-field angle, the higher the impact of ${J_{c\parallel}}$, 
and hence there exist places with the current density around ${J_{c\parallel}}$. The current density around ${J_{c\parallel}}$ creates higher magnetic moment in comparison to ${\theta=0\degree}$ where ${|\vJ|}$ is limited by ${J_{c\perp}}$. The self-field is dominant at the range of the applied field ${\pm}$5 mT{, causing a mostly perpendicular local magnetic field, and hence $|\vJ|$ is again limited to $J_{c\perp}$}. This is the reason why the magnetization is decreasing back to the same value as in the case of ${\theta=0\degree}$. We calculated the same situation with isotropic power law. 
The results of the isotropic case are the same for each angle ${\theta}$ {because the perpendicular applied field is the same {as for} ${\theta=0\degree}$ [see magnetization loops in figure \ref{hys3.fig}(b)]. Consistently, these magnetization loops also agree with the anisotropic case with ${\theta=0\degree}$, since $J_c=J_{c\perp}$ for the whole loop [figure \ref{hys3.fig}(a)]}.

\subsubsection{Magnet situation}

\begin{figure}[tbp]
\centering
{\includegraphics[trim=70 0 70 0,clip,height=8 cm]{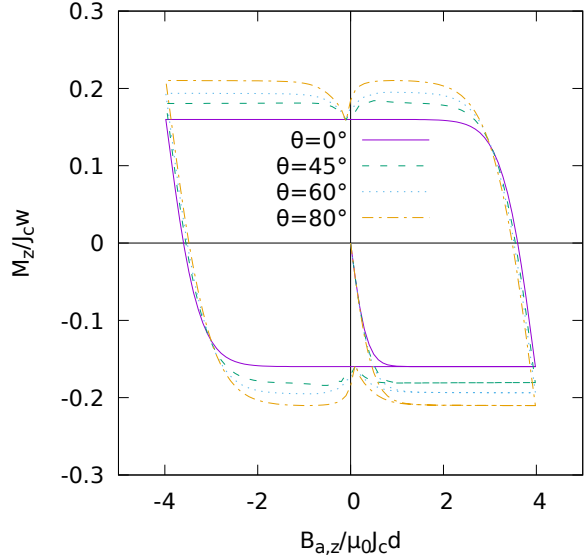}}
\caption{{For a triangular waveform of the applied field, the minimum at remanence of the hysteresis loops is very pronounced. Calculations for the force-free anisotropic power law of $n=$100 and triangular applied magnetic field with $B_{a,m}=$ } 150.0 mT, ${B_{a,zm}=}$50 mT and ${f=}$1 mHz.}
\label{hys1.fig}
\end{figure}

The next calculation assumes the same parameters and geometry as the previous cases. The difference is in {the} $n$-factor, with value 100, triangular waveform of the applied field of 1 mHz frequency 
and amplitude ${B_{a,z,m}=}$ 150 mT. This magnetization is qualitatively similar to magnet charge and discharge. The angles of applied field are the same ${\theta=0\degree, 45\degree, 60\degree, 80\degree}$. 
The high $n$-factor reduces the current density to values equal or below ${J_{c\perp}}$ or ${J_{c\parallel}}$. Another reason for reduction of current density is the very low frequency of the applied field, 
which allows higher flux relaxation. {The constant ramp rate causes that the magnetization loops are flat after the sample is fully saturated [figure \ref{hys1.fig}(b)].} The case of ${\theta=0\degree}$ induces only current density perpendicular to the applied field, and hence magnetization loop {is horizontal} at the remanent state. {Again, we see a minimum at remanence for higher $\theta$}.

\begin{figure}[tbp]
\centering
 \subfloat[][]
{\includegraphics[trim=70 0 70 0,clip,height=8 cm]{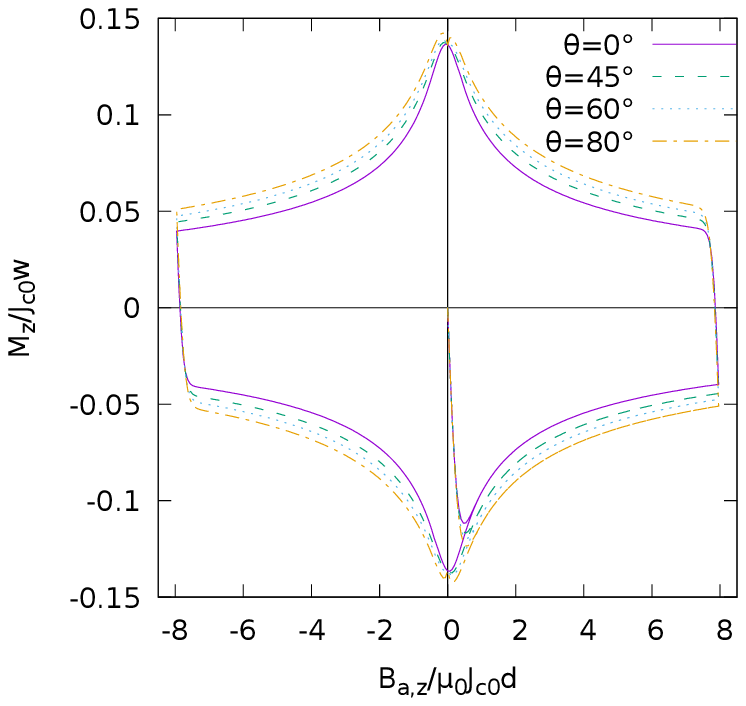}}
 \subfloat[][]
{\includegraphics[trim=60 0 60 0,clip,height=8 cm]{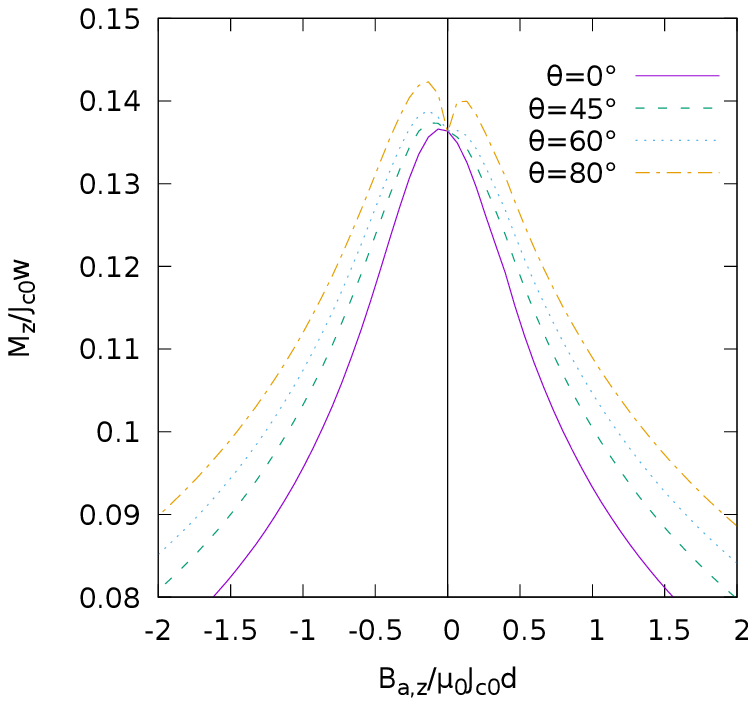}}
\caption{{(a) With a Kim-like dependence of $J_{c\parallel}$ and $J_{c\perp}$, the area of the magnetization loop increases with $\theta$ due to force-free effects. (b) This zoom at low applied field shows the minimum of the remanence. The parameters are $n=100$ and triangular applied magnetic field with $B_{az,m}=$300 mT and $f=$1 mHz.
}}
\label{hys2.fig}
\end{figure}

The last thin film example assumes anisotropic power law with two critical current densities, which depends on the magnetic field according Kim model ${J_{c\parallel}(B)}$, ${J_{c\perp}(B)}$. 
The dependence is on figure \ref{sketch1.fig}(b). The magnetic field ${\vB}$ is calculated in the same time step ${\vB(t)}$ as the functional is minimized, and hence now the remanent state 
is straightforwardly for ${\vB=0}$ as it is shown on figure figure \ref{hys2.fig}(b). The ${B_{a,z}}$ component of the maximum applied field is 300 mT and it is the same for all angles ${\theta}$. 
The magnetization of the sample [figure \ref{hys2.fig}(a)] is higher close to the remanent state, since the applied field is close to zero and the self-field only slightly decreases 
the critical current density. {With increasing the applied field from the zero-field-cool situation, } the sample {becomes} fully saturated already at 40 mT. 
{With further increase of the applied field, the Kim dependence causes a decrease in $J_{c\parallel}$,$J_{c\perp}$ and $|\vJ|$, decreasing the magnitude of the magnetization.} The highest magnetization is at the applied field with ${\theta=80\degree}$, in spite of ${|\vB_a|}$ being the largest and hence reducing the most ${J_{c\perp}}$ and ${J_{c\parallel}}$. The cause is that there still exist areas with current density around ${J_{c\parallel}}$. At the remanent state, we can see again reduction of magnetization to the level of ${\theta=0\degree}$ [figure \ref{hys2.fig}(b)]. 

\subsection{Anisotropic force-free effects in prisms}

\subsubsection{Current density in prisms.}

{In the following, we analyze the} force-free effects in a prisms. We model the prisms with the same dimensions as thin film ${12\times 12}$ mm {but} thickness 1 mm. The mesh of the sample is created by elongated cells, which we explain in \ref{s.e_cell}. The total number of cells is ${31\times 31\times 15}$, which {corresponds to} around 43000 degrees of freedom. The frequency of the applied filed is 50 Hz {and the amplitude of the $z$ component of $\vB_a$ is 50 mT for all angles ${\theta=0,\degree45\degree,60\degree,80\degree}$; and hence the total amplitude is $B_{a,m}=$50, 70.7, 100 and 287.9 mT, respectively.} The critical current densities are chosen so that the sheet critical current density {$K_c\equiv{J_c d}$} is the same as for the thin film, being ${d}$ the sample thickness. 
{Further} values are ${J_{c\perp}=3\cdot10^7}$ A/mm${^2}$, ${J_{c\parallel}=3J_{c\perp}}$ {and ${n}$=30}.     

The force-free effects are modelled with {the} anisotropic power law {and the sharp dependence of $J_{c\parallel}$ with the magnetic field of figure \ref{sketch1.fig}(a)}. {Then,} the functional is minimized with the magnetic field from the previous time step like in the case of thin film.

\begin{figure}[htp]
\centering
 \subfloat[][]
{\includegraphics[trim=60 0 60 0,clip,height=5.5 cm]{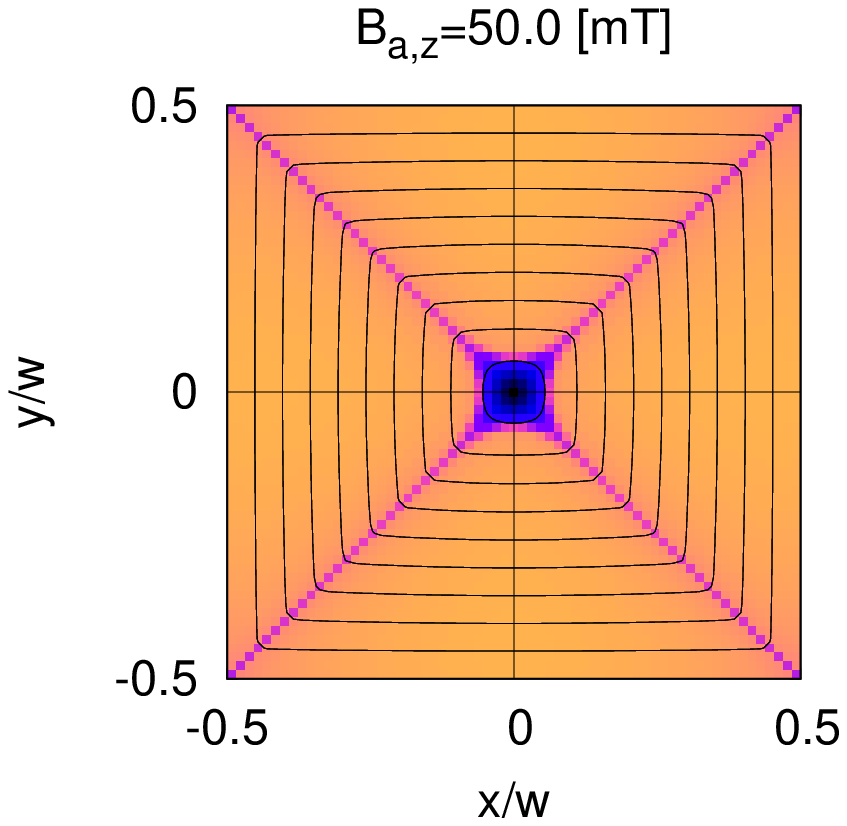}} 
 \subfloat[][]
{\includegraphics[trim=60 0 40 0,clip,height=5.5 cm]{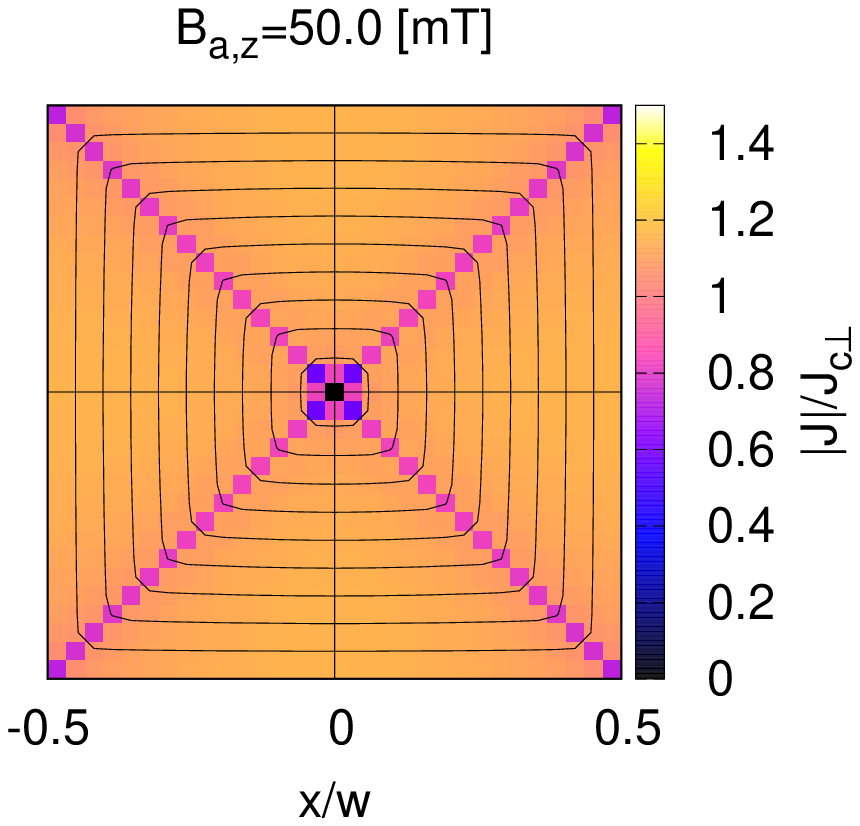}}\\
 \subfloat[][]
{\includegraphics[trim=40 0 50 0,clip,height=4.5 cm]{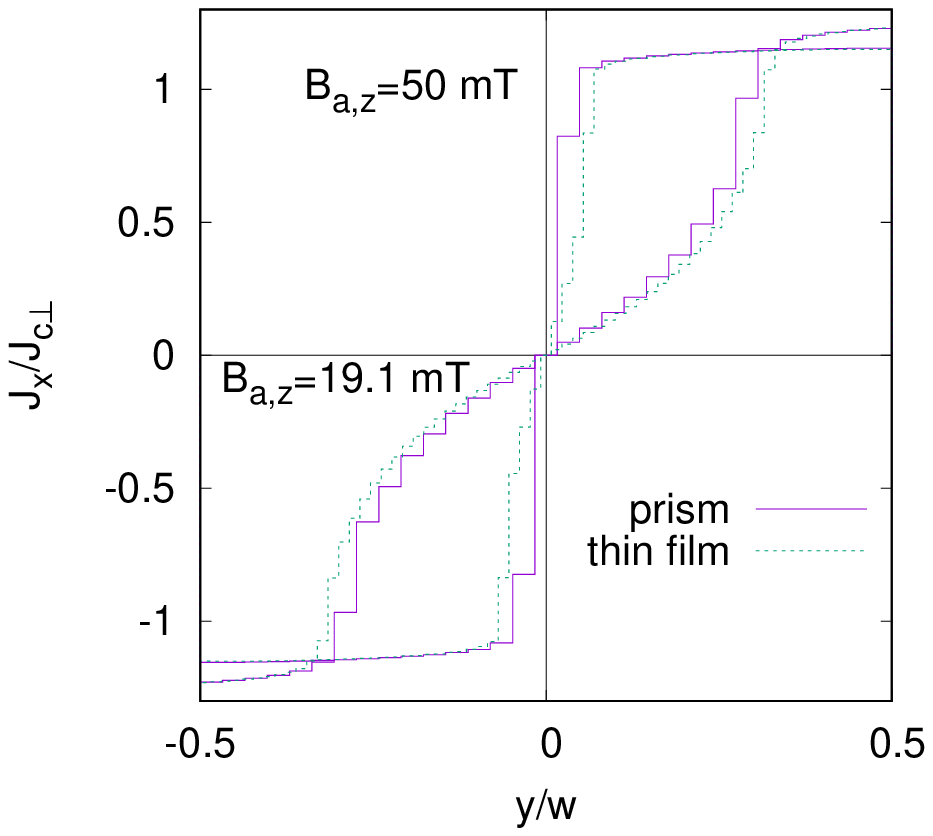}}  
 \subfloat[][]
{\includegraphics[trim=40 0 0	0 0,clip,height=4.5 cm]{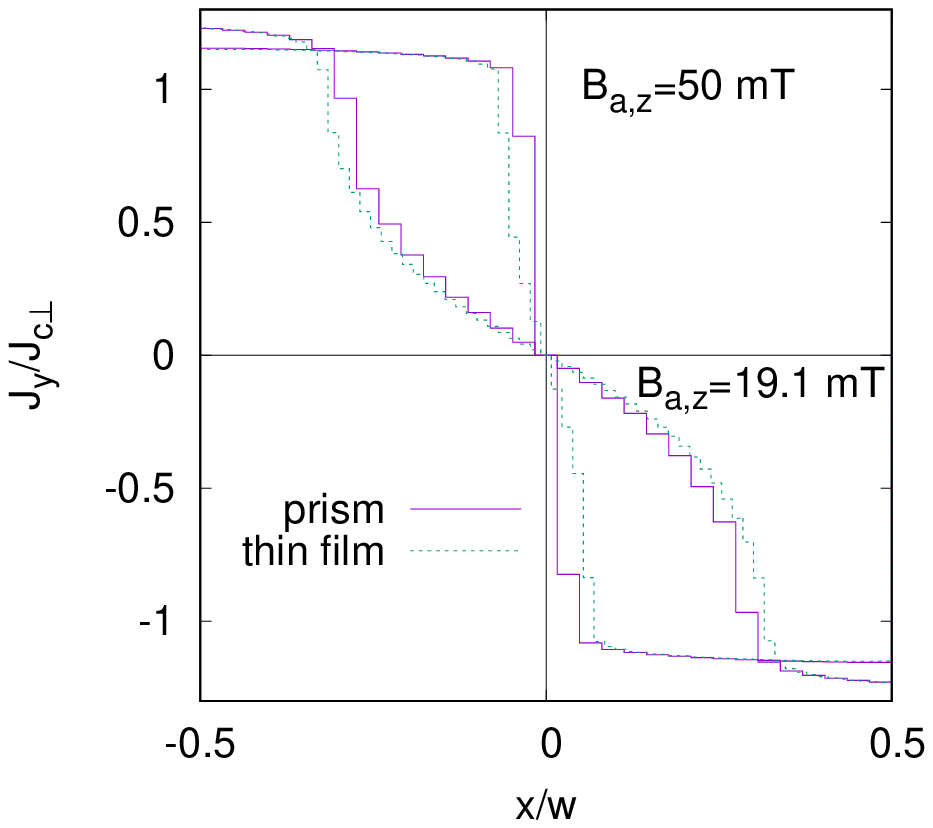}} 
\caption{{For low angles of the applied field, the thickness-averaged current density in a prim agrees with the thin film. Thickness-average current density and flux lines for a thin film (a) and prism (b). Plots (c) and (d) show the current density at the $x=0$ and $y=0$ lines, respectively. The parameters are $\theta=0$, $B_{az,m}$=50 mT and $f=$50 Hz.}}
\label{prism0.fig}
\end{figure}

The first case is with applied field ${\theta}$=0${\degree}$. We calculated the average current density over thickness. The penetration of the average current density into the prism at the peak of applied field is on figure \ref{prism0.fig}{, where we add the case of thin film for comparison.} There is {a} small difference in penetration depth of the current density, 
which can be explained by different number of elements in the {$x$ and $y$ directions}. The result of prism looks coarser, but we solved 10 times higher number of degrees of freedom compare{d to the} thin film{,} since the prism 
is a 3D object. The smaller penetration in thin film is more visible {in the profiles over the $x$ and $y$ directions on the sample center} [figure \ref{prism0.fig}(c),(d)]{, also for a lower applied field. For these planes, $J_z$ vanishes due to symmetry, although $J_z\neq 0$ at other regions \cite{Pardo17SST}.}

\begin{figure}[htp]
\centering
 \subfloat[][]
{\includegraphics[trim=60 0 60 0,clip,height=5.5 cm]{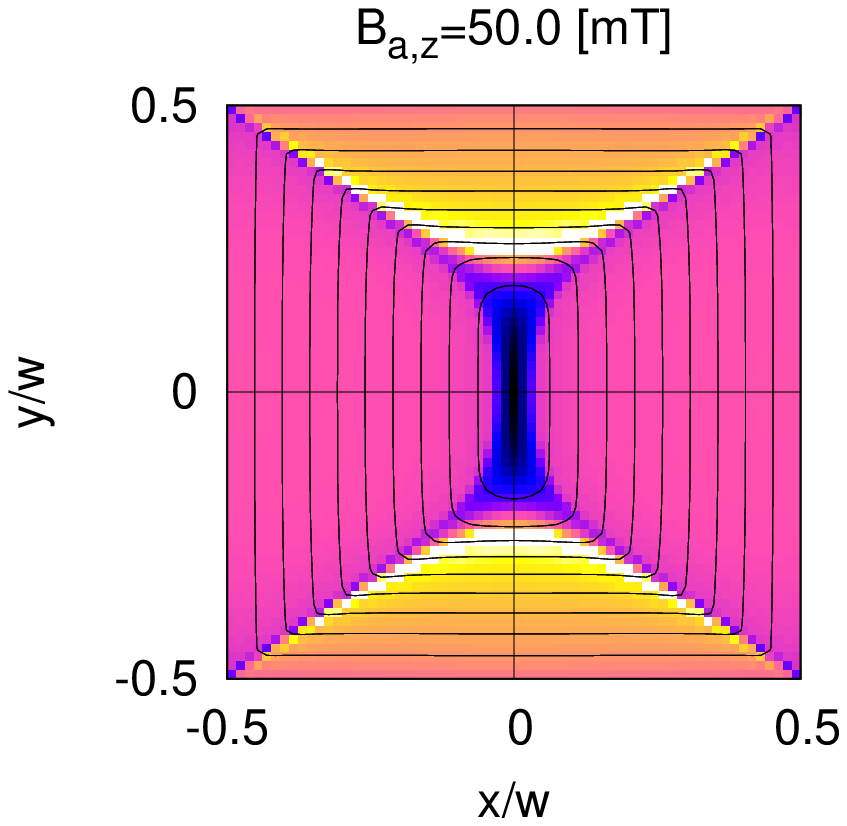}} 
 \subfloat[][]
{\includegraphics[trim=60 0 40 0,clip,height=5.5 cm]{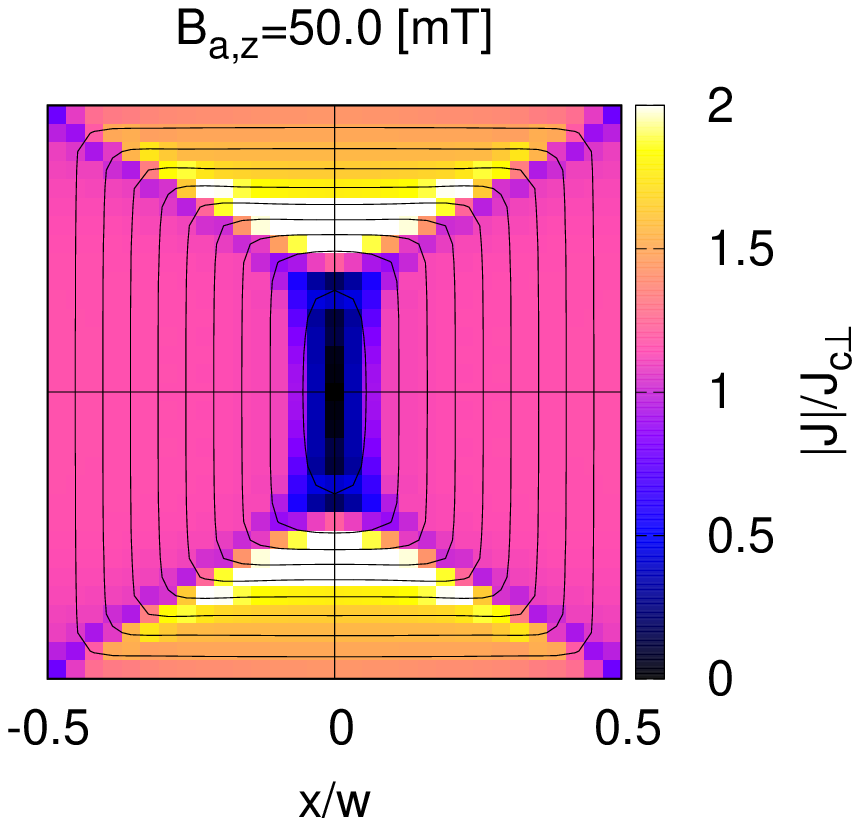}}\\
 \subfloat[][]
{\includegraphics[trim=40 0 50 0,clip,height=4.5 cm]{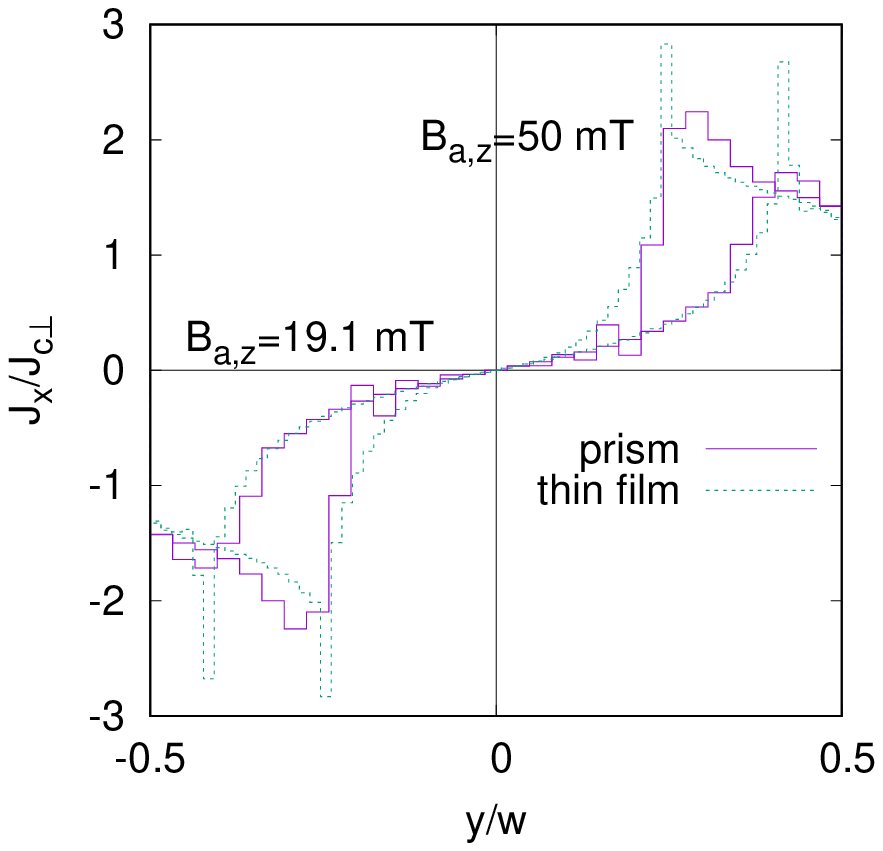}} 
 \subfloat[][]
{\includegraphics[trim=40 0 0 0,clip,height=4.5 cm]{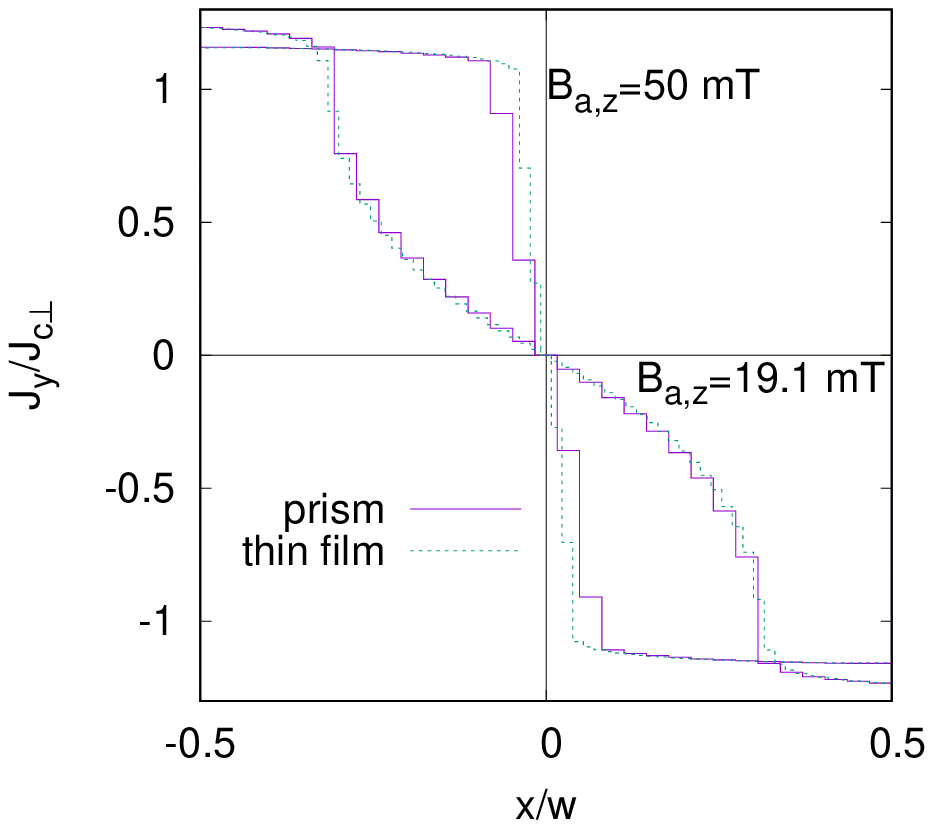}}  
\caption{{The same as figure \ref{prism0.fig} but for $\theta=45\dg$. The transverse component of the applied field amplitude of $B_{az,m}$=50 mT requires a total amplitude of $B_{am}=$70.7 mT.}}
\label{prism45.fig}
\end{figure}

The second case of prism is ${\theta=45\degree}$. The penetration depth of the average current density in the prism [figure \ref{prism45.fig}(b)] agrees with the thin film case [figure \ref{prism45.fig}(a)]. The agreement is as well in the {lines of} ${x=0}$ and ${y=0}$ [figure \ref{prism45.fig}(c),(d)]. 
The ${J_y}$ component of the current density is around ${J_{c\perp}}$ [figure \ref{prism45.fig}(d)], but ${J_x}$ is 2 times higher. {The reason of the higher magnitude of $J_x$ is that the applied field has a component in the $x$ direction, causing force-free effects. This also causes that $J_x$ at the penetration front reaches $J_{c\parallel}$ in the thin film, since $B_z$ there vanishes (figure \ref{prism45.fig}). } The penetration depth in the prism is smaller, because of the thicker cells.

\begin{figure}[htp]
\centering
 \subfloat[][]
{\includegraphics[trim=60 0 60 0,clip,height=5.5 cm]{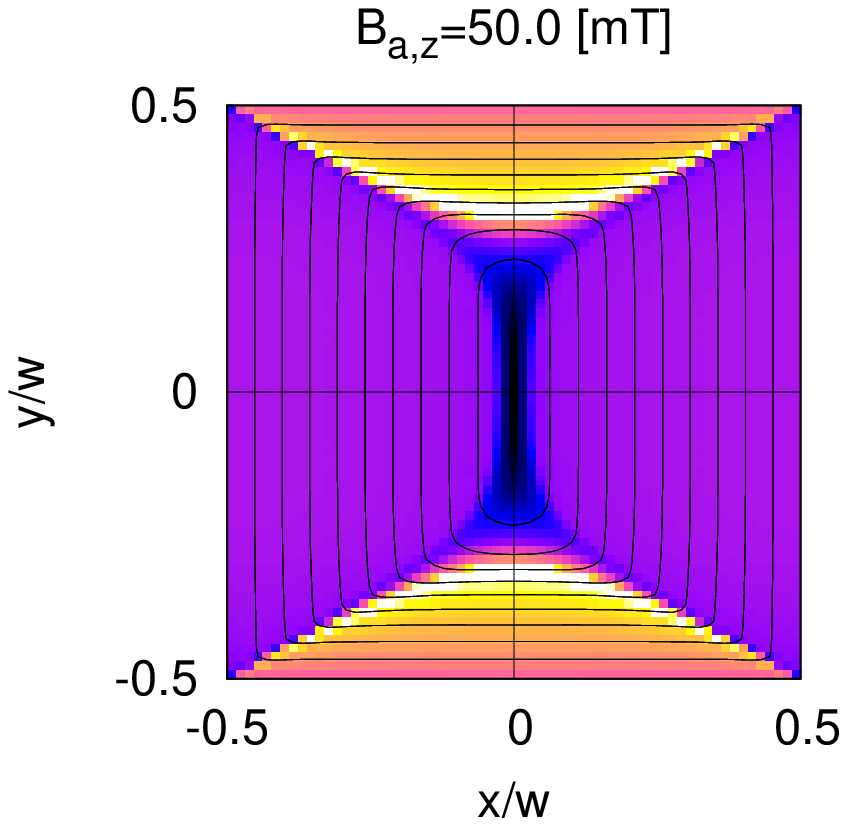}} 
 \subfloat[][]
{\includegraphics[trim=60 0 40 0,clip,height=5.5 cm]{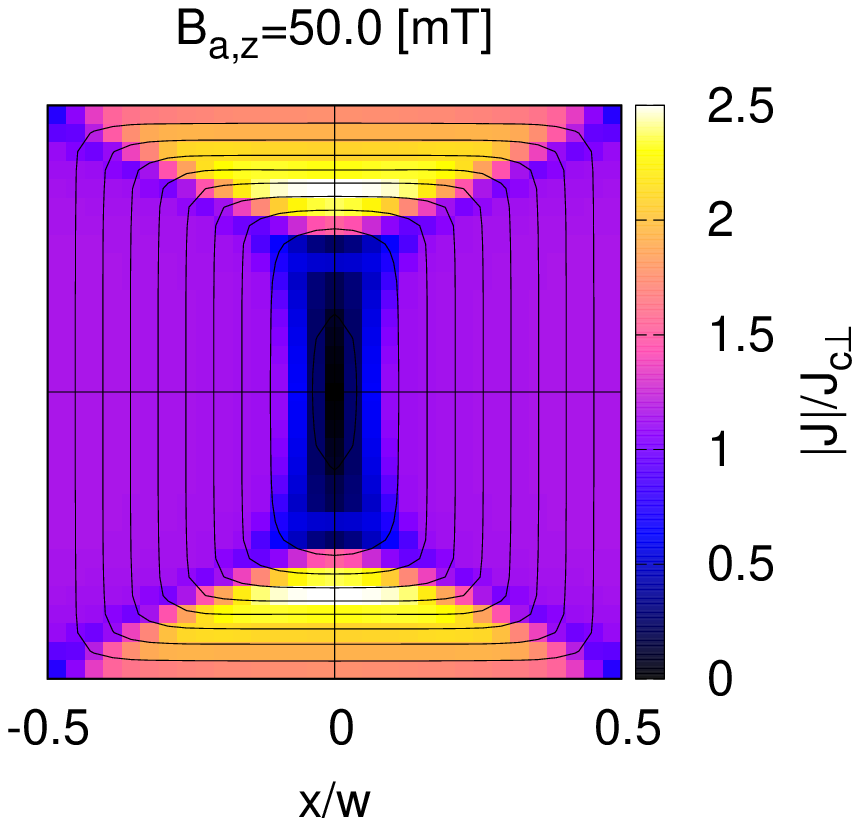}}\\
 \subfloat[][]
{\includegraphics[trim=40 0 50 0,clip,height=4.5 cm]{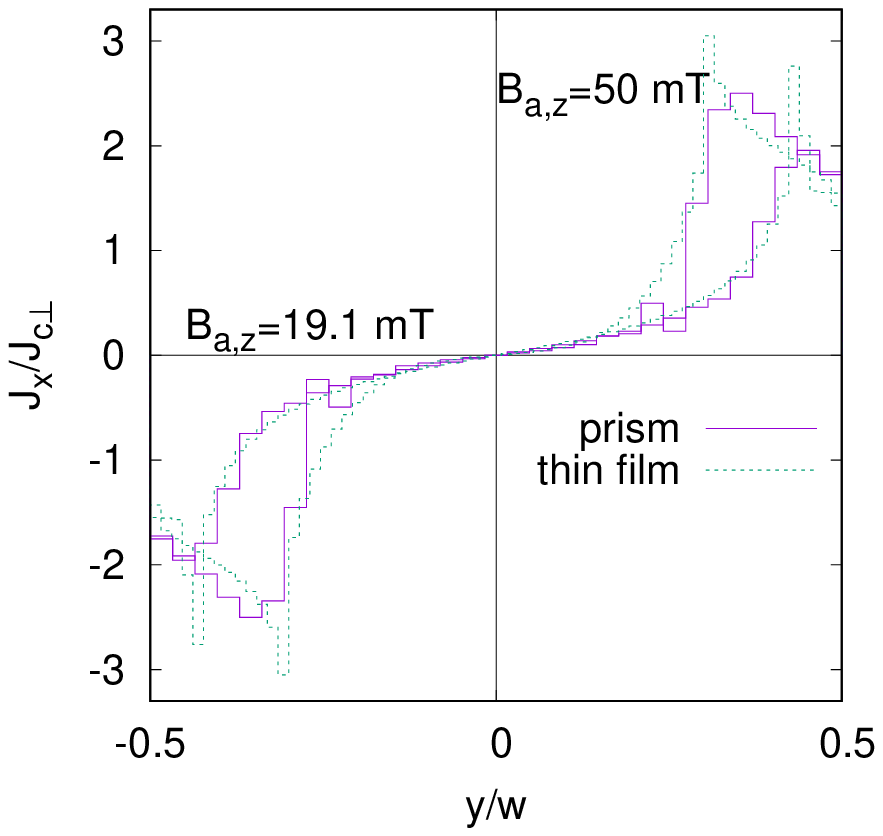}}
 \subfloat[][]
{\includegraphics[trim=40 0 0 0,clip,height=4.5 cm]{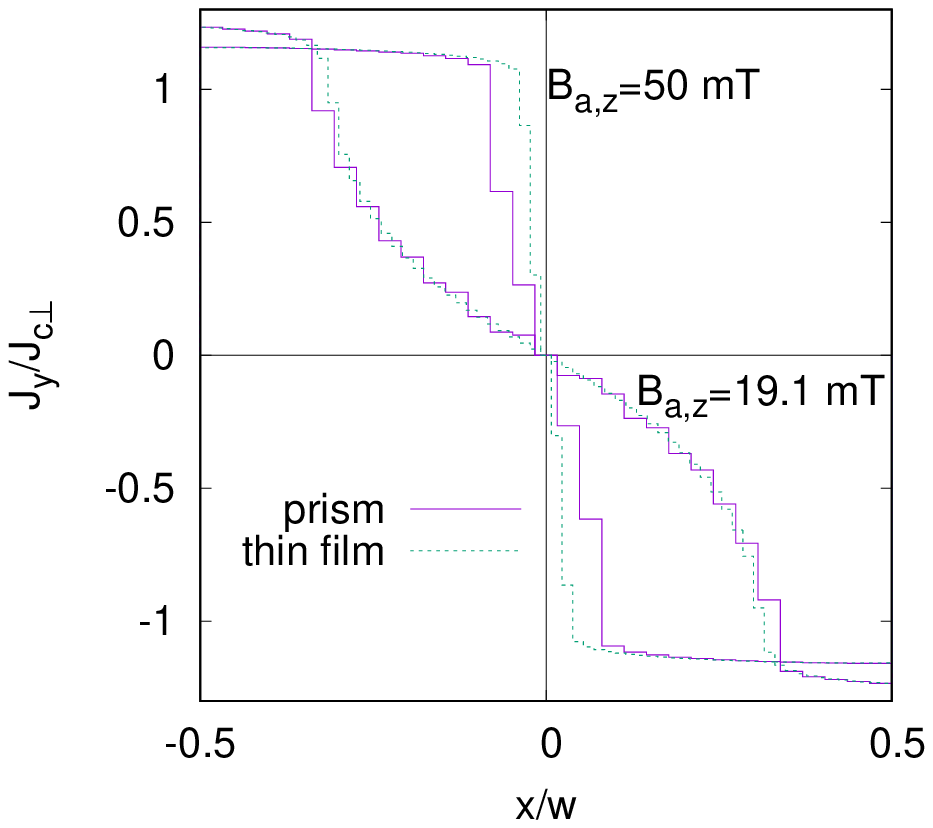}}
\caption{{The same as figure \ref{prism0.fig} but for $\theta=60\dg$ ($B_{az,m}$=50 mT and $B_{am}=$100 mT). A slightly lower critical-current penetration can be observed from the prism (b,d) compared to the film (a,c).}}
\label{prism60.fig}
\end{figure}

\begin{figure}[htp]
\centering
 \subfloat[][]
{\includegraphics[trim=60 0 60 0,clip,height=5.5 cm]{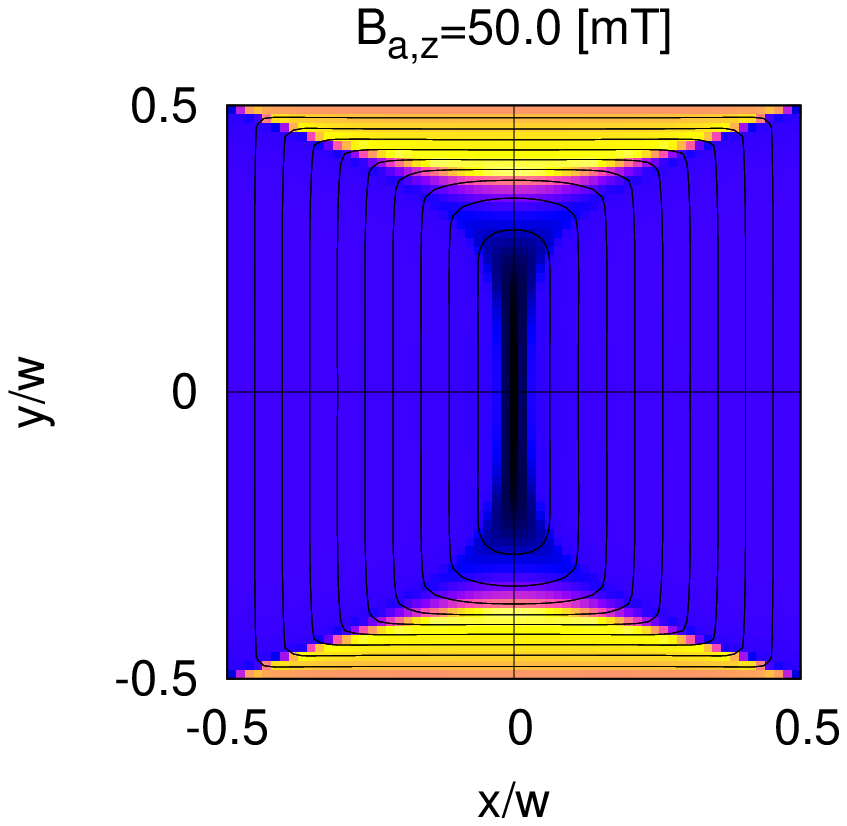}} 
 \subfloat[][]
{\includegraphics[trim=60 0 40 0,clip,height=5.5 cm]{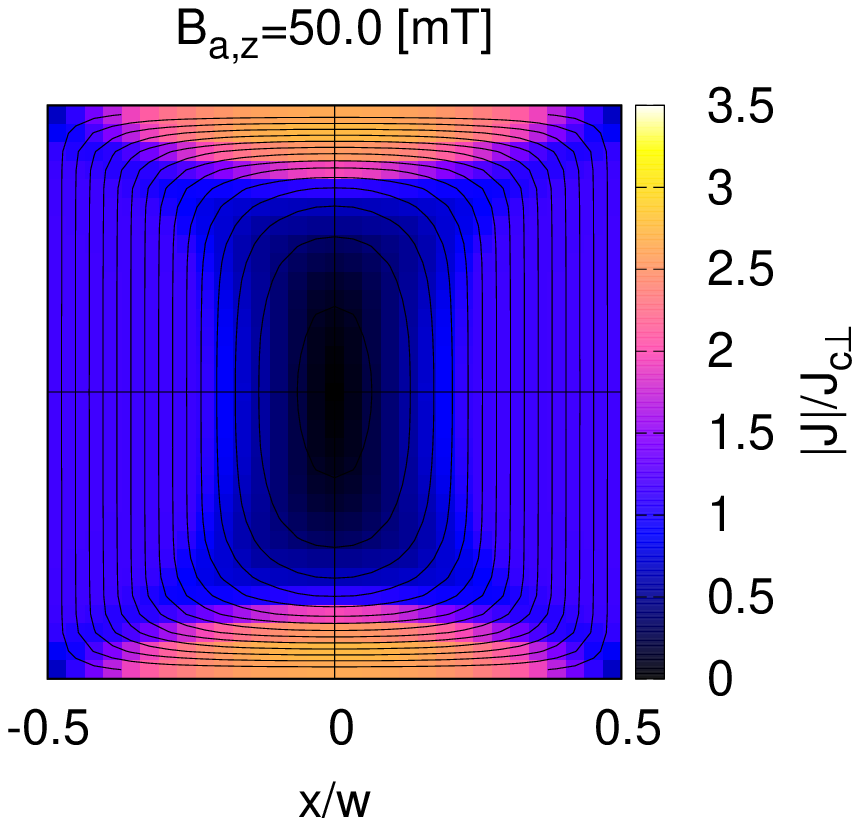}}\\
 \subfloat[][]
{\includegraphics[trim=40 0 50 0,clip,height=4.5 cm]{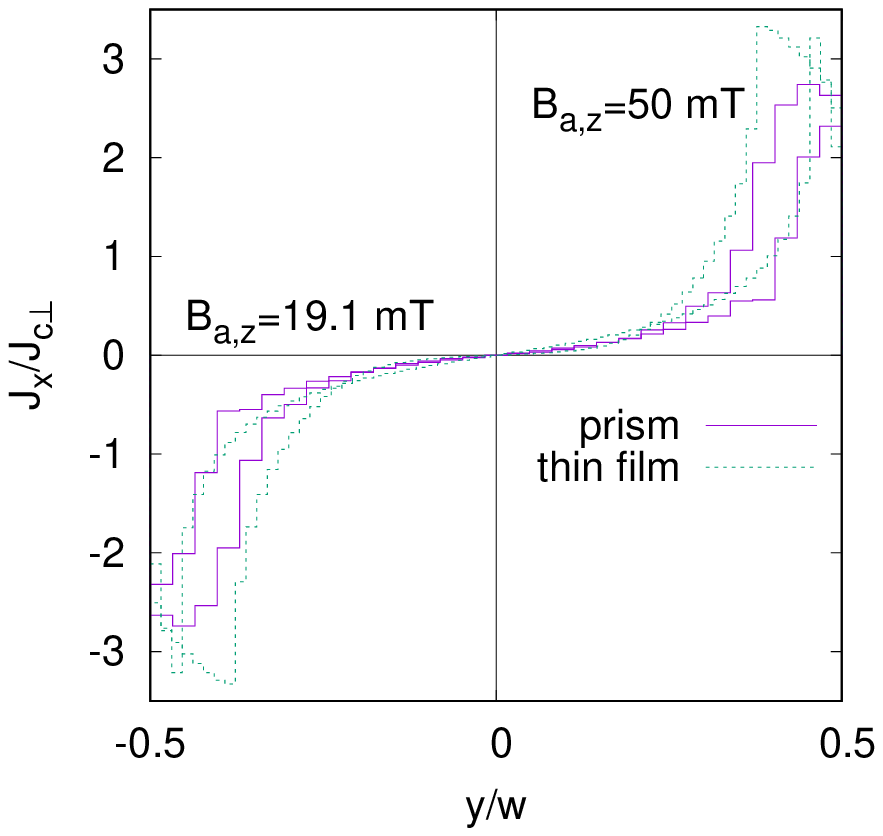}}
 \subfloat[][]
{\includegraphics[trim=40 0 0 0,clip,height=4.5 cm]{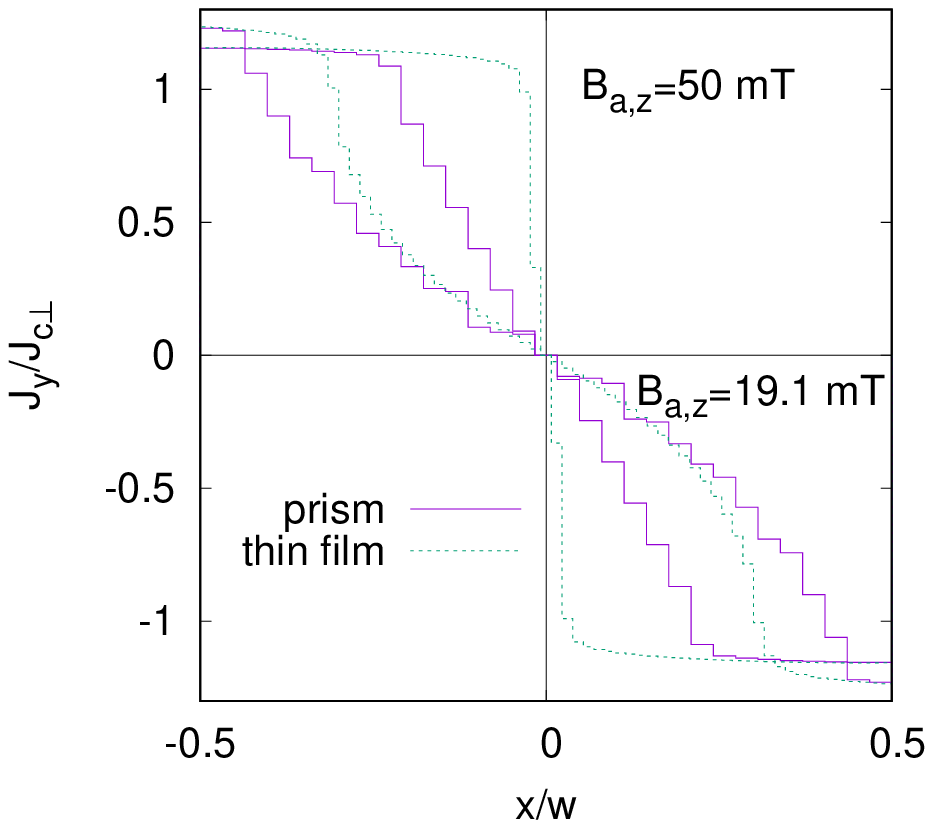}} 
\caption{{The same as figure \ref{prism0.fig} but for $\theta=80\dg$ ($B_{az,m}$=50 mT and $B_{am}=$287.9 mT). The prism (b,d) presents a substantially lower penetration of the critical-current density than the film (a,c).}}
\label{prism80.fig}
\end{figure}

The last two cases with ${\theta=60\degree,80\degree}$ are similar to the appropriate cases of thin film{, although with certain differences}. For the angle ${\theta=60\degree}$ there is lower penetration depth from the right and left sides [figure \ref{prism60.fig}(b)] than the thin film [figure \ref{prism60.fig}(a)]. The angle ${\theta=80\degree}$ has even lower penetration depth from these sides [figure \ref{prism80.fig}(b)] compared to thin film [figure \ref{prism80.fig}(a)]. {The current profiles along the $x$ and $y$ directions show the same behaviour of lower current penetration [figure \ref{prism60.fig}(c),(d), \ref{prism80.fig}(c),(d)]. The cause of lower penetration depth along both $x$ and $y$ directions is due to the prism finite thickness.}
{Since ${\theta\neq 0\degree}$, there is a significant ${J_z}$ component of the current density, which is around ${J_{c\perp}}$[figure \ref{3D80.fig}(c) ${\theta=80\degree}$].}  	

\begin{figure}[htp]
\centering
 \subfloat[][]
{\includegraphics[trim=0 0 0 0,clip,height=4.0 cm]{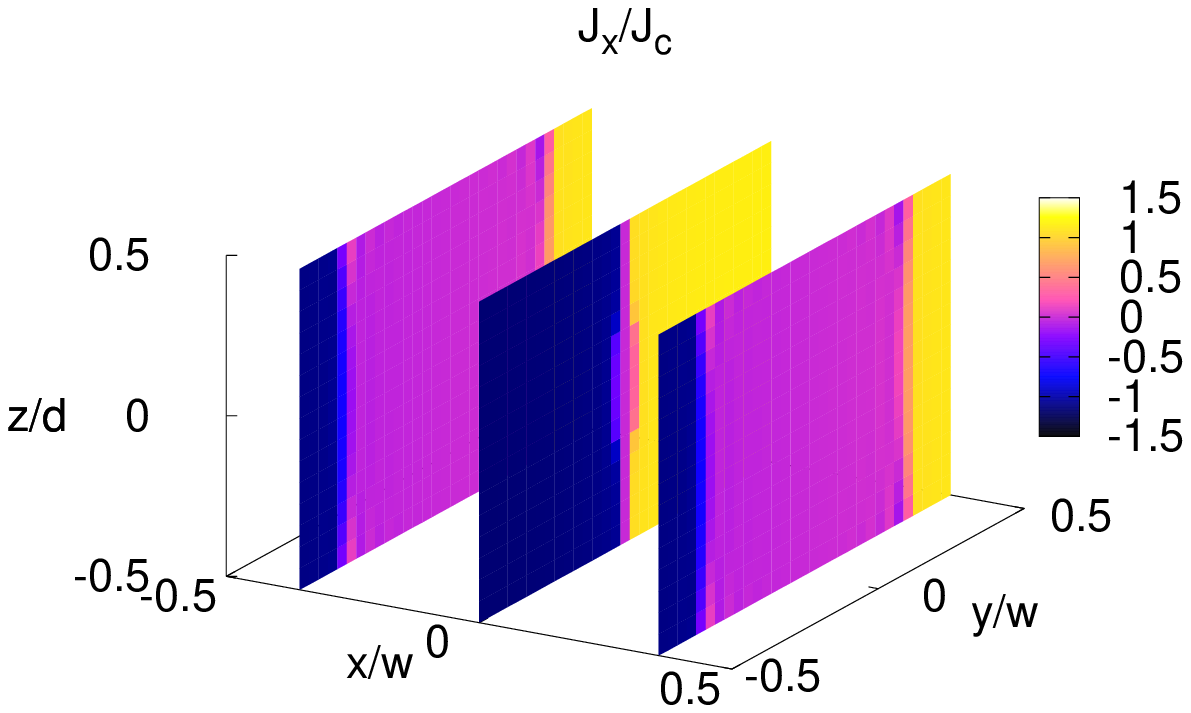}} 
 \subfloat[][]
{\includegraphics[trim=0 0 0 0,clip,height=4.0 cm]{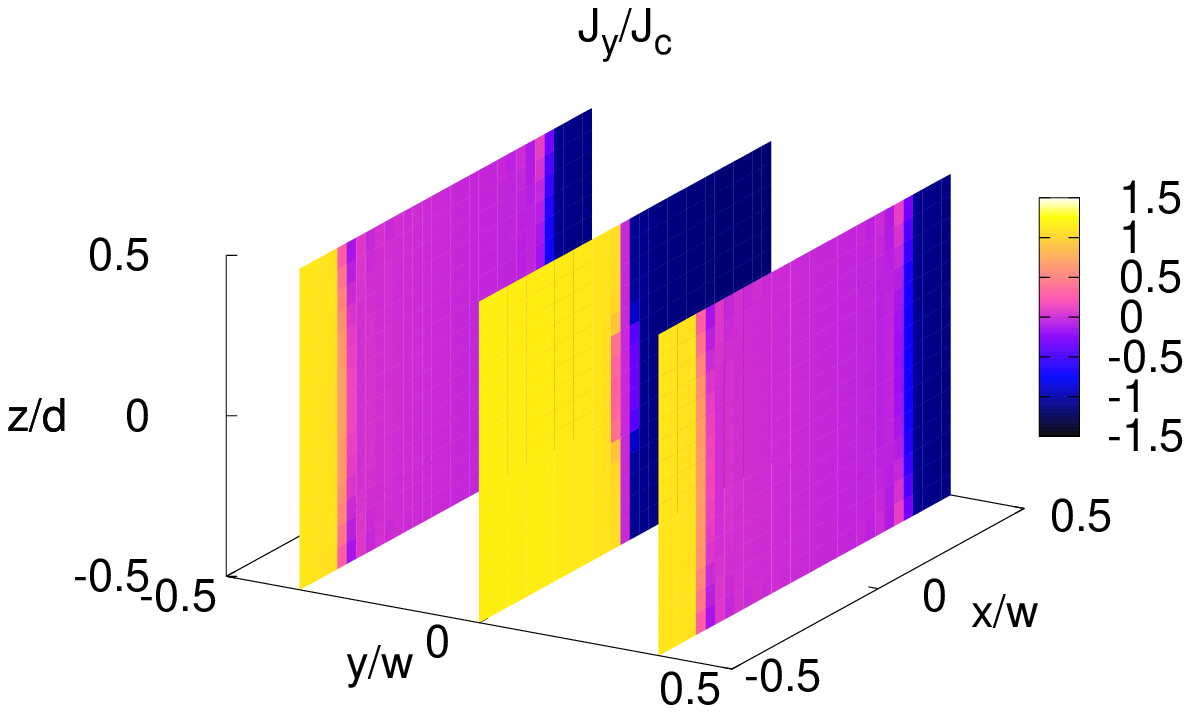}} 
 \subfloat[][]
{\includegraphics[trim=0 0 0 0,clip,height=4.0 cm]{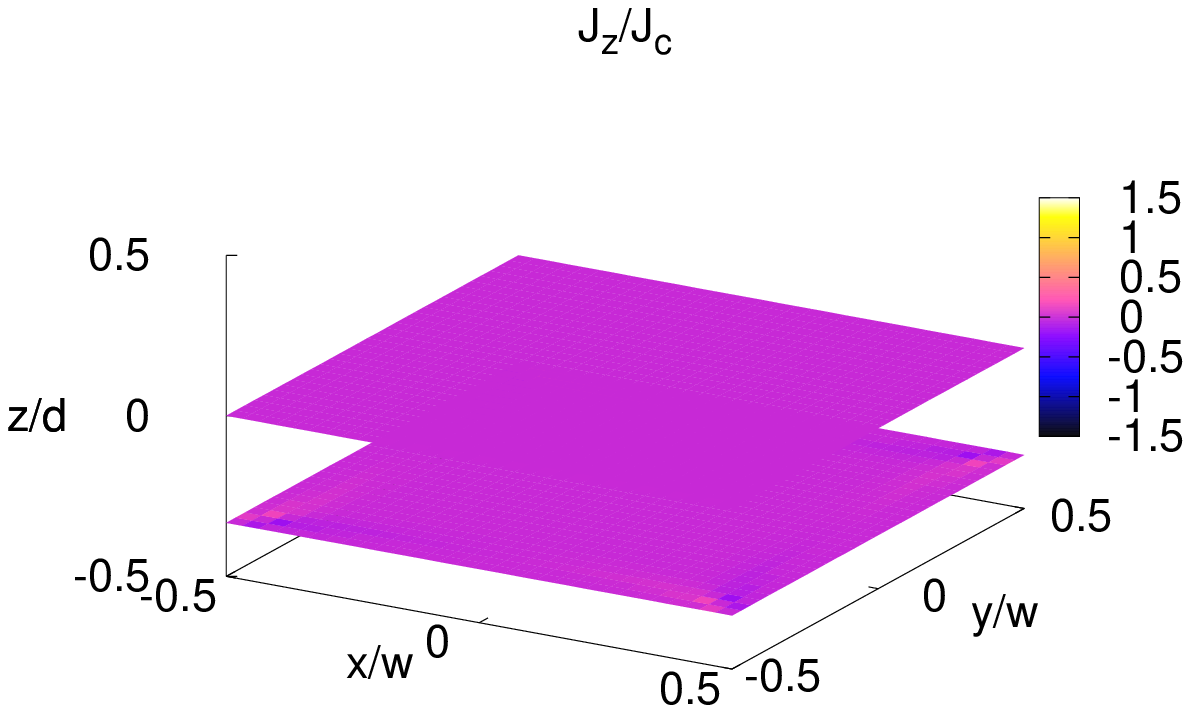}} 
\caption{{Current density in a prism with force-free anisotropy at the peak of the applied field and $w=$12 mm, $d$=1 mm, $\theta=0\dg$, and $B_{az,m}=$50 mT.} The current density components {are}: (a) ${J_x}$ (b) ${J_y}$ (c) ${J_z}$. {Note that the plotted planes in (b) are not the same as (a,c).}}
\label{3D0.fig}
\end{figure}

\begin{figure}[htp]
\centering
 \subfloat[][]
{\includegraphics[trim=0 0 0 0,clip,height=4.0 cm]{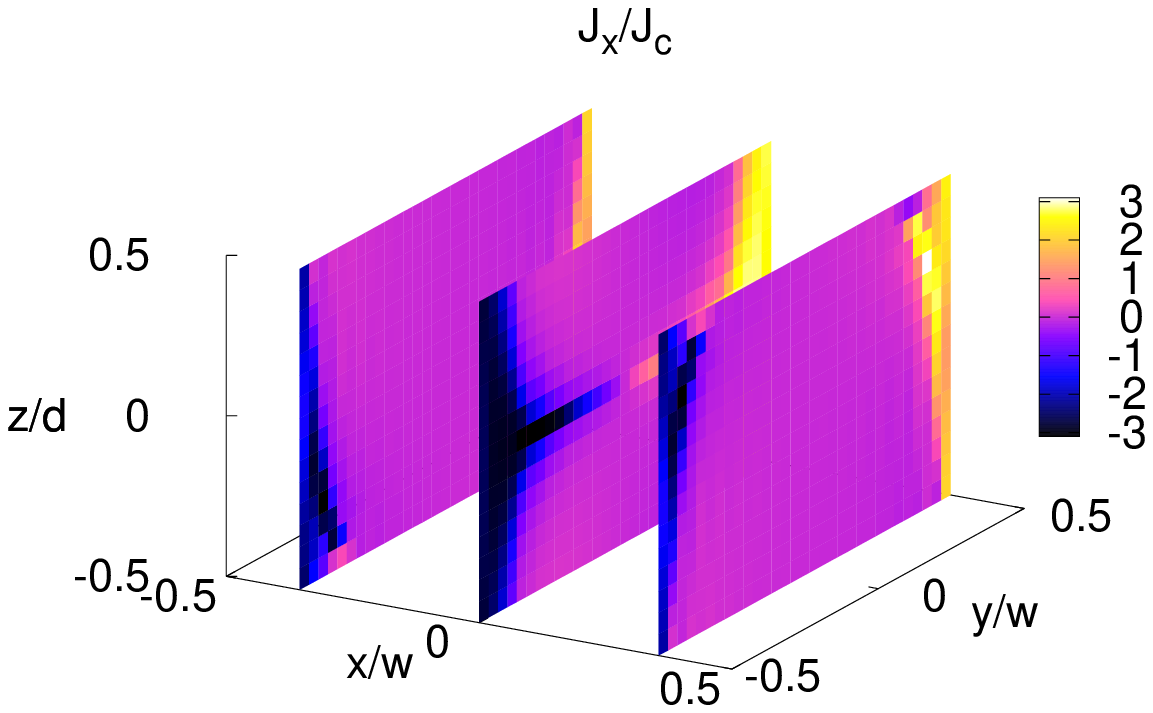}} 
 \subfloat[][]
{\includegraphics[trim=0 0 0 0,clip,height=4.0 cm]{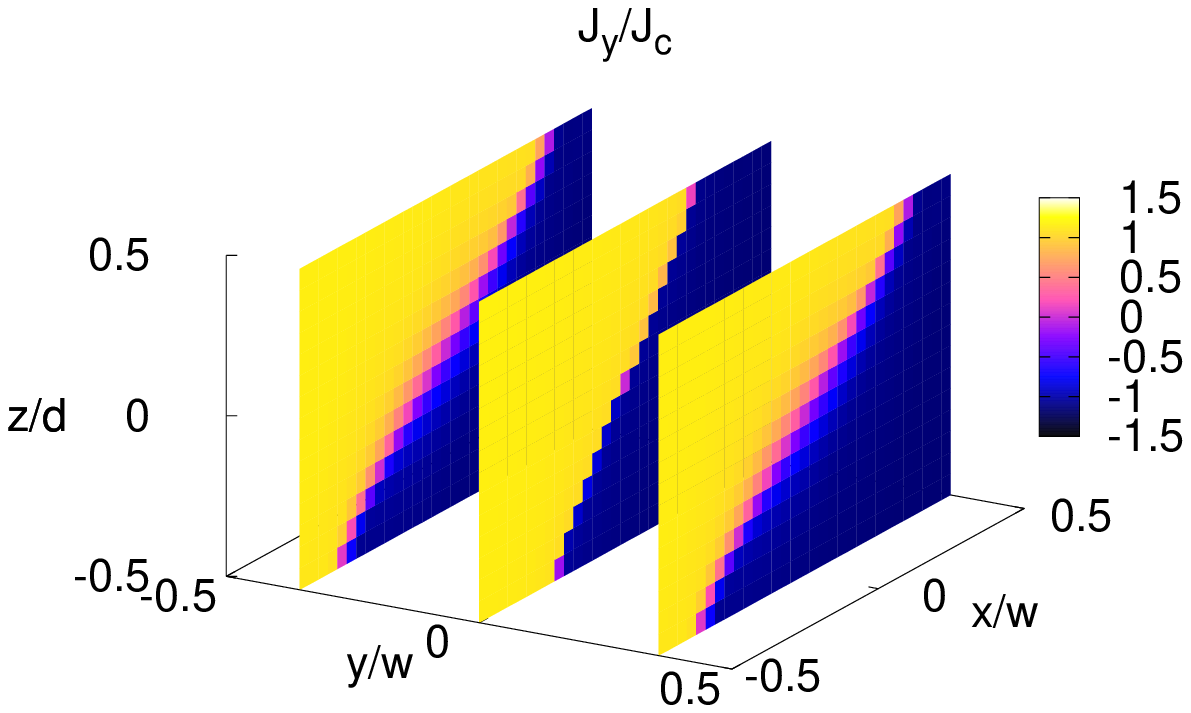}} 
 \subfloat[][]
{\includegraphics[trim=0 0 0 0,clip,height=4.0 cm]{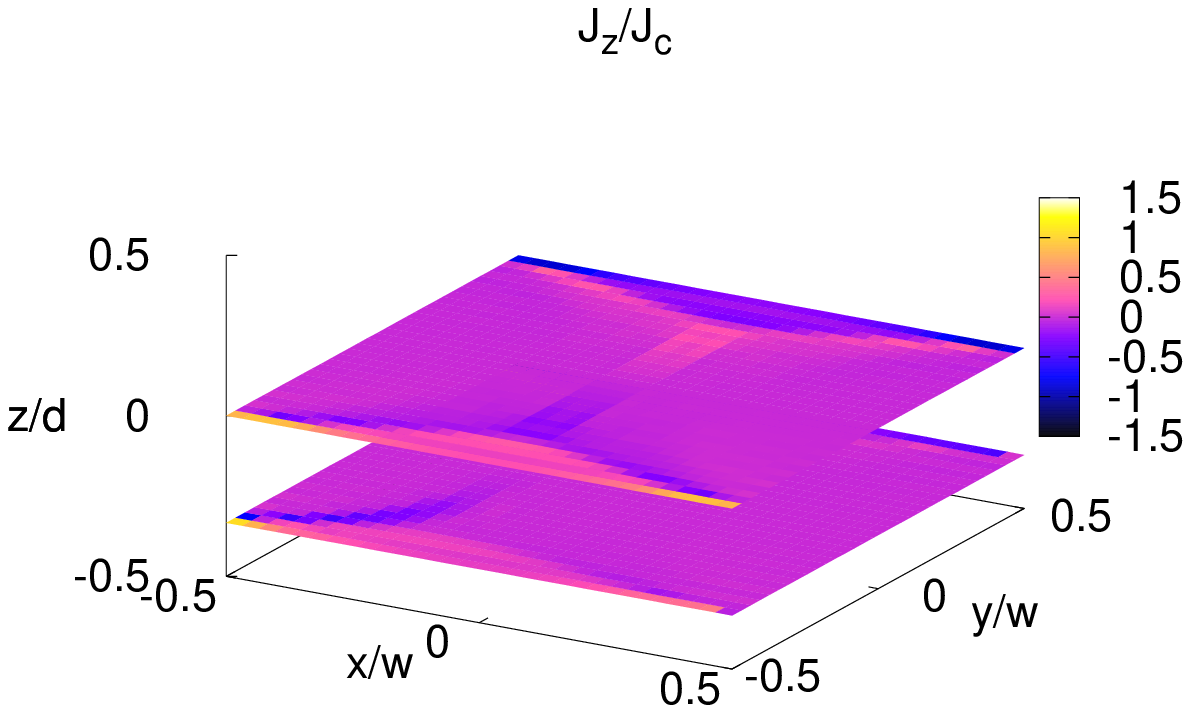}} 
\caption{{The same as figure \ref{3D0.fig} but for $\theta=80\dg$ ($B_{am}=287.9$ mT).}}
\label{3D80.fig}
\end{figure}

Finally, we compare the 3D current paths in the prism at the peak of the applied field for {the anisotropic case and} two applied field angles ${\theta=0\degree}$ and ${80\degree}$. For the first {angle}, the sample is fully saturated as seeing the mid planes perpendicular to the ${x}$ and ${y}$ axis [figure \ref{3D0.fig}(a),(b)] and hence {$J_z$ almost vanishes} [figure \ref{3D0.fig}(c)]. For the second {angle} $(\theta=80\degree)$, the ${J_y}$ component of the current density is also saturated {in most of the volume} [figure \ref{3D80.fig}(b)]. Now, the border between positive and negative ${J_y}$ component {follows roughly the direction of the applied magnetic field}. The ${J_x}$ component is not saturated in the sample [figure \ref{3D80.fig}(a)] and the highest penetration depth is at the centre of the prism. {Since the current loops are almost perpendicular to the angle of the applied field, there exists a substantial $J_z$ component [figure \ref{3D80.fig}(c)].}

\subsubsection{Magnetization loops in prisms.}

\begin{figure}[tbp]
\centering
 \subfloat[][]
{\includegraphics[trim=60 0 70 0,clip,width=8 cm]{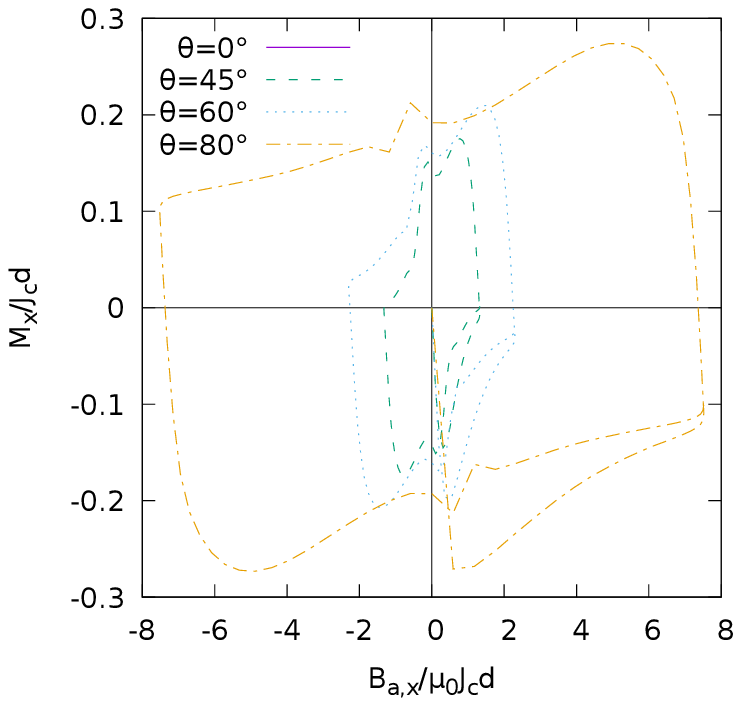}}
 \subfloat[][]
{\includegraphics[trim=60 0 70 0,clip,width=8 cm]{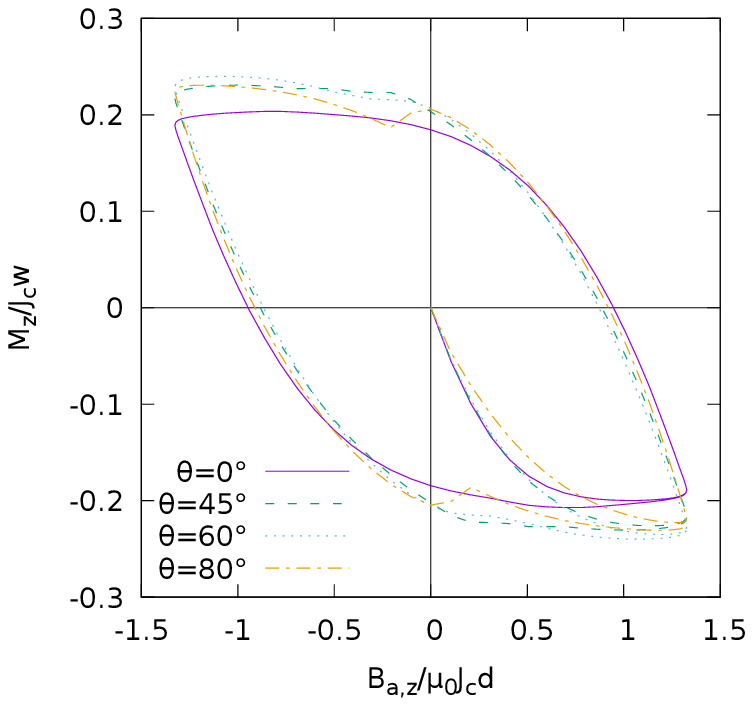}}
\caption{{The influence of the applied field angle to the magnetization loops in a prism is somewhat different than that of a thin film with the same sheet critical current density $J_cd$. Case of} anisotropic ${\vE(\vJ)}$ relation and constant ${J_{c\parallel}}$, ${J_{c\perp}}$, $n=$ 30, sinusoidal applied magnetic field {with} ${B_{a,z}=}$50.0 mT and ${f=}$50 Hz. (a) {The $x$ component of the magnetization }${M_x}$ component. (b) ${M_z}$.}
\label{loop_a.fig}
\end{figure}

\begin{figure}[tbp]
\centering
 \subfloat[][]
{\includegraphics[trim=60 0 70 0,clip,width=8 cm]{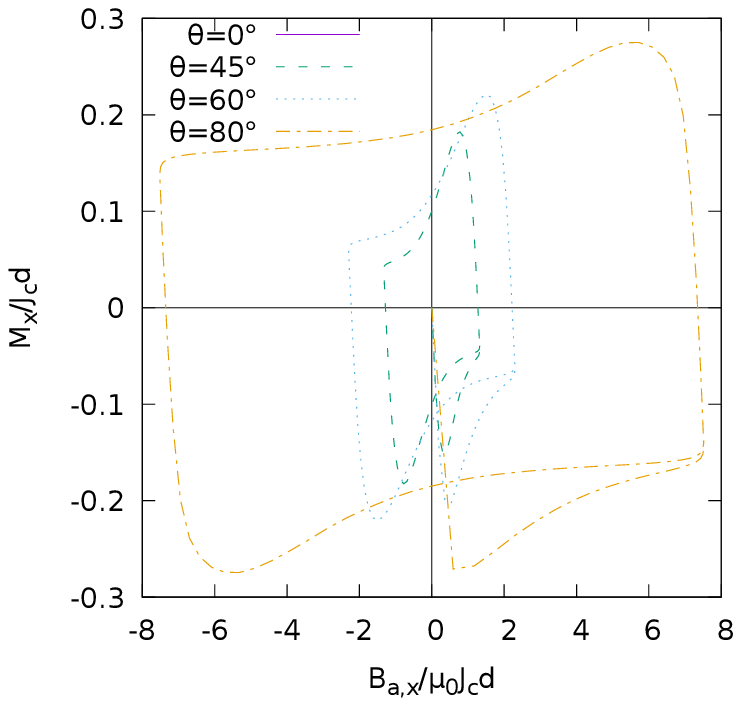}}
 \subfloat[][]
{\includegraphics[trim=60 0 70 0,clip,width=8 cm]{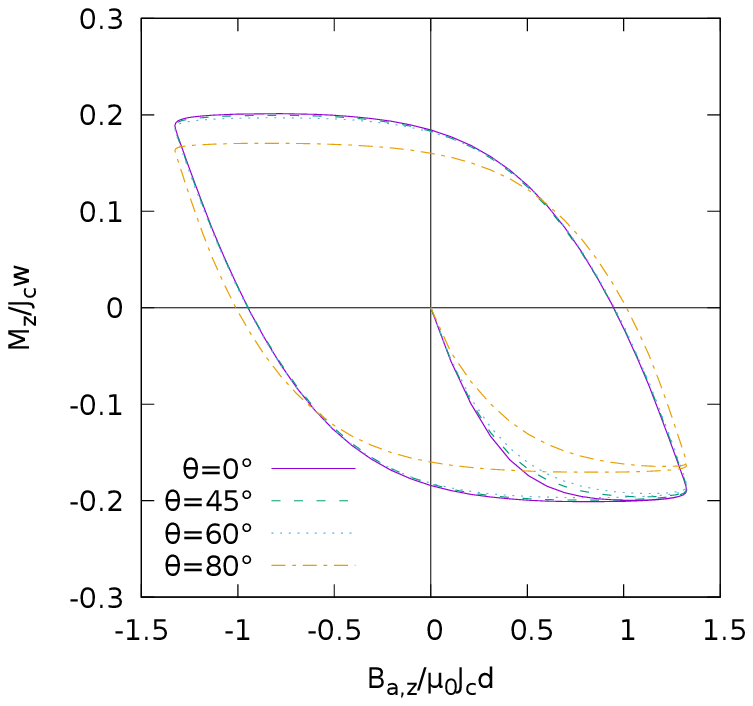}}
\caption{{The same as figure \ref{loop_a.fig} but for isotropic $\vE(\vJ)$. Several differences appear from the anisotropic case, such as the peaks close to the remanence in figure \ref{loop_a.fig}.}}
\label{loop_i.fig}
\end{figure}

We calculated the hysteresis loops for all previous cases (figure \ref{loop_a.fig}). In order to explain all effects, we {also analyzed} the same situation with isotropic power law 
(figure \ref{loop_i.fig}). The ${M_z}$ component of the magnetization is lower for higher applied magnetic field angle ${\theta}$ [figure \ref{loop_i.fig}(b)]. This is because the {path of the screening current loops} tilts away from the {$xy$} plane. The ${M_x}$ component is zero for ${\theta=0\degree}$ [figure \ref{loop_i.fig}(a)], since the current path is only in the {$xy$} plane. {This also causes and increase of the $M_x$ component with increasing $\theta$. This geometry effect can be reduced by decreasing the prism thickness. Consistently, $M_x$ vanishes at $\theta=0$ because the current loops are mainly in the $xy$ plane and the remaining bending in the $z$ direction is symmetric (see figure 5 of \cite{Pardo17SST}).}

The hysteresis loops with anisotropic ${\vE(\vJ)}$ relation have more effects. {On one hand, increasing the angle $\theta$ enlarges the region with $|\vJ|\approx J_{c\parallel}$, increasing also $M_z$. On the other hand, by increasing $\theta$, the tilt increases, reducing $M_z$. The result is an increase in $M_z$ with $\theta$ but for $\theta=80\dg$ this increase is smaller than for the thin film [figure \ref{loop_a.fig}(b)].} The magnetization in {the} ${x}$ direction [figure \ref{loop_a.fig}(a)] shows {mostly} the same behavior as isotropic case. The difference is only in {a} peak of the magnetization around the zero applied field. {This} peak of the magnetization appears {for} both components, ${M_x}$ and ${M_z}$. {The reason of this peak is the following. For very small magnetic fields, the self-field dominates. Close to the top (highest $z$) and bottom (lowest $z$) of the sample, the self-field is parallel to the surface. Then, part of the sample experiences a local magnetic field parallel to the current density, increasing $J_c$ towards $J_{c\parallel}$. For applied fields much larger than the self-field, the magnetic field follows the direction of the applied field. This applied field is not perfectly parallel to the surface, causing a lower $J_c$.}

\begin{figure}[tbp]
\centering
{\includegraphics[trim=60 0 70 0,clip,width=8 cm]{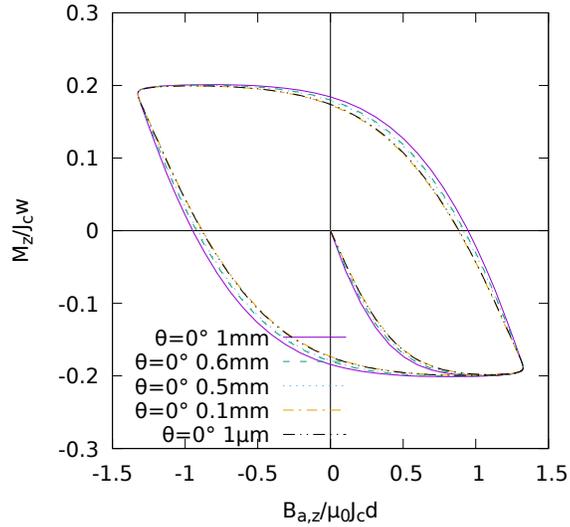}}
\caption{{For isotropic $\vE(\vJ)$ in prisms and $\theta=0$, the normalized magnetization converges to the thin film situation. The properties are $n=$30, sinusoidal applied magnetic field ${B_{a,z}=}$50.0 mT and ${f=}$50 Hz.}}
\label{loop_d1.fig}
\end{figure}

\begin{figure}[tbp]
\centering
 \subfloat[][]
{\includegraphics[trim=60 0 70 0,clip,width=8 cm]{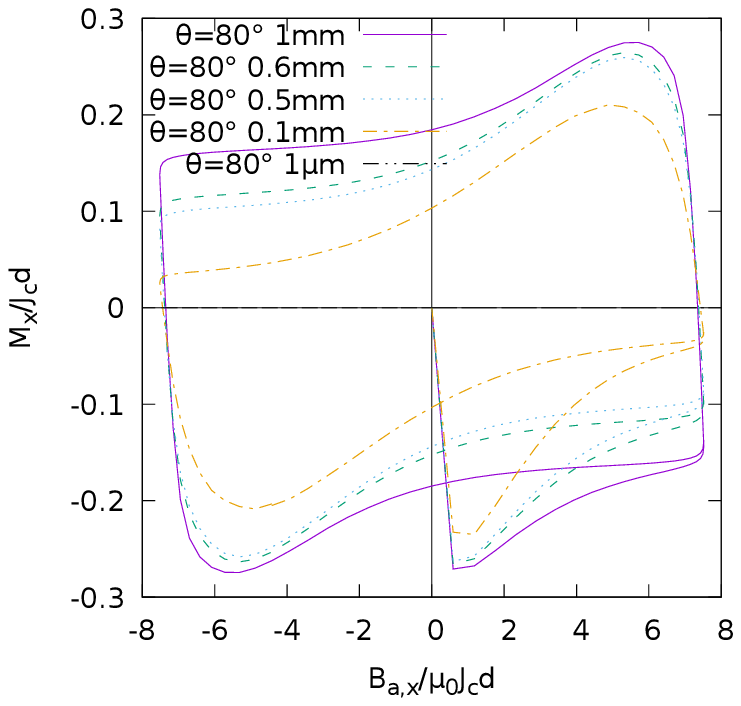}}
 \subfloat[][]
{\includegraphics[trim=60 0 70 0,clip,width=8 cm]{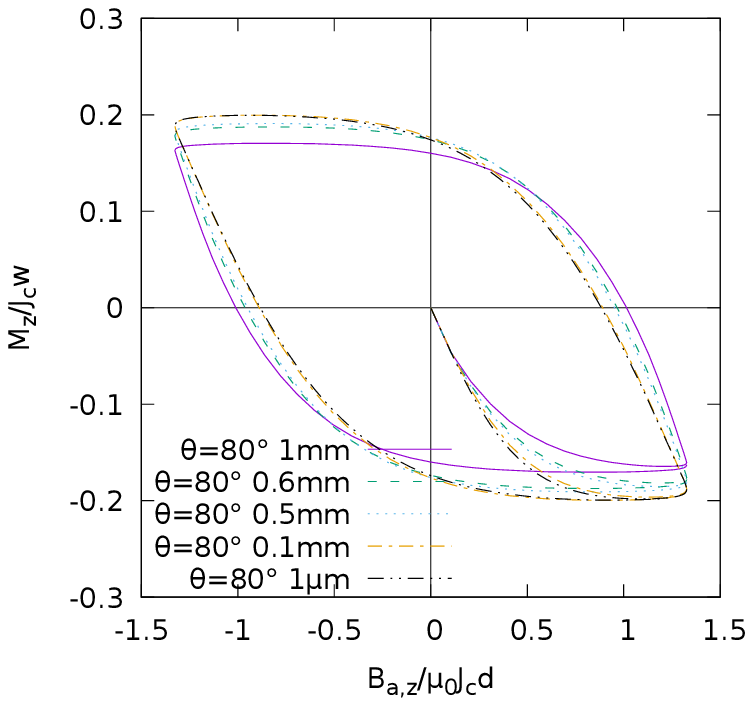}}
\caption{Hysteresis loops {of prisms with several thicknesses $d$ and the same width $w$} with isotropic ${\vE(\vJ)}$ relation and constant $J_{c}$, {$n=$30}, and sinusoidal applied magnetic field with ${B_{a,z}=}$50.0 mT and ${f=}$50 Hz. (a) ${M_x}$ component. (b) ${M_z}$ component}
\label{loop_d.fig}
\end{figure}

Another calculation with isotropic $\vE(\vJ)$ relation shows {the} geometry effects due to different thickness of the prism {($d=$1,0.6,0.5,0.1) while keeping a constant sheet current density. First we check that for only perpendicular applied field, the prism results approach to the thin film by reducing the thickness. Figure \ref{loop_d1.fig} shows that the normalized $z$ component $M_z/J_cw$ is roughly the same for all thicknesses $d$. This figure also tells us that the magnetic moment $m_z$ almost does not depend on $d$, since $M_z/J_cw=m_z/J_cdw^3$ and we keep both $J_cd$ and $w$ constant. For a magnetic field angle $\theta$ of $80\dg$, $M_z$ increases with decreasing the sample thickness, since the screening current is forced to flow closer to the $xy$ plane [figure \ref{loop_d.fig}(b)].} However, the {other normalized component, $M_x/J_cd$,} increases with {the} thickness $d$, due to the increase of the {area of the projection of the current loops in the $yz$ plane}.

\section{Conclusions}

This article systematically studied the anisotropic force-free effects in superconducting thin films and prisms under uniform applied magnetic field making on angle {$\theta$} with the surface. 
In order to better understand all effects, we performed modelling with isotropic and anisotropic ${\vE(\vJ)}$ relation due to force-free effects. 

For this purpose, we use the MEMEP 3D numerical method \cite{Pardo17JCP}. We further developed the model in order to enable elongated cells, to reduce the total number of elements or 
enable to model relatively long {or thin} structures without further increasing the total number of elements. In particular, we studied the elliptical double critical state model 
with a continuous $\vE(\vJ)$ relation\cite{Badia12SST}.   

In the thin film force-free model, we calculated the gradual penetration of the current density. We found at the remanent state that ${J}$ decreases to $J_{c\perp}$ and {the} magnetization increases 
with the angle $\theta$. The magnetization of the isotropic film is the same for all applied field angles, when comparing for the same perpendicular component of the applied magnetic field and its amplitude. 
The anisotropic model, both with and without $J_c(B)$ dependence, shows {a minimum} of the magnetization at the remanent state {for $\theta\neq 0$}. The cause is the absence of any parallel component of the local magnetic field 
to the current density, avoiding $J_c$ enhancement due to force-free effects. In superconducting prisms, we observed 3D current paths. The average current density over thickness shows good agreement 
with thin film sample. However, for high applied magnetic field angles there appear small differences. The $M_x$ component is increasing with the angle $\theta$, because of the significant increase of $J_z$. 
The $M_z$ slightly decreases due to the tilt in the screening currents. The magnetization loops show a peak after the remanent state {due to the influence of the parallel component of the self-field, increasing $J_c$ up to $J_{\parallel}$ at part of the sample. This effect is not present for the thin film approximation because the parallel component of the self-field is neglected.} {Calculations for several prism thicknesses down to 100 $\mu$m support the validity of the results}.
  
We expect that the thin film geometry may not be a good approximation for study force-free effects {in magnetization measurements. This} study confirmed that the MEMEP 3D method is suitable for any ${\vE(\vJ)}$ relation and it can be solved with {a relatively} high number of degrees of freedom {and relatively thin samples} in 3D space. Further work could be to investigate different shapes of the sample and speed up 
the calculation, maybe by multi-pole expansion.


\section*{Acknowledgements}

Authors acknowledge the use of computing resources provided by the project SIVVP, ITMS 26230120002 supported by the Research $\&$ Development Operational Programme funded by the ERDF, 
the financial support of the Grant Agency of the Ministry of Education of the Slovak Republic and the Slovak Academy of Sciences (VEGA) under contract No. 2/0097/18, 
and the support by the Slovak Research and Development Agency under the contract No. APVV-14-0438. 


\appendix

\section{Elongated cells}
\label{s.e_cell}

\begin{figure}[tbp]
\centering
 \subfloat[][]
{\includegraphics[trim=0 0 0 0,clip,height=3.5 cm]{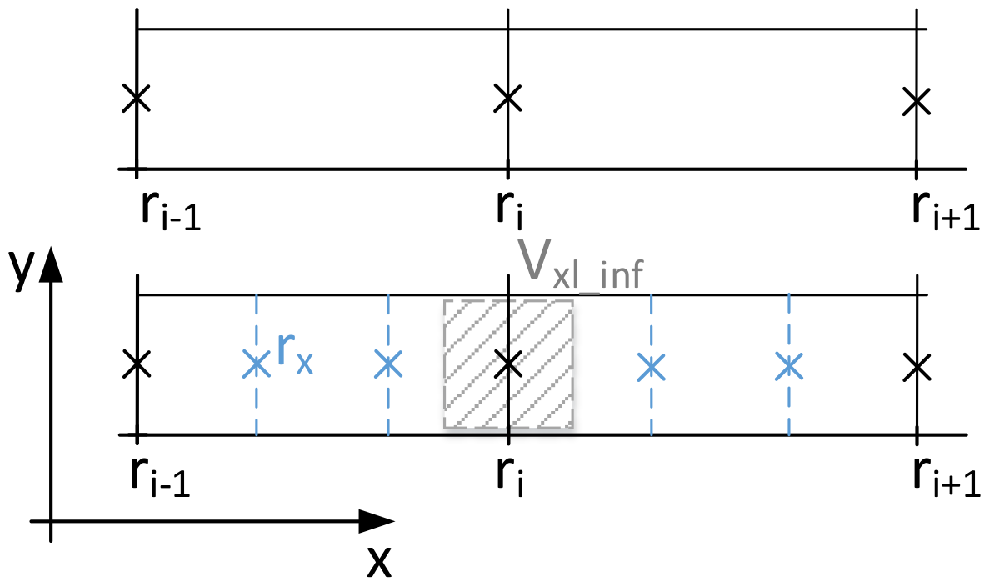}}
 \subfloat[][]
{\includegraphics[trim=-20 0 0 0,clip,height=3.5 cm]{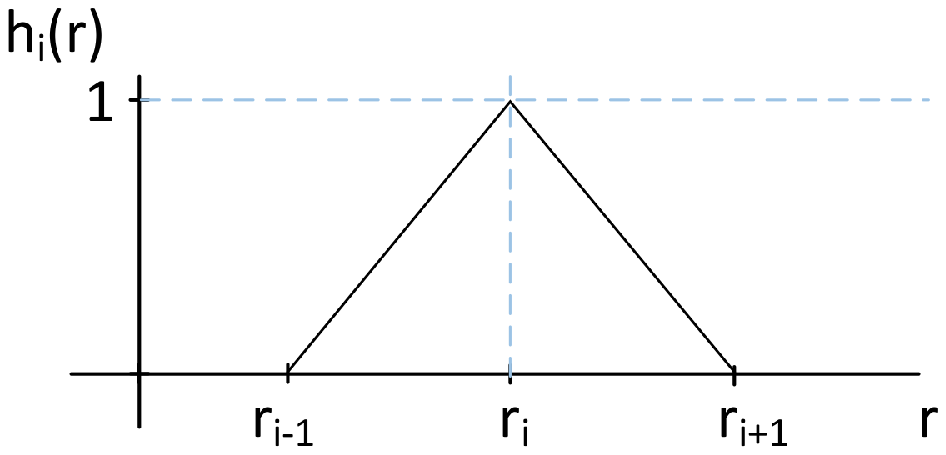}}
\caption{(a) Dividing elongated cells into sub-elements enable{s} to compute accurate interaction matrix elements. Solid lines: original elongated mesh. Dash lines: sub-elements mesh. (b) Interpolation ${h_i(r)}$ function.}
\label{elongated.fig}
\end{figure}

The elongated cells are cells with different geometry ratio than square (for 2D) or cubic (for 3D) . These cells allow to model geometries such as long thin film, or thin/thick bulk. 
The elongated cells enable to reduce the total number of elements, and hence reduce the computing time. {A key issue is the calculation of the interaction matrix between elemental surfaces (or ``surfaces").} The interaction matrix between surfaces ${i}$ and ${j}$ of type ${s\in \{x,y,z\}}$ is {generally} \cite{Pardo17JCP}

\begin{equation}
a_{sij}=\frac{\mu_0}{4\pi V_{si}V_{sj}}\int_{V_{si}}{\dvol}\int_{V_{sj}}{\dvol}'\frac{h_{si}({\bf r})h_{sj}({\bf r}')}{|{\bf r}-{\bf r}'|},
\label{}
\end{equation}
with
\begin{equation}
V_{si}\equiv\int_{V}{\dvol} h_{si}({\bf r}).
\label{}
\end{equation} 
The first-order interpolation functions ${h_{si}({\bf r})}$ are defined as {in} figure \ref{elongated.fig}(b) for coordinate ${r_s=r_x}$, vanishing outside the two neighboring cells in the ${r_s}$ direction.
  
In the case of square or cubic cells or square sub-elements, the self-interaction term ${a_{sii}}$ can be calculated by the approximated analytical formula  
\begin{equation}
a_{sii}\approx\frac{\mu_0}{4\pi V_{si}{^2}}\int_{V_{si}}{\dvol}\int_{V_{si}}{\dvol}'\frac{1}{|{\bf r}-{\bf r}'|} 
.
\label{a_ij}
\end{equation}
{The integral $\int_{V_{si}}\dvol\int_{V_{sj}}\dvol'(1/|{\bf r}-{\bf r}'|)$ on a rectangular prism is a lengthy analytical formula. For a cube and a square surface, the expression can be greatly simplified as
\begin{equation}
a_{sii}\approx\frac{\mu_0}{4\pi L_{si}}\left\{ \frac{1+\sqrt{2}-2\sqrt{3}}{5}-\frac{\pi}{3}+\ln\left[ \left( 1+\sqrt{2} \right)\left( 2+\sqrt{3} \right) \right] \right\}
\end{equation}
for a cube of side $L_{si}$ \cite{ciftja11PLA} and 
\begin{equation}
a_{sii}\approx \frac{\mu_0}{\pi L_{si}}\left\{ \frac{1-\sqrt{2}}{3}+\ln\left( 1+\sqrt{2} \right) \right\}
\end{equation}
for a thin prism \cite{ciftja10PLA} with thickness $d$ much smaller than its side $L_{si}$. For equation (\ref{a_ij}) we assumed that} the current density is {uniform} in the volume of influence, defined as the volume between surface ${i}$ of type ${s}$ and the centre of the neighbouring cells in the ${s}$ 
direction (see figure \ref{elongated.fig}(a) for ${s=x}$). The average vector potential ${a_{sij}}$ is calculated by approximation 
everywhere else, ${i\neq j}$

\begin{equation}
a_{sij}\approx\frac{\mu_0}{4\pi|r_{si}-r_{sj}|},
\label{}
\end{equation} 
where ${r_{si}}$ is the centre of surface ${i}$ of type ${s}$.

In the case of elongated cells, the interaction matrix of the vector potential, ${a_{sij}}$ needs to be calculated numerically. The numerical calculation splits the surrounded area of two surfaces 
into small square sub-elements (figure \ref{elongated.fig}(a)). The average vector potential of the two surfaces is integrated over all sub-elements, which contain surfaces again. 
The sub-elements are calculated in the same way as square cells, but sub-elements are multiplied by the linear interpolations functions ${h_{si}({\bf r}), h_{sj}({\bf r})}$ at the centre of the sub-element 
surfaces with indexes ${l,m}$, being ${{\bf r}_{sl}}$ and ${{\bf r}_{sm}}$. Elongated cells contain as many sub-elements in order to reach {as square as possible shape}. {In general, the average vector potential generated by sub-element $l$ on sub-element $m$ is
\begin{equation}
a_{sijlm}=\frac{\mu_0}{4\pi V_{sl}V_{sm}}\int_{V_{sl}}{\dvol} \int_{V_{sm}}{\dvol}' \frac{h_{si}({\bf r}_{sl})h_{sj}({\bf r}_{sm})}{|{\bf r}-{\bf r}'|},
\end{equation}
where $V_{sl}$ and $V_{sm}$ are the volume of influence of the sub-elements, as defined in figure \ref{elongated.fig}(a). For $l\neq m$, we approximate the integral above by
}
\begin{equation}
a_{sijlm}\approx\frac{\mu_0 h_{si}({\bf r}_{sl})h_{sj}({\bf r}_{sm})}{4\pi|{\bf r}_{sl}-{\bf r}_{sm}|},
\label{}
\end{equation}
where ${V_{sl}}$ is the volume of the influence of the sub-elements, defined in figure {\ref{elongated.fig}(a)}. {When $l$ corresponds to the sub-element $m$ both in position and size, we use the approximated formula for uniform current density in the sub-element
\begin{eqnarray}
a_{sijll} & \approx & \frac{\mu_0h_{si}({\bf r}_{sl})h_{sj}({\bf r}_{sl})}{4\pi V_{sl}^2}\int_{V_{sl}}{\dvol}\int_{V_{sl}}{\dvol}'\frac{1}{|{\bf r}-{\bf r}'|}.
\end{eqnarray}
Following the same steps as for equation (\ref{a_ij}), $a_{sijll}$ becomes
\begin{eqnarray}
a_{sijll} & \approx & \frac{\mu_0h_{si}({\bf r}_{sl})h_{sj}({\bf r}_{sl})}{4\pi L_{sl}} \nonumber\\
&& \cdot\left\{ \frac{1+\sqrt{2}-2\sqrt{3}}{5}-\frac{\pi}{3}+\ln\left[ \left( 1+\sqrt{2} \right)\left( 2+\sqrt{3} \right) \right] \right\}
\end{eqnarray}
for a cube of side $L_{sl}$ and 
\begin{equation}
a_{sijll} \approx \frac{\mu_0h_{si}({\bf r}_{sl})h_{sj}({\bf r}_{sl})}{\pi L_{sl}}\left\{ \frac{1-\sqrt{2}}{3}+\ln\left( 1+\sqrt{2} \right) \right\}
\end{equation}
for a thin prism of side $L_{sl}$.
}

\input Anisotropy_static.bbl

\end{document}

%% file: commands.tex
\usepackage{ulem}

\newcommand{\dvol}{{\rm d}^3{\bf r}}

\newcommand{\vJ}{{\bf J}}
\newcommand{\vE}{{\bf E}}
\newcommand{\vB}{{\bf B}}
\newcommand{\vA}{{\bf A}}

\newcommand{\vT}{{\bf T}}
\newcommand{\vM}{{\bf M}}
\newcommand{\vm}{{\bf m}}
\newcommand{\vF}{{\bf F}}
\newcommand{\rotT}{\nabla\times{\bf T}}
\newcommand{\rotDT}{\nabla\times\Delta{\bf T}}

\newcommand{\ve}{{\bf e}}

\newcommand{\dg}{^{\rm o}}


%% file: Anisotropy_static.bbl
%

%% file: Anisotropy.bbl
\begin{thebibliography}{56}%
\makeatletter
\providecommand \@ifxundefined [1]{%
 \@ifx{#1\undefined}
}%
\providecommand \@ifnum [1]{%
 \ifnum #1\expandafter \@firstoftwo
 \else \expandafter \@secondoftwo
 \fi
}%
\providecommand \@ifx [1]{%
 \ifx #1\expandafter \@firstoftwo
 \else \expandafter \@secondoftwo
 \fi
}%
\providecommand \natexlab [1]{#1}%
\providecommand \enquote  [1]{``#1''}%
\providecommand \bibnamefont  [1]{#1}%
\providecommand \bibfnamefont [1]{#1}%
\providecommand \citenamefont [1]{#1}%
\providecommand \href@noop [0]{\@secondoftwo}%
\providecommand \href [0]{\begingroup \@sanitize@url \@href}%
\providecommand \@href[1]{\@@startlink{#1}\@@href}%
\providecommand \@@href[1]{\endgroup#1\@@endlink}%
\providecommand \@sanitize@url [0]{\catcode `\\12\catcode `\$12\catcode
  `\&12\catcode `\#12\catcode `\^12\catcode `\_12\catcode `\%12\relax}%
\providecommand \@@startlink[1]{}%
\providecommand \@@endlink[0]{}%
\providecommand \url  [0]{\begingroup\@sanitize@url \@url }%
\providecommand \@url [1]{\endgroup\@href {#1}{\urlprefix }}%
\providecommand \urlprefix  [0]{URL }%
\providecommand \Eprint [0]{\href }%
\providecommand \doibase [0]{http://dx.doi.org/}%
\providecommand \selectlanguage [0]{\@gobble}%
\providecommand \bibinfo  [0]{\@secondoftwo}%
\providecommand \bibfield  [0]{\@secondoftwo}%
\providecommand \translation [1]{[#1]}%
\providecommand \BibitemOpen [0]{}%
\providecommand \bibitemStop [0]{}%
\providecommand \bibitemNoStop [0]{.\EOS\space}%
\providecommand \EOS [0]{\spacefactor3000\relax}%
\providecommand \BibitemShut  [1]{\csname bibitem#1\endcsname}%
\let\auto@bib@innerbib\@empty
\bibitem [{\citenamefont {Larbalestier}\ \emph {et~al.}(2014)\citenamefont
  {Larbalestier}, \citenamefont {J.Jiang}, \citenamefont {Trociewitz},
  \citenamefont {Kametami}, \citenamefont {Scheuerlein}, \citenamefont
  {Dalban-Canassy}, \citenamefont {Matras}, \citenamefont {Chen}, \citenamefont
  {craig}, \citenamefont {Lee},\ and\ \citenamefont
  {Hellstrom}}]{larbalestier14NaM}%
  \BibitemOpen
  \bibfield  {author} {\bibinfo {author} {\bibfnamefont {D.~C.}\ \bibnamefont
  {Larbalestier}}, \bibinfo {author} {\bibnamefont {J.Jiang}}, \bibinfo
  {author} {\bibfnamefont {U.~P.}\ \bibnamefont {Trociewitz}}, \bibinfo
  {author} {\bibfnamefont {F.}~\bibnamefont {Kametami}}, \bibinfo {author}
  {\bibfnamefont {C.}~\bibnamefont {Scheuerlein}}, \bibinfo {author}
  {\bibfnamefont {M.}~\bibnamefont {Dalban-Canassy}}, \bibinfo {author}
  {\bibfnamefont {M.}~\bibnamefont {Matras}}, \bibinfo {author} {\bibfnamefont
  {P.}~\bibnamefont {Chen}}, \bibinfo {author} {\bibfnamefont {N.~C.}\
  \bibnamefont {craig}}, \bibinfo {author} {\bibfnamefont {P.~J.}\ \bibnamefont
  {Lee}}, \ and\ \bibinfo {author} {\bibfnamefont {E.~E.}\ \bibnamefont
  {Hellstrom}},\ }\bibfield  {title} {\enquote {\bibinfo {title} {Isotropic
  round-wire multifilament cuprate superconductor for generation of magnetic
  fields above {30 T}},}\ }\href@noop {} {\bibfield  {journal} {\bibinfo
  {journal} {Nature Materials}\ }\textbf {\bibinfo {volume} {13}},\ \bibinfo
  {pages} {375--381} (\bibinfo {year} {2014})}\BibitemShut {NoStop}%
\bibitem [{\citenamefont {Kim}\ \emph {et~al.}(2017)\citenamefont {Kim},
  \citenamefont {Bhattarai}, \citenamefont {Jang}, \citenamefont {Hwang},
  \citenamefont {Kim}, \citenamefont {Yoon}, \citenamefont {Lee},\ and\
  \citenamefont {Hahn}}]{Kim17SST}%
  \BibitemOpen
  \bibfield  {author} {\bibinfo {author} {\bibfnamefont {K.}~\bibnamefont
  {Kim}}, \bibinfo {author} {\bibfnamefont {K.~R.}\ \bibnamefont {Bhattarai}},
  \bibinfo {author} {\bibfnamefont {J.~Y.}\ \bibnamefont {Jang}}, \bibinfo
  {author} {\bibfnamefont {Y.~J.}\ \bibnamefont {Hwang}}, \bibinfo {author}
  {\bibfnamefont {K.}~\bibnamefont {Kim}}, \bibinfo {author} {\bibfnamefont
  {S.}~\bibnamefont {Yoon}}, \bibinfo {author} {\bibfnamefont {S.}~\bibnamefont
  {Lee}}, \ and\ \bibinfo {author} {\bibfnamefont {S.}~\bibnamefont {Hahn}},\
  }\bibfield  {title} {\enquote {\bibinfo {title} {Design and performance
  estimation of a {35 T 40} mm no-insulation {all-REBCO} user magnet},}\
  }\href@noop {} {\bibfield  {journal} {\bibinfo  {journal} {Supercond. Sci.
  Technol.}\ }\textbf {\bibinfo {volume} {30}} (\bibinfo {year}
  {2017})}\BibitemShut {NoStop}%
\bibitem [{\citenamefont {Park}\ \emph {et~al.}(2018)\citenamefont {Park},
  \citenamefont {Bascunan}, \citenamefont {Michael}, \citenamefont {Lee},
  \citenamefont {Hahn},\ and\ \citenamefont {Iwasa}}]{Park18IES}%
  \BibitemOpen
  \bibfield  {author} {\bibinfo {author} {\bibfnamefont {D.}~\bibnamefont
  {Park}}, \bibinfo {author} {\bibfnamefont {J.}~\bibnamefont {Bascunan}},
  \bibinfo {author} {\bibfnamefont {P.}~\bibnamefont {Michael}}, \bibinfo
  {author} {\bibfnamefont {J.}~\bibnamefont {Lee}}, \bibinfo {author}
  {\bibfnamefont {S.}~\bibnamefont {Hahn}}, \ and\ \bibinfo {author}
  {\bibfnamefont {Y.}~\bibnamefont {Iwasa}},\ }\bibfield  {title} {\enquote
  {\bibinfo {title} {Construction and test results of coils 2 and 3 of a
  3-nested-coil {800-MHz} {REBCO} insert for the {MIT 1.3-GHz LTS/HTS NMR
  Magnet}},}\ }\href@noop {} {\bibfield  {journal} {\bibinfo  {journal} {IEEE
  Trans. Appl. Supercond.}\ }\textbf {\bibinfo {volume} {28}} (\bibinfo {year}
  {2018})}\BibitemShut {NoStop}%
\bibitem [{\citenamefont {Liu}\ and\ \citenamefont {Li}(2016)}]{Liu16IES}%
  \BibitemOpen
  \bibfield  {author} {\bibinfo {author} {\bibfnamefont {J.}~\bibnamefont
  {Liu}}\ and\ \bibinfo {author} {\bibfnamefont {Y.}~\bibnamefont {Li}},\
  }\bibfield  {title} {\enquote {\bibinfo {title} {{High-Field Insert With Bi-
  and Y-Based Tapes for 25-T All-Superconducting Magnet}},}\ }\href@noop {}
  {\bibfield  {journal} {\bibinfo  {journal} {IEEE Trans. Appl. Supercond.}\
  }\textbf {\bibinfo {volume} {26}} (\bibinfo {year} {2016})}\BibitemShut
  {NoStop}%
\bibitem [{\citenamefont {Masson}\ \emph {et~al.}(2007)\citenamefont {Masson},
  \citenamefont {Breschi}, \citenamefont {Pascal},\ and\ \citenamefont
  {Luongo}}]{Masson07IES}%
  \BibitemOpen
  \bibfield  {author} {\bibinfo {author} {\bibfnamefont {P.~J.}\ \bibnamefont
  {Masson}}, \bibinfo {author} {\bibfnamefont {M.}~\bibnamefont {Breschi}},
  \bibinfo {author} {\bibfnamefont {T.}~\bibnamefont {Pascal}}, \ and\ \bibinfo
  {author} {\bibfnamefont {C.}~\bibnamefont {Luongo}},\ }\bibfield  {title}
  {\enquote {\bibinfo {title} {Design of {HTS} axial flux motor for aircraft
  propulsion},}\ }\href@noop {} {\bibfield  {journal} {\bibinfo  {journal}
  {IEEE Trans. Appl. Supercond.}\ }\textbf {\bibinfo {volume} {17}} (\bibinfo
  {year} {2007})}\BibitemShut {NoStop}%
\bibitem [{\citenamefont {Masson}\ \emph {et~al.}(2013)\citenamefont {Masson},
  \citenamefont {Ratelle}, \citenamefont {Delobel}, \citenamefont {Lipardi},\
  and\ \citenamefont {Lorin}}]{masson13IES}%
  \BibitemOpen
  \bibfield  {author} {\bibinfo {author} {\bibfnamefont {P.~J.}\ \bibnamefont
  {Masson}}, \bibinfo {author} {\bibfnamefont {K.}~\bibnamefont {Ratelle}},
  \bibinfo {author} {\bibfnamefont {P.~A.}\ \bibnamefont {Delobel}}, \bibinfo
  {author} {\bibfnamefont {A.}~\bibnamefont {Lipardi}}, \ and\ \bibinfo
  {author} {\bibfnamefont {C.}~\bibnamefont {Lorin}},\ }\bibfield  {title}
  {\enquote {\bibinfo {title} {Development of a {3D} sizing model for
  all-superconducting machines for turbo-electric aircraft propulsion},}\
  }\href@noop {} {\bibfield  {journal} {\bibinfo  {journal} {IEEE Trans. Appl.
  Supercond.}\ }\textbf {\bibinfo {volume} {23}},\ \bibinfo {pages} {3600805}
  (\bibinfo {year} {2013})}\BibitemShut {NoStop}%
\bibitem [{\citenamefont {Yanamoto}\ \emph {et~al.}(2017)\citenamefont
  {Yanamoto}, \citenamefont {Izumi}, \citenamefont {Umemoto}, \citenamefont
  {Oryu}, \citenamefont {Murase},\ and\ \citenamefont
  {Kawamura}}]{Yanamoto17IES}%
  \BibitemOpen
  \bibfield  {author} {\bibinfo {author} {\bibfnamefont {T.}~\bibnamefont
  {Yanamoto}}, \bibinfo {author} {\bibfnamefont {M.}~\bibnamefont {Izumi}},
  \bibinfo {author} {\bibfnamefont {K.}~\bibnamefont {Umemoto}}, \bibinfo
  {author} {\bibfnamefont {T.}~\bibnamefont {Oryu}}, \bibinfo {author}
  {\bibfnamefont {Y.}~\bibnamefont {Murase}}, \ and\ \bibinfo {author}
  {\bibfnamefont {M.}~\bibnamefont {Kawamura}},\ }\bibfield  {title} {\enquote
  {\bibinfo {title} {Load test of {3-MW HTS} motor for ship propulsion},}\
  }\href@noop {} {\bibfield  {journal} {\bibinfo  {journal} {IEEE Trans. Appl.
  Supercond.}\ }\textbf {\bibinfo {volume} {27}} (\bibinfo {year}
  {2017})}\BibitemShut {NoStop}%
\bibitem [{\citenamefont {Gamble}\ \emph {et~al.}(2011)\citenamefont {Gamble},
  \citenamefont {Snitchler},\ and\ \citenamefont {MacDonald}}]{Gamble11IES}%
  \BibitemOpen
  \bibfield  {author} {\bibinfo {author} {\bibfnamefont {B.}~\bibnamefont
  {Gamble}}, \bibinfo {author} {\bibfnamefont {G.}~\bibnamefont {Snitchler}}, \
  and\ \bibinfo {author} {\bibfnamefont {T.}~\bibnamefont {MacDonald}},\
  }\bibfield  {title} {\enquote {\bibinfo {title} {Full power test of a 36.5
  {MW} {HTS} propulsion motor},}\ }\href@noop {} {\bibfield  {journal}
  {\bibinfo  {journal} {IEEE Trans. Appl. Supercond.}\ }\textbf {\bibinfo
  {volume} {21}},\ \bibinfo {pages} {1083--1088} (\bibinfo {year}
  {2011})}\BibitemShut {NoStop}%
\bibitem [{\citenamefont {Jeong}\ \emph {et~al.}(2017)\citenamefont {Jeong},
  \citenamefont {An}, \citenamefont {Hong}, \citenamefont {Kim},\ and\
  \citenamefont {Jo}}]{Jeong17IES}%
  \BibitemOpen
  \bibfield  {author} {\bibinfo {author} {\bibfnamefont {JS.}\ \bibnamefont
  {Jeong}}, \bibinfo {author} {\bibfnamefont {DK.}\ \bibnamefont {An}},
  \bibinfo {author} {\bibfnamefont {JP.}\ \bibnamefont {Hong}}, \bibinfo
  {author} {\bibfnamefont {HJ.}\ \bibnamefont {Kim}}, \ and\ \bibinfo {author}
  {\bibfnamefont {YS.}\ \bibnamefont {Jo}},\ }\bibfield  {title} {\enquote
  {\bibinfo {title} {Design of a {10-MW-Class HTS} homopolar generator for wind
  turbines},}\ }\href@noop {} {\bibfield  {journal} {\bibinfo  {journal} {IEEE
  Trans. Appl. Supercond.}\ }\textbf {\bibinfo {volume} {27}} (\bibinfo {year}
  {2017})}\BibitemShut {NoStop}%
\bibitem [{\citenamefont {Abrahamsen}\ \emph {et~al.}(2010)\citenamefont
  {Abrahamsen}, \citenamefont {Mijatovic}, \citenamefont {Seiler},
  \citenamefont {Zirngibl}, \citenamefont {Tr{\ae}holt}, \citenamefont
  {N{\o}rg{\aa}rd}, \citenamefont {Pedersen}, \citenamefont {Andersen},\ and\
  \citenamefont {{\O}stergaard}}]{abrahamsen10SST}%
  \BibitemOpen
  \bibfield  {author} {\bibinfo {author} {\bibfnamefont {A.B.}\ \bibnamefont
  {Abrahamsen}}, \bibinfo {author} {\bibfnamefont {N.}~\bibnamefont
  {Mijatovic}}, \bibinfo {author} {\bibfnamefont {E.}~\bibnamefont {Seiler}},
  \bibinfo {author} {\bibfnamefont {T.}~\bibnamefont {Zirngibl}}, \bibinfo
  {author} {\bibfnamefont {C.}~\bibnamefont {Tr{\ae}holt}}, \bibinfo {author}
  {\bibfnamefont {P.B.}\ \bibnamefont {N{\o}rg{\aa}rd}}, \bibinfo {author}
  {\bibfnamefont {N.F.}\ \bibnamefont {Pedersen}}, \bibinfo {author}
  {\bibfnamefont {N.H.}\ \bibnamefont {Andersen}}, \ and\ \bibinfo {author}
  {\bibfnamefont {J.}~\bibnamefont {{\O}stergaard}},\ }\bibfield  {title}
  {\enquote {\bibinfo {title} {Superconducting wind turbine generators},}\
  }\href@noop {} {\bibfield  {journal} {\bibinfo  {journal} {Supercond. Sci.
  Technol.}\ }\textbf {\bibinfo {volume} {23}},\ \bibinfo {pages} {034019}
  (\bibinfo {year} {2010})}\BibitemShut {NoStop}%
\bibitem [{sup()}]{suprapower}%
  \BibitemOpen
  \href@noop {} {}\bibinfo {note} {SUPRAPOWER-EU project. Superconducting light
  generator for large offshore wind turbines.
  http://www.suprapower-fp7.eu/}\BibitemShut {NoStop}%
\bibitem [{\citenamefont {Volkov}\ \emph {et~al.}(2012)\citenamefont {Volkov},
  \citenamefont {Vysotsky},\ and\ \citenamefont {Firsov}}]{Volkov16PCS}%
  \BibitemOpen
  \bibfield  {author} {\bibinfo {author} {\bibfnamefont {E.}~\bibnamefont
  {Volkov}}, \bibinfo {author} {\bibfnamefont {V.}~\bibnamefont {Vysotsky}}, \
  and\ \bibinfo {author} {\bibfnamefont {V.}~\bibnamefont {Firsov}},\
  }\bibfield  {title} {\enquote {\bibinfo {title} {First russian long length
  {HTS} power cable},}\ }\href@noop {} {\bibfield  {journal} {\bibinfo
  {journal} {Physica C-Superc. and its apl.}\ }\textbf {\bibinfo {volume}
  {482}},\ \bibinfo {pages} {87--91} (\bibinfo {year} {2012})}\BibitemShut
  {NoStop}%
\bibitem [{\citenamefont {Yagi}\ \emph {et~al.}(2015)\citenamefont {Yagi},
  \citenamefont {Liu}, \citenamefont {Mukoyama}, \citenamefont {Mitsuhashi},
  \citenamefont {Teng}, \citenamefont {Hayakawa}, \citenamefont {Wang},
  \citenamefont {Ishiyama}, \citenamefont {Amemiya}, \citenamefont {Hasegawa},
  \citenamefont {Saitoh}, \citenamefont {Maruyama},\ and\ \citenamefont
  {Ohkuma}}]{Yagi15IES}%
  \BibitemOpen
  \bibfield  {author} {\bibinfo {author} {\bibfnamefont {M.}~\bibnamefont
  {Yagi}}, \bibinfo {author} {\bibfnamefont {J.}~\bibnamefont {Liu}}, \bibinfo
  {author} {\bibfnamefont {S.}~\bibnamefont {Mukoyama}}, \bibinfo {author}
  {\bibfnamefont {T.}~\bibnamefont {Mitsuhashi}}, \bibinfo {author}
  {\bibfnamefont {J.}~\bibnamefont {Teng}}, \bibinfo {author} {\bibfnamefont
  {N.}~\bibnamefont {Hayakawa}}, \bibinfo {author} {\bibfnamefont
  {W.}~\bibnamefont {Wang}}, \bibinfo {author} {\bibfnamefont {A.}~\bibnamefont
  {Ishiyama}}, \bibinfo {author} {\bibfnamefont {N.}~\bibnamefont {Amemiya}},
  \bibinfo {author} {\bibfnamefont {T.}~\bibnamefont {Hasegawa}}, \bibinfo
  {author} {\bibfnamefont {T.}~\bibnamefont {Saitoh}}, \bibinfo {author}
  {\bibfnamefont {O.}~\bibnamefont {Maruyama}}, \ and\ \bibinfo {author}
  {\bibfnamefont {T.}~\bibnamefont {Ohkuma}},\ }\bibfield  {title} {\enquote
  {\bibinfo {title} {Experimental results of {275-kV 3-kA REBCO HTS} power
  cable},}\ }\href@noop {} {\bibfield  {journal} {\bibinfo  {journal} {IEEE
  Trans. Appl. Supercond.}\ }\textbf {\bibinfo {volume} {25}} (\bibinfo {year}
  {2015})}\BibitemShut {NoStop}%
\bibitem [{\citenamefont {Hellmann}\ \emph {et~al.}(2017)\citenamefont
  {Hellmann}, \citenamefont {Abplanalp}, \citenamefont {Hofstetter},\ and\
  \citenamefont {Noe}}]{Hellmann17IES}%
  \BibitemOpen
  \bibfield  {author} {\bibinfo {author} {\bibfnamefont {S.}~\bibnamefont
  {Hellmann}}, \bibinfo {author} {\bibfnamefont {M.}~\bibnamefont {Abplanalp}},
  \bibinfo {author} {\bibfnamefont {L.}~\bibnamefont {Hofstetter}}, \ and\
  \bibinfo {author} {\bibfnamefont {M.}~\bibnamefont {Noe}},\ }\bibfield
  {title} {\enquote {\bibinfo {title} {Manufacturing of a {1-MVA-Class}
  superconducting fault current limiting transformer with recovery-under-load
  capabilities},}\ }\href@noop {} {\bibfield  {journal} {\bibinfo  {journal}
  {IEEE Trans. Appl. Supercond.}\ }\textbf {\bibinfo {volume} {27}} (\bibinfo
  {year} {2017})}\BibitemShut {NoStop}%
\bibitem [{\citenamefont {Glasson}\ \emph {et~al.}(2017)\citenamefont
  {Glasson}, \citenamefont {Staines}, \citenamefont {Allpress}, \citenamefont
  {Pannu}, \citenamefont {Tanchon}, \citenamefont {Pardo}, \citenamefont
  {Badcock},\ and\ \citenamefont {Buckley}}]{Glasson17IES}%
  \BibitemOpen
  \bibfield  {author} {\bibinfo {author} {\bibfnamefont {N.}~\bibnamefont
  {Glasson}}, \bibinfo {author} {\bibfnamefont {M.}~\bibnamefont {Staines}},
  \bibinfo {author} {\bibfnamefont {N.}~\bibnamefont {Allpress}}, \bibinfo
  {author} {\bibfnamefont {M.}~\bibnamefont {Pannu}}, \bibinfo {author}
  {\bibfnamefont {J.}~\bibnamefont {Tanchon}}, \bibinfo {author} {\bibfnamefont
  {E.}~\bibnamefont {Pardo}}, \bibinfo {author} {\bibfnamefont
  {R.}~\bibnamefont {Badcock}}, \ and\ \bibinfo {author} {\bibfnamefont
  {R.}~\bibnamefont {Buckley}},\ }\bibfield  {title} {\enquote {\bibinfo
  {title} {Test results and conclusions from a 1 {MVA} superconducting
  transformer featuring {2G HTS} roebel cable},}\ }\href@noop {} {\bibfield
  {journal} {\bibinfo  {journal} {IEEE Trans. Appl. Supercond.}\ }\textbf
  {\bibinfo {volume} {27}} (\bibinfo {year} {2017})}\BibitemShut {NoStop}%
\bibitem [{\citenamefont {Schwenterly}\ \emph {et~al.}(1999)\citenamefont
  {Schwenterly}, \citenamefont {McConnell}, \citenamefont {Demko},
  \citenamefont {Fadnek}, \citenamefont {Hsu}, \citenamefont {List},
  \citenamefont {Walker}, \citenamefont {Hazelton}, \citenamefont {Murray},
  \citenamefont {Rice}, \citenamefont {Trautwein}, \citenamefont {Shi},
  \citenamefont {Farrell}, \citenamefont {Bascunan}, \citenamefont {Hintz},
  \citenamefont {Mehta}, \citenamefont {Aversa}, \citenamefont {Ebert},
  \citenamefont {Bednar}, \citenamefont {Neder}, \citenamefont {McIlheran},
  \citenamefont {Michel}, \citenamefont {Nemec}, \citenamefont {Pleva},
  \citenamefont {Swenton}, \citenamefont {Swets}, \citenamefont {Longsworth},
  \citenamefont {Longsworth}, \citenamefont {Johnson}, \citenamefont {Jones},
  \citenamefont {Nelson}, \citenamefont {Degeneff},\ and\ \citenamefont
  {Salon}}]{Schwenterly99IES}%
  \BibitemOpen
  \bibfield  {author} {\bibinfo {author} {\bibfnamefont {S.}~\bibnamefont
  {Schwenterly}}, \bibinfo {author} {\bibfnamefont {B.}~\bibnamefont
  {McConnell}}, \bibinfo {author} {\bibfnamefont {J.}~\bibnamefont {Demko}},
  \bibinfo {author} {\bibfnamefont {A.}~\bibnamefont {Fadnek}}, \bibinfo
  {author} {\bibfnamefont {J.}~\bibnamefont {Hsu}}, \bibinfo {author}
  {\bibfnamefont {F.}~\bibnamefont {List}}, \bibinfo {author} {\bibfnamefont
  {M.}~\bibnamefont {Walker}}, \bibinfo {author} {\bibfnamefont
  {D.}~\bibnamefont {Hazelton}}, \bibinfo {author} {\bibfnamefont
  {F.}~\bibnamefont {Murray}}, \bibinfo {author} {\bibfnamefont
  {J.}~\bibnamefont {Rice}}, \bibinfo {author} {\bibfnamefont {C.}~\bibnamefont
  {Trautwein}}, \bibinfo {author} {\bibfnamefont {X.}~\bibnamefont {Shi}},
  \bibinfo {author} {\bibfnamefont {R.}~\bibnamefont {Farrell}}, \bibinfo
  {author} {\bibfnamefont {J.}~\bibnamefont {Bascunan}}, \bibinfo {author}
  {\bibfnamefont {R.}~\bibnamefont {Hintz}}, \bibinfo {author} {\bibfnamefont
  {S.}~\bibnamefont {Mehta}}, \bibinfo {author} {\bibfnamefont
  {N.}~\bibnamefont {Aversa}}, \bibinfo {author} {\bibfnamefont
  {J.}~\bibnamefont {Ebert}}, \bibinfo {author} {\bibfnamefont
  {B.}~\bibnamefont {Bednar}}, \bibinfo {author} {\bibfnamefont
  {D.}~\bibnamefont {Neder}}, \bibinfo {author} {\bibfnamefont
  {A.}~\bibnamefont {McIlheran}}, \bibinfo {author} {\bibfnamefont
  {P.}~\bibnamefont {Michel}}, \bibinfo {author} {\bibfnamefont
  {J.}~\bibnamefont {Nemec}}, \bibinfo {author} {\bibfnamefont
  {E.}~\bibnamefont {Pleva}}, \bibinfo {author} {\bibfnamefont
  {A.}~\bibnamefont {Swenton}}, \bibinfo {author} {\bibfnamefont
  {N.}~\bibnamefont {Swets}}, \bibinfo {author} {\bibfnamefont
  {R.}~\bibnamefont {Longsworth}}, \bibinfo {author} {\bibfnamefont {RC.}\
  \bibnamefont {Longsworth}}, \bibinfo {author} {\bibfnamefont
  {R.}~\bibnamefont {Johnson}}, \bibinfo {author} {\bibfnamefont
  {R.}~\bibnamefont {Jones}}, \bibinfo {author} {\bibfnamefont
  {J.}~\bibnamefont {Nelson}}, \bibinfo {author} {\bibfnamefont
  {R.}~\bibnamefont {Degeneff}}, \ and\ \bibinfo {author} {\bibfnamefont
  {S.}~\bibnamefont {Salon}},\ }\bibfield  {title} {\enquote {\bibinfo {title}
  {Performance of a {1-MVA HTS} demonstration transformer},}\ }\href@noop {}
  {\bibfield  {journal} {\bibinfo  {journal} {IEEE Trans. Appl. Supercond.}\
  }\textbf {\bibinfo {volume} {9}},\ \bibinfo {pages} {680--684} (\bibinfo
  {year} {1999})}\BibitemShut {NoStop}%
\bibitem [{\citenamefont {Mehta}(2011)}]{Mehta11PhC}%
  \BibitemOpen
  \bibfield  {author} {\bibinfo {author} {\bibfnamefont {S.}~\bibnamefont
  {Mehta}},\ }\bibfield  {title} {\enquote {\bibinfo {title} {{US} effort on
  {HTS} power transformers},}\ }\href@noop {} {\bibfield  {journal} {\bibinfo
  {journal} {Physica C}\ }\textbf {\bibinfo {volume} {471}},\ \bibinfo {pages}
  {1364--1366} (\bibinfo {year} {2011})}\BibitemShut {NoStop}%
\bibitem [{\citenamefont {Pascal}\ \emph {et~al.}(2017)\citenamefont {Pascal},
  \citenamefont {Badel}, \citenamefont {Auran},\ and\ \citenamefont
  {Pereira}}]{Pascal17IES}%
  \BibitemOpen
  \bibfield  {author} {\bibinfo {author} {\bibfnamefont {T.}~\bibnamefont
  {Pascal}}, \bibinfo {author} {\bibfnamefont {A.}~\bibnamefont {Badel}},
  \bibinfo {author} {\bibfnamefont {G.}~\bibnamefont {Auran}}, \ and\ \bibinfo
  {author} {\bibfnamefont {GS.}\ \bibnamefont {Pereira}},\ }\bibfield  {title}
  {\enquote {\bibinfo {title} {Superconducting fault current limiter for ship
  grid simulation and demonstration},}\ }\href@noop {} {\bibfield  {journal}
  {\bibinfo  {journal} {IEEE Trans. Appl. Supercond.}\ }\textbf {\bibinfo
  {volume} {27}} (\bibinfo {year} {2017})}\BibitemShut {NoStop}%
\bibitem [{\citenamefont {Xin}\ \emph {et~al.}(2013)\citenamefont {Xin},
  \citenamefont {Gong}, \citenamefont {Sun}, \citenamefont {Cui}, \citenamefont
  {Hong}, \citenamefont {Niu}, \citenamefont {Wang}, \citenamefont {Wang},
  \citenamefont {Li}, \citenamefont {Zhang}, \citenamefont {Wei}, \citenamefont
  {Liu}, \citenamefont {Yang},\ and\ \citenamefont {Zhu}}]{Xin13IES}%
  \BibitemOpen
  \bibfield  {author} {\bibinfo {author} {\bibfnamefont {Y.}~\bibnamefont
  {Xin}}, \bibinfo {author} {\bibfnamefont {W.}~\bibnamefont {Gong}}, \bibinfo
  {author} {\bibfnamefont {Y.}~\bibnamefont {Sun}}, \bibinfo {author}
  {\bibfnamefont {J.}~\bibnamefont {Cui}}, \bibinfo {author} {\bibfnamefont
  {H.}~\bibnamefont {Hong}}, \bibinfo {author} {\bibfnamefont {X.}~\bibnamefont
  {Niu}}, \bibinfo {author} {\bibfnamefont {H.}~\bibnamefont {Wang}}, \bibinfo
  {author} {\bibfnamefont {L.}~\bibnamefont {Wang}}, \bibinfo {author}
  {\bibfnamefont {Q.}~\bibnamefont {Li}}, \bibinfo {author} {\bibfnamefont
  {J.}~\bibnamefont {Zhang}}, \bibinfo {author} {\bibfnamefont
  {Z.}~\bibnamefont {Wei}}, \bibinfo {author} {\bibfnamefont {L.}~\bibnamefont
  {Liu}}, \bibinfo {author} {\bibfnamefont {H.}~\bibnamefont {Yang}}, \ and\
  \bibinfo {author} {\bibfnamefont {X.}~\bibnamefont {Zhu}},\ }\bibfield
  {title} {\enquote {\bibinfo {title} {Factory and field tests of a {220 kV/300
  MVA} statured iron-core superconducting fault current limiter},}\ }\href@noop
  {} {\bibfield  {journal} {\bibinfo  {journal} {IEEE Trans. Appl. Supercond.}\
  }\textbf {\bibinfo {volume} {23}} (\bibinfo {year} {2013})}\BibitemShut
  {NoStop}%
\bibitem [{\citenamefont {Morandi}(2013)}]{Morandi13PhC}%
  \BibitemOpen
  \bibfield  {author} {\bibinfo {author} {\bibfnamefont {A.}~\bibnamefont
  {Morandi}},\ }\bibfield  {title} {\enquote {\bibinfo {title} {State of the
  art of superconducting fault current limiters and their application to the
  electric power system},}\ }\href@noop {} {\bibfield  {journal} {\bibinfo
  {journal} {Physica C}\ }\textbf {\bibinfo {volume} {484}},\ \bibinfo {pages}
  {242--247} (\bibinfo {year} {2013})}\BibitemShut {NoStop}%
\bibitem [{\citenamefont {{\v{S}}ouc}\ \emph {et~al.}(2012)\citenamefont
  {{\v{S}}ouc}, \citenamefont {G{\"o}m{\"o}ry},\ and\ \citenamefont
  {Vojen{\v{c}}iak}}]{souc12SST}%
  \BibitemOpen
  \bibfield  {author} {\bibinfo {author} {\bibfnamefont {J.}~\bibnamefont
  {{\v{S}}ouc}}, \bibinfo {author} {\bibfnamefont {F.}~\bibnamefont
  {G{\"o}m{\"o}ry}}, \ and\ \bibinfo {author} {\bibfnamefont {M.}~\bibnamefont
  {Vojen{\v{c}}iak}},\ }\bibfield  {title} {\enquote {\bibinfo {title} {Coated
  conductor arrangement for reduced {AC} losses in a resistive-type
  superconducting fault current limiter},}\ }\href@noop {} {\bibfield
  {journal} {\bibinfo  {journal} {Supercond. Sci. Technol.}\ }\textbf {\bibinfo
  {volume} {25}},\ \bibinfo {pages} {014005} (\bibinfo {year}
  {2012})}\BibitemShut {NoStop}%
\bibitem [{\citenamefont {Kozak}\ \emph {et~al.}(2016)\citenamefont {Kozak},
  \citenamefont {Majka}, \citenamefont {Blazejczyk},\ and\ \citenamefont
  {Berowski}}]{Kozak16SST}%
  \BibitemOpen
  \bibfield  {author} {\bibinfo {author} {\bibfnamefont {J.}~\bibnamefont
  {Kozak}}, \bibinfo {author} {\bibfnamefont {M.}~\bibnamefont {Majka}},
  \bibinfo {author} {\bibfnamefont {T.}~\bibnamefont {Blazejczyk}}, \ and\
  \bibinfo {author} {\bibfnamefont {P.}~\bibnamefont {Berowski}},\ }\bibfield
  {title} {\enquote {\bibinfo {title} {Tests of the {15-kV} class coreless
  superconducting fault current limiter},}\ }\href@noop {} {\bibfield
  {journal} {\bibinfo  {journal} {Supercond. Sci. Technol.}\ }\textbf {\bibinfo
  {volume} {26}} (\bibinfo {year} {2016})}\BibitemShut {NoStop}%
\bibitem [{\citenamefont {Durrell}\ \emph {et~al.}(2003)\citenamefont
  {Durrell}, \citenamefont {Mennema}, \citenamefont {Jooss}, \citenamefont
  {Gibson}, \citenamefont {Barber}, \citenamefont {Zandbergen},\ and\
  \citenamefont {Evetts}}]{durrell03JAP}%
  \BibitemOpen
  \bibfield  {author} {\bibinfo {author} {\bibfnamefont {J.H.}\ \bibnamefont
  {Durrell}}, \bibinfo {author} {\bibfnamefont {S.H.}\ \bibnamefont {Mennema}},
  \bibinfo {author} {\bibfnamefont {C.}~\bibnamefont {Jooss}}, \bibinfo
  {author} {\bibfnamefont {G.}~\bibnamefont {Gibson}}, \bibinfo {author}
  {\bibfnamefont {Z.~H.}\ \bibnamefont {Barber}}, \bibinfo {author}
  {\bibfnamefont {H.W.}\ \bibnamefont {Zandbergen}}, \ and\ \bibinfo {author}
  {\bibfnamefont {J.E.}\ \bibnamefont {Evetts}},\ }\bibfield  {title} {\enquote
  {\bibinfo {title} {Flux line lattice structure and behavior in antiphase
  boundary free vicinal {YBa$_2$Cu$_3$O}$_{7-\delta}$ thin films},}\
  }\href@noop {} {\bibfield  {journal} {\bibinfo  {journal} {J. Appl. Phys.}\
  }\textbf {\bibinfo {volume} {93}},\ \bibinfo {pages} {9869--9874} (\bibinfo
  {year} {2003})}\BibitemShut {NoStop}%
\bibitem [{\citenamefont {Lao}\ \emph {et~al.}(2017)\citenamefont {Lao},
  \citenamefont {Hecher}, \citenamefont {Sieger}, \citenamefont {Pahlke},
  \citenamefont {Huhne},\ and\ \citenamefont {Eisterer}}]{Lao17SST}%
  \BibitemOpen
  \bibfield  {author} {\bibinfo {author} {\bibfnamefont {M.}~\bibnamefont
  {Lao}}, \bibinfo {author} {\bibfnamefont {J.}~\bibnamefont {Hecher}},
  \bibinfo {author} {\bibfnamefont {M.}~\bibnamefont {Sieger}}, \bibinfo
  {author} {\bibfnamefont {P.}~\bibnamefont {Pahlke}}, \bibinfo {author}
  {\bibfnamefont {M.~Bauer~R.}\ \bibnamefont {Huhne}}, \ and\ \bibinfo {author}
  {\bibfnamefont {M.}~\bibnamefont {Eisterer}},\ }\bibfield  {title} {\enquote
  {\bibinfo {title} {Planar current anisotropy and field dependence of {J$_c$}
  in coated conductors assessed by scanning hall probe microscopy},}\
  }\href@noop {} {\bibfield  {journal} {\bibinfo  {journal} {Supercond. Sci.
  Technol.}\ }\textbf {\bibinfo {volume} {30}},\ \bibinfo {pages} {9} (\bibinfo
  {year} {2017})}\BibitemShut {NoStop}%
\bibitem [{\citenamefont {Blatter}\ \emph {et~al.}(1994)\citenamefont
  {Blatter}, \citenamefont {Feigelman}, \citenamefont {Geshkenbein},
  \citenamefont {Larkin},\ and\ \citenamefont {Vinokur}}]{blatter94RMP}%
  \BibitemOpen
  \bibfield  {author} {\bibinfo {author} {\bibfnamefont {G.}~\bibnamefont
  {Blatter}}, \bibinfo {author} {\bibfnamefont {M.~V.}\ \bibnamefont
  {Feigelman}}, \bibinfo {author} {\bibfnamefont {V.~B.}\ \bibnamefont
  {Geshkenbein}}, \bibinfo {author} {\bibfnamefont {A.~I.}\ \bibnamefont
  {Larkin}}, \ and\ \bibinfo {author} {\bibfnamefont {V.~M.}\ \bibnamefont
  {Vinokur}},\ }\bibfield  {title} {\enquote {\bibinfo {title} {Vortices in
  high-temperature superconductors},}\ }\href@noop {} {\bibfield  {journal}
  {\bibinfo  {journal} {Rev. Mod. Phys.}\ }\textbf {\bibinfo {volume} {66}},\
  \bibinfo {pages} {1125} (\bibinfo {year} {1994})}\BibitemShut {NoStop}%
\bibitem [{\citenamefont {Vlasko-Vlasov}\ \emph {et~al.}(2015)\citenamefont
  {Vlasko-Vlasov}, \citenamefont {Koshelev}, \citenamefont {Glatz},
  \citenamefont {Phillips}, \citenamefont {Welp},\ and\ \citenamefont
  {Kwok}}]{Vlasko15FL}%
  \BibitemOpen
  \bibfield  {author} {\bibinfo {author} {\bibfnamefont {V.}~\bibnamefont
  {Vlasko-Vlasov}}, \bibinfo {author} {\bibfnamefont {A.}~\bibnamefont
  {Koshelev}}, \bibinfo {author} {\bibfnamefont {A.}~\bibnamefont {Glatz}},
  \bibinfo {author} {\bibfnamefont {C.}~\bibnamefont {Phillips}}, \bibinfo
  {author} {\bibfnamefont {U.}~\bibnamefont {Welp}}, \ and\ \bibinfo {author}
  {\bibfnamefont {K.}~\bibnamefont {Kwok}},\ }\bibfield  {title} {\enquote
  {\bibinfo {title} {Flux cutting in {high-T$_c$} superconductors},}\
  }\href@noop {} {\bibfield  {journal} {\bibinfo  {journal} {Phys. Rev. B}\ }
  (\bibinfo {year} {2015})}\BibitemShut {NoStop}%
\bibitem [{\citenamefont {Clem}(1982)}]{clem82PRB}%
  \BibitemOpen
  \bibfield  {author} {\bibinfo {author} {\bibfnamefont {J.R.}\ \bibnamefont
  {Clem}},\ }\bibfield  {title} {\enquote {\bibinfo {title} {Flux-line-cutting
  losses in {type-II} superconductors},}\ }\href@noop {} {\bibfield  {journal}
  {\bibinfo  {journal} {Phys. Rev. B}\ }\textbf {\bibinfo {volume} {26}},\
  \bibinfo {pages} {2463} (\bibinfo {year} {1982})}\BibitemShut {NoStop}%
\bibitem [{\citenamefont {Clem}\ \emph {et~al.}(2011)\citenamefont {Clem},
  \citenamefont {Weigand}, \citenamefont {Durrell},\ and\ \citenamefont
  {Campbell}}]{clem11SST}%
  \BibitemOpen
  \bibfield  {author} {\bibinfo {author} {\bibfnamefont {J.R.}\ \bibnamefont
  {Clem}}, \bibinfo {author} {\bibfnamefont {M.}~\bibnamefont {Weigand}},
  \bibinfo {author} {\bibfnamefont {J.~H.}\ \bibnamefont {Durrell}}, \ and\
  \bibinfo {author} {\bibfnamefont {A.~M.}\ \bibnamefont {Campbell}},\
  }\bibfield  {title} {\enquote {\bibinfo {title} {Theory and experiment
  testing flux-line cutting physics},}\ }\href@noop {} {\bibfield  {journal}
  {\bibinfo  {journal} {Supercond. Sci. Technol.}\ }\textbf {\bibinfo {volume}
  {24}},\ \bibinfo {pages} {062002} (\bibinfo {year} {2011})}\BibitemShut
  {NoStop}%
\bibitem [{\citenamefont {Clem}\ and\ \citenamefont
  {Perez-Gonzalez}(1984)}]{clem84PRB}%
  \BibitemOpen
  \bibfield  {author} {\bibinfo {author} {\bibfnamefont {J.R.}\ \bibnamefont
  {Clem}}\ and\ \bibinfo {author} {\bibfnamefont {A.}~\bibnamefont
  {Perez-Gonzalez}},\ }\bibfield  {title} {\enquote {\bibinfo {title}
  {Flux-line-cutting and flux-pinning losses in {type-II} superconductors in
  rotating magnetic fields},}\ }\href@noop {} {\bibfield  {journal} {\bibinfo
  {journal} {Phys. Rev. B}\ }\textbf {\bibinfo {volume} {30}},\ \bibinfo
  {pages} {5041} (\bibinfo {year} {1984})}\BibitemShut {NoStop}%
\bibitem [{\citenamefont {Brandt}\ and\ \citenamefont
  {Mikitik}(2007)}]{Brandt07PRB}%
  \BibitemOpen
  \bibfield  {author} {\bibinfo {author} {\bibfnamefont {E.~H.}\ \bibnamefont
  {Brandt}}\ and\ \bibinfo {author} {\bibfnamefont {G.~P.}\ \bibnamefont
  {Mikitik}},\ }\bibfield  {title} {\enquote {\bibinfo {title} {Unusual
  critical states in {type-II} superconductors},}\ }\href@noop {} {\bibfield
  {journal} {\bibinfo  {journal} {Phys. Rev. B}\ }\textbf {\bibinfo {volume}
  {76}} (\bibinfo {year} {2007})}\BibitemShut {NoStop}%
\bibitem [{\citenamefont {Romero-Salazar}\ and\ \citenamefont
  {P{\'e}rez-Rodr{\'\i}guez}(2003)}]{romerosalazar03APL}%
  \BibitemOpen
  \bibfield  {author} {\bibinfo {author} {\bibfnamefont {C.}~\bibnamefont
  {Romero-Salazar}}\ and\ \bibinfo {author} {\bibfnamefont {F.}~\bibnamefont
  {P{\'e}rez-Rodr{\'\i}guez}},\ }\bibfield  {title} {\enquote {\bibinfo {title}
  {Elliptic flux-line-cutting critical-state model},}\ }\href@noop {}
  {\bibfield  {journal} {\bibinfo  {journal} {Appl. Phys. Lett.}\ }\textbf
  {\bibinfo {volume} {83}},\ \bibinfo {pages} {5256} (\bibinfo {year}
  {2003})}\BibitemShut {NoStop}%
\bibitem [{\citenamefont {Chepikov}\ \emph {et~al.}(2017)\citenamefont
  {Chepikov}, \citenamefont {Mineev}, \citenamefont {Degtyarenko},
  \citenamefont {Lee}, \citenamefont {Petrykin}, \citenamefont {Ovcharov},
  \citenamefont {Vasiliev}, \citenamefont {Kaul}, \citenamefont {Amelichev},
  \citenamefont {Kamenev}, \citenamefont {Molodyk},\ and\ \citenamefont
  {Samoilenkov}}]{Chepikov17SST}%
  \BibitemOpen
  \bibfield  {author} {\bibinfo {author} {\bibfnamefont {V.}~\bibnamefont
  {Chepikov}}, \bibinfo {author} {\bibfnamefont {N.}~\bibnamefont {Mineev}},
  \bibinfo {author} {\bibfnamefont {P.}~\bibnamefont {Degtyarenko}}, \bibinfo
  {author} {\bibfnamefont {S.}~\bibnamefont {Lee}}, \bibinfo {author}
  {\bibfnamefont {V.}~\bibnamefont {Petrykin}}, \bibinfo {author}
  {\bibfnamefont {A.}~\bibnamefont {Ovcharov}}, \bibinfo {author}
  {\bibfnamefont {A.}~\bibnamefont {Vasiliev}}, \bibinfo {author}
  {\bibfnamefont {A.}~\bibnamefont {Kaul}}, \bibinfo {author} {\bibfnamefont
  {V.}~\bibnamefont {Amelichev}}, \bibinfo {author} {\bibfnamefont
  {A.}~\bibnamefont {Kamenev}}, \bibinfo {author} {\bibfnamefont
  {A.}~\bibnamefont {Molodyk}}, \ and\ \bibinfo {author} {\bibfnamefont
  {S.}~\bibnamefont {Samoilenkov}},\ }\bibfield  {title} {\enquote {\bibinfo
  {title} {Introduction of {BaSnO$_3$} and {BaZrO$_3$} artificial pinning
  centres into {2G HTS} wires based on {PLD GdBCO} films. {Phase I} of the
  industrial {R$\&$D} programme at {SuperOx}},}\ }\href@noop {} {\bibfield
  {journal} {\bibinfo  {journal} {Supercond. Sci. Technol.}\ }\textbf {\bibinfo
  {volume} {30}} (\bibinfo {year} {2017})}\BibitemShut {NoStop}%
\bibitem [{\citenamefont {Iijima}\ \emph {et~al.}(2015)\citenamefont {Iijima},
  \citenamefont {Adachi}, \citenamefont {Fujita}, \citenamefont {Igarashi},
  \citenamefont {Kakimoto}, \citenamefont {Ohsugi}, \citenamefont {Nakamura},
  \citenamefont {Hanyu}, \citenamefont {Kikutake}, \citenamefont {Daibo},
  \citenamefont {Nagata}, \citenamefont {Tateno},\ and\ \citenamefont
  {Itoh}}]{Lijima15IES}%
  \BibitemOpen
  \bibfield  {author} {\bibinfo {author} {\bibfnamefont {Y.}~\bibnamefont
  {Iijima}}, \bibinfo {author} {\bibfnamefont {Y.}~\bibnamefont {Adachi}},
  \bibinfo {author} {\bibfnamefont {S.}~\bibnamefont {Fujita}}, \bibinfo
  {author} {\bibfnamefont {M.}~\bibnamefont {Igarashi}}, \bibinfo {author}
  {\bibfnamefont {K.}~\bibnamefont {Kakimoto}}, \bibinfo {author}
  {\bibfnamefont {M.}~\bibnamefont {Ohsugi}}, \bibinfo {author} {\bibfnamefont
  {N.}~\bibnamefont {Nakamura}}, \bibinfo {author} {\bibfnamefont
  {S.}~\bibnamefont {Hanyu}}, \bibinfo {author} {\bibfnamefont
  {R.}~\bibnamefont {Kikutake}}, \bibinfo {author} {\bibfnamefont
  {M.}~\bibnamefont {Daibo}}, \bibinfo {author} {\bibfnamefont
  {M.}~\bibnamefont {Nagata}}, \bibinfo {author} {\bibfnamefont
  {F.}~\bibnamefont {Tateno}}, \ and\ \bibinfo {author} {\bibfnamefont
  {M.}~\bibnamefont {Itoh}},\ }\bibfield  {title} {\enquote {\bibinfo {title}
  {Development for mass production of homogeneous {RE123} coated conductors by
  hot-wall {PLD} process on {IBAD} template technique},}\ }\href@noop {}
  {\bibfield  {journal} {\bibinfo  {journal} {IEEE Trans. Appl. Supercond.}\
  }\textbf {\bibinfo {volume} {25}} (\bibinfo {year} {2015})}\BibitemShut
  {NoStop}%
\bibitem [{\citenamefont {Lee}\ \emph {et~al.}(2014)\citenamefont {Lee},
  \citenamefont {Mean}, \citenamefont {Kim}, \citenamefont {Kim}, \citenamefont
  {Cheon}, \citenamefont {Kim}, \citenamefont {Park}, \citenamefont {Song},
  \citenamefont {Kim}, \citenamefont {Chung}, \citenamefont {Lee},\ and\
  \citenamefont {Moon}}]{Lee14IES}%
  \BibitemOpen
  \bibfield  {author} {\bibinfo {author} {\bibfnamefont {J.}~\bibnamefont
  {Lee}}, \bibinfo {author} {\bibfnamefont {B.}~\bibnamefont {Mean}}, \bibinfo
  {author} {\bibfnamefont {T.}~\bibnamefont {Kim}}, \bibinfo {author}
  {\bibfnamefont {Y.}~\bibnamefont {Kim}}, \bibinfo {author} {\bibfnamefont
  {K.}~\bibnamefont {Cheon}}, \bibinfo {author} {\bibfnamefont
  {T.}~\bibnamefont {Kim}}, \bibinfo {author} {\bibfnamefont {D.}~\bibnamefont
  {Park}}, \bibinfo {author} {\bibfnamefont {D.}~\bibnamefont {Song}}, \bibinfo
  {author} {\bibfnamefont {H.}~\bibnamefont {Kim}}, \bibinfo {author}
  {\bibfnamefont {W.}~\bibnamefont {Chung}}, \bibinfo {author} {\bibfnamefont
  {H.}~\bibnamefont {Lee}}, \ and\ \bibinfo {author} {\bibfnamefont
  {S.}~\bibnamefont {Moon}},\ }\bibfield  {title} {\enquote {\bibinfo {title}
  {Vision inspection methods for uniformity enhancement in long-length 2g hts
  wire production},}\ }\href@noop {} {\bibfield  {journal} {\bibinfo  {journal}
  {IEEE Trans. Appl. Supercond.}\ }\textbf {\bibinfo {volume} {24}} (\bibinfo
  {year} {2014})}\BibitemShut {NoStop}%
\bibitem [{\citenamefont {Rossi}\ \emph {et~al.}(2016)\citenamefont {Rossi},
  \citenamefont {Hu}, \citenamefont {Kametani}, \citenamefont {Abraimov},
  \citenamefont {Polyanskii}, \citenamefont {Jaroszynski},\ and\ \citenamefont
  {Larbalestier}}]{Rosii16SST}%
  \BibitemOpen
  \bibfield  {author} {\bibinfo {author} {\bibfnamefont {L.}~\bibnamefont
  {Rossi}}, \bibinfo {author} {\bibfnamefont {X.}~\bibnamefont {Hu}}, \bibinfo
  {author} {\bibfnamefont {F.}~\bibnamefont {Kametani}}, \bibinfo {author}
  {\bibfnamefont {D.}~\bibnamefont {Abraimov}}, \bibinfo {author}
  {\bibfnamefont {A.}~\bibnamefont {Polyanskii}}, \bibinfo {author}
  {\bibfnamefont {J.}~\bibnamefont {Jaroszynski}}, \ and\ \bibinfo {author}
  {\bibfnamefont {DC.}\ \bibnamefont {Larbalestier}},\ }\bibfield  {title}
  {\enquote {\bibinfo {title} {Sample and length-dependent variability of 77
  and 4.2 {K} properties in nominally identical {RE123} coated conductors},}\
  }\href@noop {} {\bibfield  {journal} {\bibinfo  {journal} {Supercond. Sci.
  Technol.}\ }\textbf {\bibinfo {volume} {29}} (\bibinfo {year}
  {2016})}\BibitemShut {NoStop}%
\bibitem [{\citenamefont {Xu}\ \emph {et~al.}(2017)\citenamefont {Xu},
  \citenamefont {Zhang}, \citenamefont {Gharahcheshmeh}, \citenamefont
  {Delgado}, \citenamefont {Khatri}, \citenamefont {Liu}, \citenamefont
  {Galstyan},\ and\ \citenamefont {Selvamanickam}}]{Xu17IES}%
  \BibitemOpen
  \bibfield  {author} {\bibinfo {author} {\bibfnamefont {A.}~\bibnamefont
  {Xu}}, \bibinfo {author} {\bibfnamefont {Y.}~\bibnamefont {Zhang}}, \bibinfo
  {author} {\bibfnamefont {M.}~\bibnamefont {Gharahcheshmeh}}, \bibinfo
  {author} {\bibfnamefont {L.}~\bibnamefont {Delgado}}, \bibinfo {author}
  {\bibfnamefont {N.}~\bibnamefont {Khatri}}, \bibinfo {author} {\bibfnamefont
  {Y.}~\bibnamefont {Liu}}, \bibinfo {author} {\bibfnamefont {E.}~\bibnamefont
  {Galstyan}}, \ and\ \bibinfo {author} {\bibfnamefont {V.}~\bibnamefont
  {Selvamanickam}},\ }\bibfield  {title} {\enquote {\bibinfo {title} {Relevant
  pinning for ab-plane {J(c)} enhancement of {MOCVD REBCO} coated
  conductors},}\ }\href@noop {} {\bibfield  {journal} {\bibinfo  {journal}
  {IEEE Trans. Appl. Supercond.}\ }\textbf {\bibinfo {volume} {27}} (\bibinfo
  {year} {2017})}\BibitemShut {NoStop}%
\bibitem [{\citenamefont {Lin}\ \emph {et~al.}(2017)\citenamefont {Lin},
  \citenamefont {Liu}, \citenamefont {Cui}, \citenamefont {Bai}, \citenamefont
  {Lu}, \citenamefont {Fan}, \citenamefont {Guo}, \citenamefont {Liu},\ and\
  \citenamefont {Cai}}]{Lin17AIM}%
  \BibitemOpen
  \bibfield  {author} {\bibinfo {author} {\bibfnamefont {JX.}\ \bibnamefont
  {Lin}}, \bibinfo {author} {\bibfnamefont {XM.}\ \bibnamefont {Liu}}, \bibinfo
  {author} {\bibfnamefont {CW.}\ \bibnamefont {Cui}}, \bibinfo {author}
  {\bibfnamefont {CY.}\ \bibnamefont {Bai}}, \bibinfo {author} {\bibfnamefont
  {YM.}\ \bibnamefont {Lu}}, \bibinfo {author} {\bibfnamefont {F.}~\bibnamefont
  {Fan}}, \bibinfo {author} {\bibfnamefont {YQ.}\ \bibnamefont {Guo}}, \bibinfo
  {author} {\bibfnamefont {ZY.}\ \bibnamefont {Liu}}, \ and\ \bibinfo {author}
  {\bibfnamefont {CB.}\ \bibnamefont {Cai}},\ }\bibfield  {title} {\enquote
  {\bibinfo {title} {A review of thickness-induced evolutions of microstructure
  and superconducting performance of {REBa$_2$Cu$_3$O$_7$}-delta coated
  conductor},}\ }\href@noop {} {\bibfield  {journal} {\bibinfo  {journal}
  {Advances in Manufacturing.}\ }\textbf {\bibinfo {volume} {5}},\ \bibinfo
  {pages} {165--176} (\bibinfo {year} {2017})}\BibitemShut {NoStop}%
\bibitem [{\citenamefont {Miura}\ \emph {et~al.}(2017)\citenamefont {Miura},
  \citenamefont {Tsuchiya}, \citenamefont {Yoshida}, \citenamefont {Ichino},
  \citenamefont {Awaji}, \citenamefont {Matsumoto}, \citenamefont {Ibi},\ and\
  \citenamefont {Izumi}}]{Miura17SST}%
  \BibitemOpen
  \bibfield  {author} {\bibinfo {author} {\bibfnamefont {S.}~\bibnamefont
  {Miura}}, \bibinfo {author} {\bibfnamefont {Y.}~\bibnamefont {Tsuchiya}},
  \bibinfo {author} {\bibfnamefont {Y.}~\bibnamefont {Yoshida}}, \bibinfo
  {author} {\bibfnamefont {Y.}~\bibnamefont {Ichino}}, \bibinfo {author}
  {\bibfnamefont {S.}~\bibnamefont {Awaji}}, \bibinfo {author} {\bibfnamefont
  {K.}~\bibnamefont {Matsumoto}}, \bibinfo {author} {\bibfnamefont
  {A.}~\bibnamefont {Ibi}}, \ and\ \bibinfo {author} {\bibfnamefont
  {T.}~\bibnamefont {Izumi}},\ }\bibfield  {title} {\enquote {\bibinfo {title}
  {Strong flux pinning at {4.2 K} in {SmBa$_2$Cu$_3$O$_y$} coated conductors
  with {BaHfO$_3$} nanorods controlled by low growth temperature},}\
  }\href@noop {} {\bibfield  {journal} {\bibinfo  {journal} {Supercond. Sci.
  Technol.}\ }\textbf {\bibinfo {volume} {30}} (\bibinfo {year}
  {2017})}\BibitemShut {NoStop}%
\bibitem [{\citenamefont {Ayai}\ \emph {et~al.}(2008)\citenamefont {Ayai},
  \citenamefont {Kobayashi}, \citenamefont {Kikuchi}, \citenamefont {Ishida},
  \citenamefont {Fujikami}, \citenamefont {Yamazaki}, \citenamefont {Yamade},
  \citenamefont {Tatamidani}, \citenamefont {Hayashi}, \citenamefont {Sato},
  \citenamefont {Kitaguchi}, \citenamefont {Kumakura}, \citenamefont {Osamura},
  \citenamefont {Shimoyama}, \citenamefont {Kamijyo},\ and\ \citenamefont
  {Fukumoto}}]{Ayai08PCS}%
  \BibitemOpen
  \bibfield  {author} {\bibinfo {author} {\bibfnamefont {N.}~\bibnamefont
  {Ayai}}, \bibinfo {author} {\bibfnamefont {S.}~\bibnamefont {Kobayashi}},
  \bibinfo {author} {\bibfnamefont {M.}~\bibnamefont {Kikuchi}}, \bibinfo
  {author} {\bibfnamefont {T.}~\bibnamefont {Ishida}}, \bibinfo {author}
  {\bibfnamefont {J.}~\bibnamefont {Fujikami}}, \bibinfo {author}
  {\bibfnamefont {K.}~\bibnamefont {Yamazaki}}, \bibinfo {author}
  {\bibfnamefont {S.}~\bibnamefont {Yamade}}, \bibinfo {author} {\bibfnamefont
  {K.}~\bibnamefont {Tatamidani}}, \bibinfo {author} {\bibfnamefont
  {K.}~\bibnamefont {Hayashi}}, \bibinfo {author} {\bibfnamefont
  {K.}~\bibnamefont {Sato}}, \bibinfo {author} {\bibfnamefont {H.}~\bibnamefont
  {Kitaguchi}}, \bibinfo {author} {\bibfnamefont {H.}~\bibnamefont {Kumakura}},
  \bibinfo {author} {\bibfnamefont {K.}~\bibnamefont {Osamura}}, \bibinfo
  {author} {\bibfnamefont {J.}~\bibnamefont {Shimoyama}}, \bibinfo {author}
  {\bibfnamefont {H.}~\bibnamefont {Kamijyo}}, \ and\ \bibinfo {author}
  {\bibfnamefont {Y.}~\bibnamefont {Fukumoto}},\ }\bibfield  {title} {\enquote
  {\bibinfo {title} {Progress in performance of {DI-BSCCO} family},}\
  }\href@noop {} {\bibfield  {journal} {\bibinfo  {journal} {Physica C-Superc.
  and its apl.}\ }\textbf {\bibinfo {volume} {468}},\ \bibinfo {pages}
  {1747--1752} (\bibinfo {year} {2008})}\BibitemShut {NoStop}%
\bibitem [{\citenamefont {Goyal}\ \emph {et~al.}(1997)\citenamefont {Goyal},
  \citenamefont {Norton}, \citenamefont {Kroeger}, \citenamefont {Christen},
  \citenamefont {Paranthaman}, \citenamefont {Specht}, \citenamefont {Budai},
  \citenamefont {He}, \citenamefont {Saffian}, \citenamefont {List},
  \citenamefont {Lee}, \citenamefont {Hatfield}, \citenamefont {Martin},
  \citenamefont {Klabunde}, \citenamefont {Mathis},\ and\ \citenamefont
  {Park}}]{Goyal97JMR}%
  \BibitemOpen
  \bibfield  {author} {\bibinfo {author} {\bibfnamefont {A.}~\bibnamefont
  {Goyal}}, \bibinfo {author} {\bibfnamefont {DP.}\ \bibnamefont {Norton}},
  \bibinfo {author} {\bibfnamefont {DM.}\ \bibnamefont {Kroeger}}, \bibinfo
  {author} {\bibfnamefont {DK.}\ \bibnamefont {Christen}}, \bibinfo {author}
  {\bibfnamefont {M.}~\bibnamefont {Paranthaman}}, \bibinfo {author}
  {\bibfnamefont {ED.}\ \bibnamefont {Specht}}, \bibinfo {author}
  {\bibfnamefont {JD.}\ \bibnamefont {Budai}}, \bibinfo {author} {\bibfnamefont
  {Q.}~\bibnamefont {He}}, \bibinfo {author} {\bibfnamefont {B.}~\bibnamefont
  {Saffian}}, \bibinfo {author} {\bibfnamefont {FA.}\ \bibnamefont {List}},
  \bibinfo {author} {\bibfnamefont {DF.}\ \bibnamefont {Lee}}, \bibinfo
  {author} {\bibfnamefont {E.}~\bibnamefont {Hatfield}}, \bibinfo {author}
  {\bibfnamefont {PM.}\ \bibnamefont {Martin}}, \bibinfo {author}
  {\bibfnamefont {CE.}\ \bibnamefont {Klabunde}}, \bibinfo {author}
  {\bibfnamefont {J.}~\bibnamefont {Mathis}}, \ and\ \bibinfo {author}
  {\bibfnamefont {C.}~\bibnamefont {Park}},\ }\bibfield  {title} {\enquote
  {\bibinfo {title} {Conductors with controlled grain boundaries: An approach
  to the next generation, high temperature superconducting wire},}\ }\href@noop
  {} {\bibfield  {journal} {\bibinfo  {journal} {J. Mar. Res.}\ }\textbf
  {\bibinfo {volume} {12}},\ \bibinfo {pages} {2924--2940} (\bibinfo {year}
  {1997})}\BibitemShut {NoStop}%
\bibitem [{\citenamefont {Pallecchi}\ \emph {et~al.}(2015)\citenamefont
  {Pallecchi}, \citenamefont {Malagoli},\ and\ \citenamefont
  {Putti}}]{Pallecchi15SST}%
  \BibitemOpen
  \bibfield  {author} {\bibinfo {author} {\bibfnamefont {I.}~\bibnamefont
  {Pallecchi}}, \bibinfo {author} {\bibfnamefont {M.~Eisterer~A.}\ \bibnamefont
  {Malagoli}}, \ and\ \bibinfo {author} {\bibfnamefont {M.}~\bibnamefont
  {Putti}},\ }\bibfield  {title} {\enquote {\bibinfo {title} {Application
  potential of {Fe-based} superconductors},}\ }\href@noop {} {\bibfield
  {journal} {\bibinfo  {journal} {Supercond. Sci. Technol.}\ }\textbf {\bibinfo
  {volume} {28}} (\bibinfo {year} {2015})}\BibitemShut {NoStop}%
\bibitem [{\citenamefont {Yi}\ \emph {et~al.}(2017)\citenamefont {Yi},
  \citenamefont {Wu},\ and\ \citenamefont {Sun}}]{Yi17APS}%
  \BibitemOpen
  \bibfield  {author} {\bibinfo {author} {\bibfnamefont {W.}~\bibnamefont
  {Yi}}, \bibinfo {author} {\bibfnamefont {Q.}~\bibnamefont {Wu}}, \ and\
  \bibinfo {author} {\bibfnamefont {LL.}\ \bibnamefont {Sun}},\ }\bibfield
  {title} {\enquote {\bibinfo {title} {Superconductivities of pressurized iron
  pnictide superconductors},}\ }\href@noop {} {\bibfield  {journal} {\bibinfo
  {journal} {Acta Physica Sinica.}\ }\textbf {\bibinfo {volume} {66}} (\bibinfo
  {year} {2017})}\BibitemShut {NoStop}%
\bibitem [{\citenamefont {Ma}\ \emph {et~al.}(2017)\citenamefont {Ma},
  \citenamefont {Ji}, \citenamefont {Hu}, \citenamefont {Gao}, \citenamefont
  {Li}, \citenamefont {Mu},\ and\ \citenamefont {Xie}}]{Ma17SST}%
  \BibitemOpen
  \bibfield  {author} {\bibinfo {author} {\bibfnamefont {YH.}\ \bibnamefont
  {Ma}}, \bibinfo {author} {\bibfnamefont {QC.}\ \bibnamefont {Ji}}, \bibinfo
  {author} {\bibfnamefont {KK.}\ \bibnamefont {Hu}}, \bibinfo {author}
  {\bibfnamefont {B.}~\bibnamefont {Gao}}, \bibinfo {author} {\bibfnamefont
  {W.}~\bibnamefont {Li}}, \bibinfo {author} {\bibfnamefont {G.}~\bibnamefont
  {Mu}}, \ and\ \bibinfo {author} {\bibfnamefont {XM.}\ \bibnamefont {Xie}},\
  }\bibfield  {title} {\enquote {\bibinfo {title} {Strong anisotropy effect in
  an iron-based superconductor {CaFe0: 882Co0: 118AsF}},}\ }\href@noop {}
  {\bibfield  {journal} {\bibinfo  {journal} {Supercond. Sci. Technol.}\
  }\textbf {\bibinfo {volume} {30}} (\bibinfo {year} {2017})}\BibitemShut
  {NoStop}%
\bibitem [{\citenamefont {Hecher}\ \emph {et~al.}(2018)\citenamefont {Hecher},
  \citenamefont {Ishida}, \citenamefont {Song}, \citenamefont {Ogino},
  \citenamefont {Iyo}, \citenamefont {Eisaki}, \citenamefont {Nakajima},
  \citenamefont {Kagerbauer},\ and\ \citenamefont {Eisterer}}]{Hecher18PRB}%
  \BibitemOpen
  \bibfield  {author} {\bibinfo {author} {\bibfnamefont {J.}~\bibnamefont
  {Hecher}}, \bibinfo {author} {\bibfnamefont {S.}~\bibnamefont {Ishida}},
  \bibinfo {author} {\bibfnamefont {D.}~\bibnamefont {Song}}, \bibinfo {author}
  {\bibfnamefont {H.}~\bibnamefont {Ogino}}, \bibinfo {author} {\bibfnamefont
  {A.}~\bibnamefont {Iyo}}, \bibinfo {author} {\bibfnamefont {H.}~\bibnamefont
  {Eisaki}}, \bibinfo {author} {\bibfnamefont {M.}~\bibnamefont {Nakajima}},
  \bibinfo {author} {\bibfnamefont {D.}~\bibnamefont {Kagerbauer}}, \ and\
  \bibinfo {author} {\bibfnamefont {M.}~\bibnamefont {Eisterer}},\ }\bibfield
  {title} {\enquote {\bibinfo {title} {Direct observation of in-plane
  anisotropy of the superconducting critical current density in
  {Ba(Fe1-xCox)(2)As-2} crystals},}\ }\href@noop {} {\bibfield  {journal}
  {\bibinfo  {journal} {Phys. Rev. B}\ }\textbf {\bibinfo {volume} {97}}
  (\bibinfo {year} {2018})}\BibitemShut {NoStop}%
\bibitem [{HTS()}]{HTSdatabase}%
  \BibitemOpen
  \href@noop {} {}\bibinfo {note} {A high-temperature superconducting (HTS)
  wire critical current database. https://figshare.com/collections/A high
  temperature superconducting HTS wire critical current
  database/2861821}\BibitemShut {NoStop}%
\bibitem [{\citenamefont {Pardo}\ \emph {et~al.}(2011)\citenamefont {Pardo},
  \citenamefont {Vojenciak}, \citenamefont {Gomory},\ and\ \citenamefont
  {Souc}}]{Pardo11SST}%
  \BibitemOpen
  \bibfield  {author} {\bibinfo {author} {\bibfnamefont {E.}~\bibnamefont
  {Pardo}}, \bibinfo {author} {\bibfnamefont {M.}~\bibnamefont {Vojenciak}},
  \bibinfo {author} {\bibfnamefont {F.}~\bibnamefont {Gomory}}, \ and\ \bibinfo
  {author} {\bibfnamefont {J.}~\bibnamefont {Souc}},\ }\bibfield  {title}
  {\enquote {\bibinfo {title} {Low-magnetic-field dependence and anisotropy of
  the critical current density in coated conductors},}\ }\href@noop {}
  {\bibfield  {journal} {\bibinfo  {journal} {Supercond. Sci. Technol.}\
  }\textbf {\bibinfo {volume} {24}},\ \bibinfo {pages} {10} (\bibinfo {year}
  {2011})}\BibitemShut {NoStop}%
\bibitem [{\citenamefont {Zermeno}\ \emph {et~al.}(2016)\citenamefont
  {Zermeno}, \citenamefont {Quaiyum},\ and\ \citenamefont
  {Grilli}}]{Zermeno16IES}%
  \BibitemOpen
  \bibfield  {author} {\bibinfo {author} {\bibfnamefont {VMR.}\ \bibnamefont
  {Zermeno}}, \bibinfo {author} {\bibfnamefont {S.}~\bibnamefont {Quaiyum}}, \
  and\ \bibinfo {author} {\bibfnamefont {F.}~\bibnamefont {Grilli}},\
  }\bibfield  {title} {\enquote {\bibinfo {title} {Open-source codes for
  computing the critical current of superconducting devices},}\ }\href@noop {}
  {\bibfield  {journal} {\bibinfo  {journal} {IEEE Trans. Appl. Supercond.}\
  }\textbf {\bibinfo {volume} {26}} (\bibinfo {year} {2016})}\BibitemShut
  {NoStop}%
\bibitem [{\citenamefont {Pardo}\ and\ \citenamefont
  {Kapolka}(2017{\natexlab{a}})}]{Pardo17JCP}%
  \BibitemOpen
  \bibfield  {author} {\bibinfo {author} {\bibfnamefont {E.}~\bibnamefont
  {Pardo}}\ and\ \bibinfo {author} {\bibfnamefont {M.}~\bibnamefont
  {Kapolka}},\ }\bibfield  {title} {\enquote {\bibinfo {title} {{3D}
  computation of non-linear eddy currents: variational method and
  superconducting cubic bulk},}\ }\href@noop {} {\bibfield  {journal} {\bibinfo
   {journal} {J. Comput. Phys.}\ } (\bibinfo {year}
  {2017}{\natexlab{a}})}\BibitemShut {NoStop}%
\bibitem [{\citenamefont {Bad{\'\i}a-Maj{\'o}s}\ and\ \citenamefont
  {L{\'o}pez}(2015)}]{badia15SST}%
  \BibitemOpen
  \bibfield  {author} {\bibinfo {author} {\bibfnamefont {A.}~\bibnamefont
  {Bad{\'\i}a-Maj{\'o}s}}\ and\ \bibinfo {author} {\bibfnamefont
  {C.}~\bibnamefont {L{\'o}pez}},\ }\bibfield  {title} {\enquote {\bibinfo
  {title} {Modelling current voltage characteristics of practical
  superconductors},}\ }\href@noop {} {\bibfield  {journal} {\bibinfo  {journal}
  {Supercond. Sci. Technol.}\ }\textbf {\bibinfo {volume} {28}},\ \bibinfo
  {pages} {024003} (\bibinfo {year} {2015})}\BibitemShut {NoStop}%
\bibitem [{\citenamefont {Pardo}\ and\ \citenamefont
  {Kapolka}(2017{\natexlab{b}})}]{Pardo17SST}%
  \BibitemOpen
  \bibfield  {author} {\bibinfo {author} {\bibfnamefont {E.}~\bibnamefont
  {Pardo}}\ and\ \bibinfo {author} {\bibfnamefont {M.}~\bibnamefont
  {Kapolka}},\ }\bibfield  {title} {\enquote {\bibinfo {title} {{3D}
  magnetization currents, magnetization loop, and saturation field in
  superconducting rectangular prisms},}\ }\href@noop {} {\bibfield  {journal}
  {\bibinfo  {journal} {Supercond. Sci. Technol.}\ }\textbf {\bibinfo {volume}
  {30}},\ \bibinfo {pages} {11} (\bibinfo {year}
  {2017}{\natexlab{b}})}\BibitemShut {NoStop}%
\bibitem [{\citenamefont {Bean}(1962)}]{bean62PRL}%
  \BibitemOpen
  \bibfield  {author} {\bibinfo {author} {\bibfnamefont {C.~P.}\ \bibnamefont
  {Bean}},\ }\bibfield  {title} {\enquote {\bibinfo {title} {Magnetizatin of
  hard superconductors},}\ }\href@noop {} {\bibfield  {journal} {\bibinfo
  {journal} {Phys. Rev. Lett.}\ }\textbf {\bibinfo {volume} {8}},\ \bibinfo
  {pages} {250--253} (\bibinfo {year} {1962})}\BibitemShut {NoStop}%
\bibitem [{\citenamefont {London}(1963)}]{london63PhL}%
  \BibitemOpen
  \bibfield  {author} {\bibinfo {author} {\bibfnamefont {H.}~\bibnamefont
  {London}},\ }\bibfield  {title} {\enquote {\bibinfo {title} {Alternating
  current losses in superconductors of the second kind},}\ }\href@noop {}
  {\bibfield  {journal} {\bibinfo  {journal} {Phys. Letters}\ }\textbf
  {\bibinfo {volume} {6}},\ \bibinfo {pages} {162--165} (\bibinfo {year}
  {1963})}\BibitemShut {NoStop}%
\bibitem [{\citenamefont {Perezgonzalez}\ and\ \citenamefont
  {Clem}(1985)}]{Perezgonzalez85JAP}%
  \BibitemOpen
  \bibfield  {author} {\bibinfo {author} {\bibfnamefont {A.}~\bibnamefont
  {Perezgonzalez}}\ and\ \bibinfo {author} {\bibfnamefont {J.~R.}\ \bibnamefont
  {Clem}},\ }\bibfield  {title} {\enquote {\bibinfo {title} {Magnetic response
  of {type-II} superconductors subjected to large-amplitude parallel
  magnetic-fields varying in both magnitude and direction},}\ }\href@noop {}
  {\bibfield  {journal} {\bibinfo  {journal} {J. Appl. Phys.}\ }\textbf
  {\bibinfo {volume} {58}},\ \bibinfo {pages} {4326--4335} (\bibinfo {year}
  {1985})}\BibitemShut {NoStop}%
\bibitem [{\citenamefont {Bad{\'\i}a-Maj{\'o}s}\ and\ \citenamefont
  {L{\'o}pez}(2012)}]{Badia12SST}%
  \BibitemOpen
  \bibfield  {author} {\bibinfo {author} {\bibfnamefont {A.}~\bibnamefont
  {Bad{\'\i}a-Maj{\'o}s}}\ and\ \bibinfo {author} {\bibfnamefont
  {C.}~\bibnamefont {L{\'o}pez}},\ }\bibfield  {title} {\enquote {\bibinfo
  {title} {Electromagnetics close beyond the critical state: thermodynamic
  prospect},}\ }\href@noop {} {\bibfield  {journal} {\bibinfo  {journal}
  {Supercond. Sci. Technol.}\ }\textbf {\bibinfo {volume} {25}},\ \bibinfo
  {pages} {104004} (\bibinfo {year} {2012})}\BibitemShut {NoStop}%
\bibitem [{\citenamefont {Ciftja}(2011)}]{ciftja11PLA}%
  \BibitemOpen
  \bibfield  {author} {\bibinfo {author} {\bibfnamefont {O.}~\bibnamefont
  {Ciftja}},\ }\bibfield  {title} {\enquote {\bibinfo {title} {Coulomb
  self-energy of a uniformly charged three-dimensional cube},}\ }\href@noop {}
  {\bibfield  {journal} {\bibinfo  {journal} {Physics Letters A}\ }\textbf
  {\bibinfo {volume} {375}},\ \bibinfo {pages} {766--767} (\bibinfo {year}
  {2011})}\BibitemShut {NoStop}%
\bibitem [{\citenamefont {Ciftja}(2010)}]{ciftja10PLA}%
  \BibitemOpen
  \bibfield  {author} {\bibinfo {author} {\bibfnamefont {O.}~\bibnamefont
  {Ciftja}},\ }\bibfield  {title} {\enquote {\bibinfo {title} {Coulomb
  self-energy and electrostatic potential of a uniformly charged square in two
  dimensions},}\ }\href@noop {} {\bibfield  {journal} {\bibinfo  {journal}
  {Physics Letters A}\ }\textbf {\bibinfo {volume} {374}},\ \bibinfo {pages}
  {981--983} (\bibinfo {year} {2010})}\BibitemShut {NoStop}%
\end{thebibliography}
